\documentclass[linenumbers]{aastex631}

\newcommand*\rot{\rotatebox{90}}
\usepackage[caption=false]{subfig}
\usepackage{tikz}
\usetikzlibrary{shapes.geometric, arrows}
\usetikzlibrary{fit}

\tikzstyle{io} = [trapezium, 
trapezium stretches=true, 
trapezium left angle=70, 
trapezium right angle=110, 
minimum width=3cm, 
minimum height=1cm, text centered, 
draw=black, fill=blue!30]

\tikzstyle{process} = [rectangle, 
minimum width=3cm, 
minimum height=1cm, 
text centered, 
text width=3cm, 
draw=black, 
fill=orange!30]

\tikzstyle{decision} = [diamond, 
minimum width=3cm, 
minimum height=1cm, 
text centered, 
draw=black, 
fill=green!30]
\tikzstyle{arrow} = [thick,->,>=stealth]

\begin{document}

\title{Millisecond Cadence Radio Frequency Interference Filters}

\author[0000-0002-3354-3859]{Joseph W. Kania}
\affiliation{Jodrell Bank Centre for Astrophysics, \\
Department of Physics and Astronomy, The University of Manchester,\\
Oxford Road, Manchester M13 9PL, United Kingdom
}
\affiliation{Department of Physics and Astronomy, \\
West Virginia University, \\
PO Box 6315, Morgantown, WV 26506, USA}
\affiliation{Center for Gravitational Waves and Cosmology, \\
West Virginia University, \\
Chestnut Ridge Research Building, \\
Morgantown, WV 26505, USA}

\author[0000-0003-3772-2798]{Kevin Bandura}
\affiliation{Lane Department of Computer Science and Electrical Engineering, 
\\
1220 Evansdale Drive, PO Box 6109, \\
Morgantown, WV 26506, USA}
\affiliation{Center for Gravitational Waves and Cosmology, \\
West Virginia University, \\
Chestnut Ridge Research Building, \\
Morgantown, WV 26505, USA}

\author[0000-0003-1301-966X]{Duncan R. Lorimer}
\affiliation{Department of Physics and Astronomy, \\
West Virginia University, \\
PO Box 6315, Morgantown, WV 26506, USA}
\affiliation{Center for Gravitational Waves and Cosmology, \\
West Virginia University, \\
Chestnut Ridge Research Building, \\
Morgantown, WV 26505, USA}

\author{Richard Prestage}
\altaffiliation{Deceased}



\begin{abstract}

Radio Frequency Interference (RFI) greatly reduces sensitivity of radio observations  to astrophysical signals and creates false positive candidates in
searches for radio transients. Real signals are missed while considerable computational and human resources are needed to remove RFI candidates. Effective RFI removal is vital to carry out successful searches for fast radio bursts
and pulsars. Mitigation techniques that excise RFI on short timescales account for a changing radio frequency and pulse environments. We
evaluate the effectiveness of three filters, as well as a novel composite of the three, that excises RFI at the cadence that the data
is recorded. Each of these filters operates in a different domain and thus excises as a different RFI morphology. The composite filter removes RFI not accessible to other filtering methods. We analyze the performance of these four filters in three different situations: (i) synthetic pulses in Gaussian noise; (ii) synthetic pulses injected into observed spectra; (iii) test observations of four pulsars. From these tests, we gain insight into how the filters affect both the pulse and the noise level. This allows us to
outline which and how the filters should be used based on the RFI present and the characteristics of the source signal. 
These
filters both increase sensitivity and reduce the number of false positives.
By flagging less than 5\% of the spectrum, we demonstrate a 53\% increase in detected pulses and 34\% decrease
in the number of RFI candidates relative to the  Heimdall-Fetch-Your search pipeline. 
\end{abstract}

\keywords{Light pollution (2318) --- Radio Pulsars (1353) --- Radio transient sources (2008) --- Astronomy software (1855)}

\bigskip

\section{Introduction}

Radio frequency interference (RFI) refers to anthropomorphic radio energy received by radio telescopes. These signals can be significantly brighter than those of astrophysical origin and therefore a major impediment to the science goals of an observation. The effects of RFI are of grave concern for transient radio astronomy, where some events do not necessarily repeat. Excellent examples of such source classes are fast radio bursts \citep[FRBs;][]{Lorimer-2007} and a subset of the pulsar population, the rotating radio transients \citep[RRATs;][]{McLaughlin-2006}. For both  classes, where there is a limited number of pulses, RFI can often mean that vital information remains irretrievable if individual pulses are missed. This contrasts with persistent sources such as pulsars, where longer observations can increase the signal-to-noise ratio (S/N). Effective RFI mitigation is still important for these sources because it can save valuable telescope time and, as is well known for pulsar searches \citep[see, e.g.,][]{Burke_Spolaor_2011, iqrm}, can reduce the number of false positive candidates that 
ultimately limit the overall sensitivity of a survey. Currently, there is significant interest in transient astrophysical sources. A larger sample of FRBs, for example, will allow their use as cosmological probes \citep[][]{Walters-2018} and permit better constraints on progenitor models, which in turn will provide insights into the physics of the FRB emission mechanism \citep{frb_overview, Bhandari}. Pulsars provide insights into different states of matter, the interstellar medium, galactic dynamics, general relativity, and gravitational waves \citep{pulsar_handbook}. 

RFI can originate from anything that has an electrical current. This means there is an expansive variety of RFI sources that differ between telescope locations, observing times, telescope pointings, and observing frequencies. Different sources produce RFI with different temporal and spectral morphologies. Emission from spark plugs or bad connections in power lines will cause a broadband response across all channels. Digital communications will be limited to a few channels, and the power (for our purposes) will be effectively randomly distributed. Transitional radar will likewise be limited to a few channels, but will be perfectly periodic. We also encounter so-called ``salt-and-pepper noise'' \citep{gonzalez}, i.e., bright random RFI that does not have a clear source.

RFI mitigation is a multi-pronged approach that, ideally, takes place at every step in the signal path \citep{Baan:2010y6, An_2017, Baan_2019}. Spectrum management attempts to reduce conflict between human-made signals and frequencies of astronomical interest \citep[][]{Pankonin-Spectrum-management}. Increasing spectrum usage \citep[][]{liszt_2019} and larger telescope bandwidths \citep[][]{Deng_2020, Velazco-2019} are leading to greater conflict. Radio observatories are typically built in isolated locations, protecting them from household electronics (e.g., microwave ovens) or television broadcasts. These locations are often legally protected by radio-quiet zones, which limit the use of transmitters around the instruments. This isolation does not protect them from satellites or aircraft radar, the latter being bright enough that radio telescopes can see them in other star systems \citep[][]{seti-2015}. Telescopes are now also being designed to be less sensitive to RFI, for example the Green Bank Telescope (GBT) has an off-axis feed \citep[][]{Norrod-GBT-design}. Persistently bad frequencies are often blocked with band-stop filters. After the signal is digitized, several techniques can be employed to flag undesirable sections of data. Criteria include usually bright regions \citep[][]{greenburst, Offranga-2008}, statistical tests for Gaussianity \citep[][]{Ransom-2011, Gary-2007, iqrm}, and projecting the dynamic spectra into a new basis where the RFI can be removed more easily \citep[][]{Mann-2021, Kocz-2012}. Interferometers can use adaptive pointing to steer their beams away from RFI sources \citep[][]{adaptive_steering}.

Most of the above post-detection techniques focus on using a single filter on sections of dynamic spectra, with notable exceptions \citep[see, e.g.,][]{Lazarus_2015}. In this paper, we evaluate three filters. One to remove unusually bright points in the dynamic spectra, another to remove narrow-band periodic signals, and finally, one to remove broad-band signals. Each of these are based on filters found in the literature, where their effectiveness on removing RFI or finding previously unseen astronomical signals has already been shown. We modify these filters in an attempt better preserve burst characteristics or more flexible in removing faint RFI. These filters could be useful in obtaining better pulse statistics, removing short timescale RFI while keeping more of the astronomical signal than methods that flag data with higher cadences. We show that by flagging a small amount of data, these filters can drastically decrease the overall noise level. However, the filtering effects on pulse statistics is complicated, depending on the statistic of interest, the filter, and pulse characteristics. 
Each of the three filters can filter a unique RFI morphology. 
The first two filters remove narrowband signals, allowing the last
filter to better remove broadband signals. This approach allows us
to improve upon the broadband removal technique as described by \cite{Eatough-2009}.
These filters are a part of \textsc{JESS} \citep[][]{jess}\footnote{\url{https://github.com/josephwkania/jess}}, a \textsc{Python} package we are writing that includes many RFI filters as well as utilities to better understand the quality of time domain radio astronomy data. \textsc{jess} is written in \textsc{Python 3} \citep[][]{python3} and uses
\textsc{CuPy}  \citep[][]{cupy}, \textsc{Matplotlib} \citep[][]{matplotlib}, \textsc{NumPy} \citep[][]{numpy}, 
\textsc{Rich} \citep[][]{rich}, \textsc{scikit-learn} \citep[][]{scikit-learn, sklearn_api}, \textsc{SciPy} \citep[][]{2020SciPy-NMeth}, \textsc{Sphinx}, and \textsc{Your} \citep[][]{your}. The rest of this paper is organized as follows. In Section~\ref{sec:filters} we  describe the three filters. In Section~\ref{sec:filter_effects} we discuss the effects of our filters on synthetic pulses in ideal and real noise and show the effectiveness of the filters on four pulsars. In Section~\ref{sec:performance} we discuss the performance of the filters and how they can be extended to multibeam systems. Finally, in Section~\ref{sec:conclude}, we summarize our findings and speculate on future improvements.

\section{Three High Cadence Filters}\label{sec:filters}

We present three filters in this work. These three filters improve on existing filtering concepts by attempting to make them more robust to different burst morphologies or RFI types; the final filter will be a combination of the three. The first is a median absolute deviation filter (see below for a definition) that operates in the time domain, the second is a Fourier domain median absolute deviation filter that removes bright narrow-band periodic signals. Finally, we use a high-pass filter that removes wideband RFI. The first two filters work by calculating a robust to outliers measure of scale and a central value in their respective domains. These two values are then used to find outliers (i.e., data points that are significantly far from the central value, here median. Outliers are then replaced with values that are less destructive to the observation. The high-pass filter removes slowly varying signals that are present across most of the frequencies at a given sampling time. If many sources of RFI  are present, these three filters can be used together to remove RFI that are more prominent in their respective filtering domains, synergistically cleaning the observation. When we construct the composite filter, the two MAD filters will remove bright outliers that will allow us to use the high-pass filter to remove broadband RFI.

Astronomical signals are expected to produce Gaussian distributed voltages \citep[][]{Gary-2007}. These complex samples are square-law detected (i.e., multiplied by their complex conjugate and summed). This results in $\chi^2$ distributed power samples. As large numbers of power samples are averaged, the resulting distribution looks more Gaussian, as expected by the central limit theorem. In this paper, we excised outliers that are above four sigma away from the 
median.

\subsection{Robust Measures of Scale}

To find outliers, we require both a central location and a measure of scale. The central location tells us the expected value of our data and is a useful replacement value for bad data, there we will use the median. The measure of scale tells us the dispersion of the distribution, standard deviation being the most common measure of scale. RFI can be orders of magnitude brighter than the signal of interest, so we must use estimators that are robust to large outliers. There are many possible choices that can be made; see \citet[][]{fridman-measures-of-scale} for an overview from the radio astronomy perspective and \citet[][]{robustregression, maronna2006robust} for overviews of the field. We choose Median Absolute Deviation (MAD) because of its overall simplicity and robustness to RFI. For a set of samples, $X$, we define,
\begin{equation}
   \text{MAD} = \text{median}(|X_i - \text{median}(X)|).
	\label{eq:mad}
\end{equation}
\citet[][]{Rousseeuw} calculates the coefficient $\hat{k}=1.4826$ that relates MAD to standard deviation, $\sigma$, as simply
\begin{equation}
   \sigma = \hat{k}\cdot \text{MAD}.
	\label{eq:mad-to-sigma}
\end{equation}
As we can infer from Eq.~\ref{eq:mad}, MAD has a ``breakdown point'' of 0.5, meaning that half the samples can be RFI corrupted and MAD will still give the correct estimate of the noise. A breakdown point of 0.5 is, in general, the best we can achieve \citep[][]{maronna2006robust}.  This stability gives MAD an advantage over other robust to outliers measures of scale
that use data higher in the distribution. We considered other robust measures: interquartile range \citep{Rousseeuw};  estimators $S_n$ and $Q_n$ and biweight midvariance \citep{biweight} but they either
did not provide as stable results as MAD, had high computation cost, or proved difficult to implement.

\subsection{Time Domain MAD}
\label{sec:mad_filter}
The time domain MAD filter works to remove bright points in the dynamic spectra while preserving
the astrophysical signal. Implementations of this filter have been used successfully in several experiments.
Several examples of these can be found in the literature \citep[see, e.g.,][]{chime_frb_2018, MAD-Buch, ramey, apertif, Rajwade_332}, although we note that not all of these use robust statistics, and that some of these implementations are bespoke
to a particular instrument. Our aim is a more generic implementation that will work on high time resolution
data from a variety of telescopes and sources, with the ability to deal with various formats: \textsc{psrfits} \citep{psrfits}, \textsc{sigproc} filterbanks \citep{sigproc}, or \textsc{NumPy} arrays \citep{numpy}. 

\begin{figure*}
   \includegraphics[width=\textwidth]{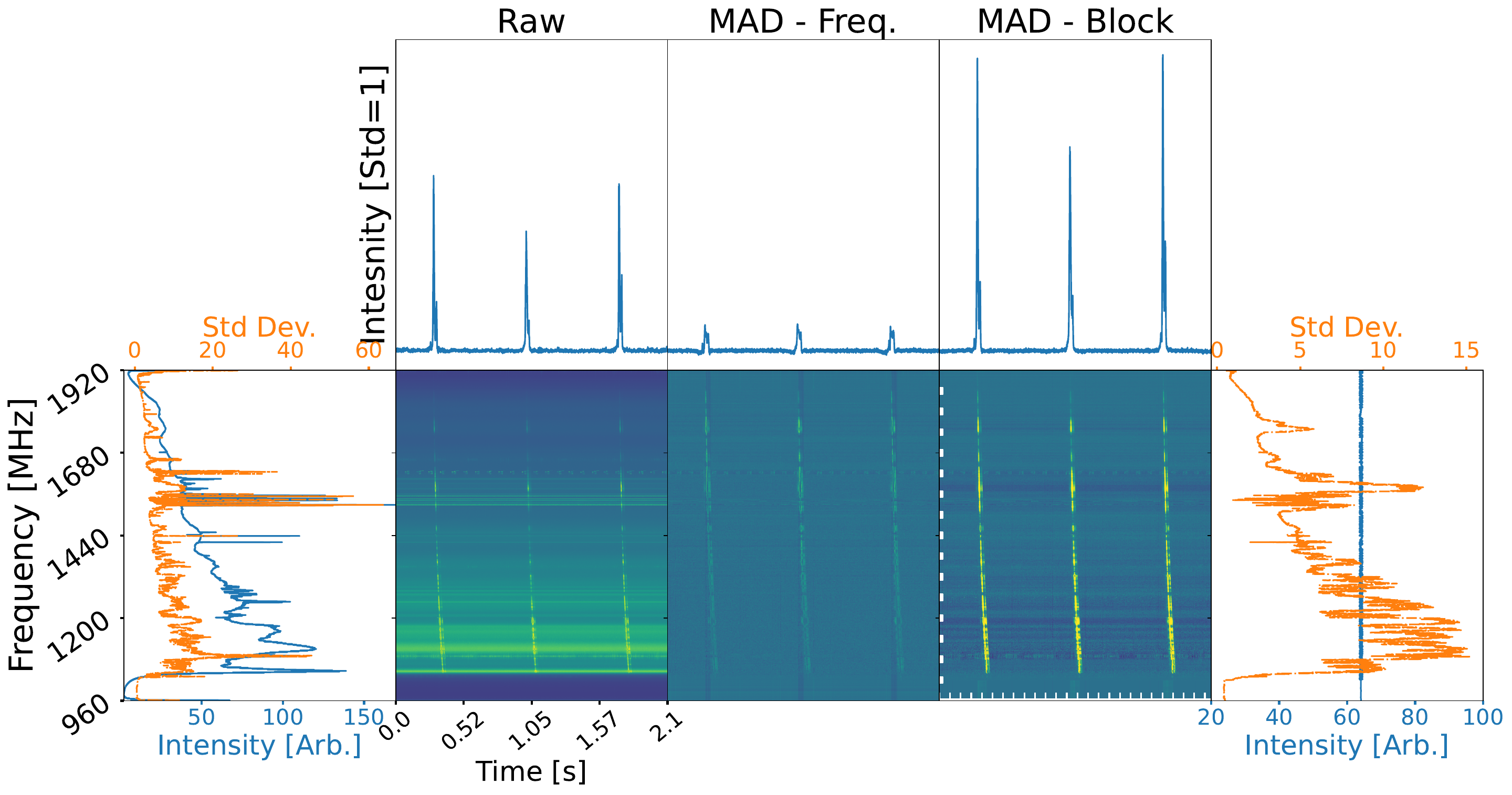}
    \caption{GREENBURST dynamic spectra with three pulses from PSR~B0329+54. The top row shows three time series de-dispersed at the pulsar's DM, these plots share the 
    same scale. 
    The bottom  
    left plot shows the power (blue) and standard deviation (orange)
    of unfiltered dynamic spectra. We see very bright spikes in both metrics between 1.5 and
    1.6~GHz from satellite communications and radar. Some channels have a standard deviation close to 60, adding  a correlated signal that is ten
    times larger than the thermal background. 
    The next three plots show
    dynamic spectra that are: (i) unfiltered; (ii) filtered using MAD across all 
    frequencies (13.2\% flagged); (iii) filtered using MAD on frequency-time blocks (7.2\% flagged). The range of the 
    dynamic spectra are set at +5$\sigma$ to --3$\sigma$ around the median (with the standard deviation, $\sigma$, calculated via MAD in Eq.~\ref{eq:mad-to-sigma}).
    The MAD filter across all frequencies significantly reduces the pulse intensity. Using MAD on frequency-time blocks of 256 channels by 32 samples helps preserve the pulse. The block 
    shape is indicated by white spaces, the time axis blocks are difficult to 
    distinguish, so every tenth block is longer.  The rightmost plot shows the 
    channel statistics after cleaning\. The band-pass has been flattened and changes in standard deviation are now smooth. 
    }
    \label{fig:green_burst}
\end{figure*}

Fig.~\ref{fig:green_burst} shows a section of dynamic spectra that includes three bursts
from PSR~B0329+54 recorded using the GREENBURST data acquisition system, a real time single-pulse search experiment using the GBT \citep[][]{greenburst_sunis, greenburst}. While GREENBURST is designed to search for transients at 1.4~GHz, 
the filtering performance here is typical for other telescopes and systems.
We want to remove the bright terrestrial sources while preserving the fidelity of the astrophysical pulses. This can be challenging because bright astrophysical pulses can be comparable in intensity to RFI, as shown in the left most dynamic spectra of Fig.~\ref{fig:green_burst} Bottom row under Raw. There the vertical pulses are as bright as the horizontal RFI. A straightforward strategy for outlier filtering is to pick the time or frequency axis, calculate the 
center, and scale  across all times or frequencies. This approach suffers from two problems. Firstly, when the pulse occupies a small fraction of the area used to calculate the measure of scale, the measure of scale does not reflect the pulse brightness, and large sections of the pulse are removed by the filter. This case is shown in the center of Fig.~\ref{fig:green_burst}. \citet[][]{Rajwade_332} discuss how this causes the filter to have a selection effect against extremely bright or narrow pules. The second issue arises from changes in the noise level across the band-pass, as shown in the right and left most plots of Fig.~\ref{fig:green_burst}. For the GBT, this means higher noise levels for lower frequencies. This broadband change is due to optics and radio sky temperature \citep[][]{gbt_man,gbt_performances}. Band-stop filters and band roll-off can produce areas of near zero noise. A section of low sensitivity is shown below 1~GHz in Fig.~\ref{fig:green_burst}. Together, these effects mean that a measure of scale calculated across all frequencies may be a poor fit for a smaller range of frequencies. The filter will be too aggressive in areas of higher noise and not effective in areas of low noise. We need to choose sections of dynamic spectra which are large compared to the RFI, so the percentage of RFI is below the breakdown point, but
small compared to the pulse or changes in sensitivity.  Such a block of data will provide robust statistics to remove and replace the RFI, while the
pulse affects the statistics and is preserved. 
We will operate on two-dimensional sub-blocks of the dynamic spectra, using blocks of 256 channels by 32 time samples,
as shown in the middle right plot of Fig.~\ref{fig:green_burst}. Our filters accept this as a user 
defined input. 

The analogue receiver chain can add substantial frequency structure to the 
dynamic spectra. An example of this structure is shown in the bottom left plot of 
Fig.~\ref{fig:green_burst}, labeled as Intensity.
To perform outlier tests on two-dimensional chunks of dynamic spectra, we must remove this structure, while preserving the
 outlying RFI. If done correctly, this separates the RFI from the shape of the frequency response across the observing band. The RFI
will then be removed as a non-Gaussian outlier.

\begin{figure}
\centering
   \includegraphics[width=0.6\columnwidth]{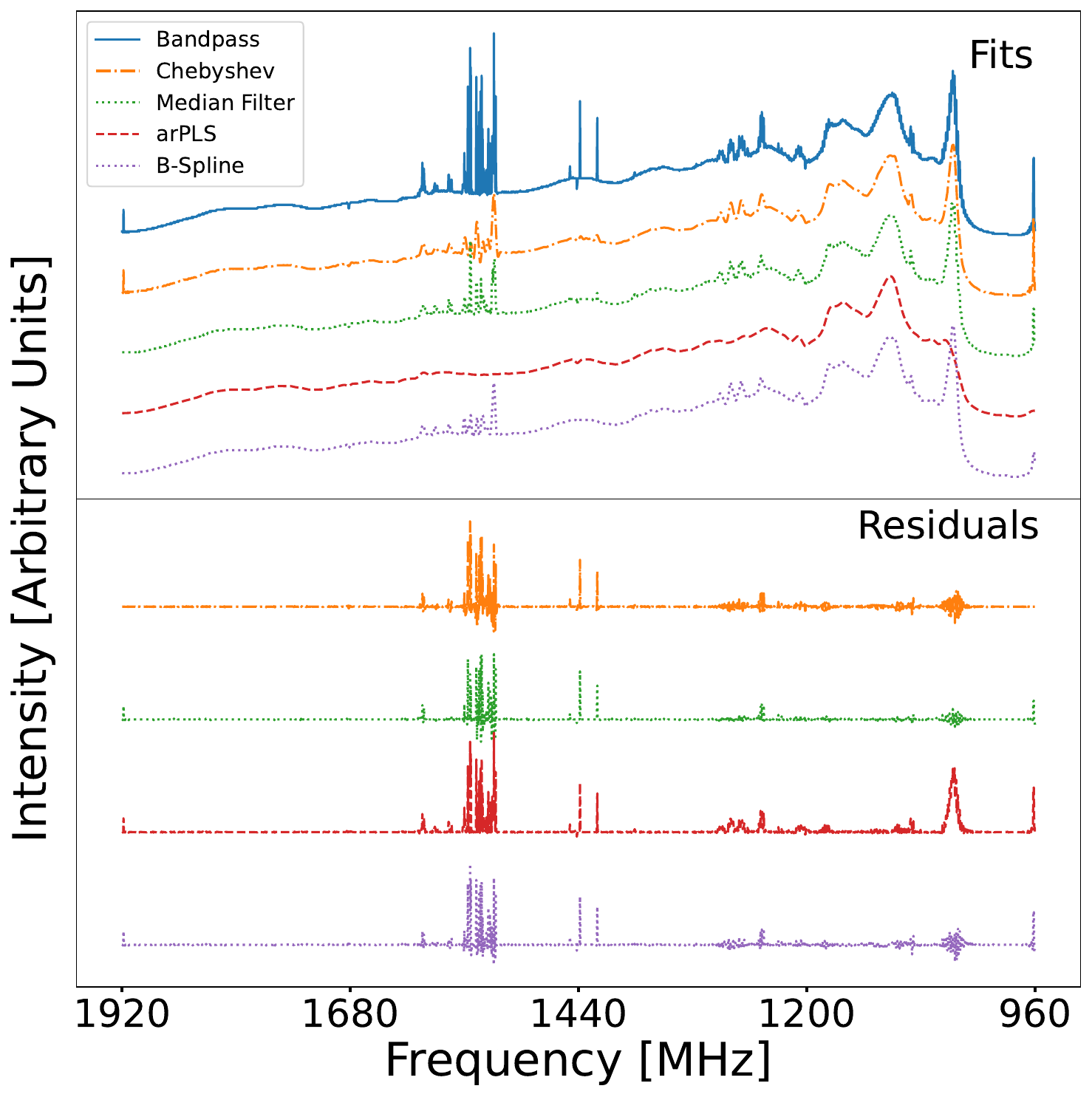}
    \caption{The frequency structure of a typical GREENBURST observation, computed by taking the median along the time axis.
    The median is a robust to outliers estimator, however some channels (around frequencies 1550~MHz and 1030~MHz) are still significantly affected by
    RFI. The four offset curves are of robust to outlying channels fitting schemes we investigated.}
    \label{fig:fit}
\end{figure} 

We considered several fitting methodologies, some of which are shown in Fig.~\ref{fig:fit}. The non-linear nature of RFI precludes filters such as high-pass
or Savitzky–Golay \citep[][]{savitzky}. First considered is iterative
Chebyshev polynomial \citep[][]{chebyshev} fitting, where an initial fit to all points is used to exclude large outliers
for a subsequent fit. Next to discuss are Basis splines \citep{deboor} fitted with a Huber loss function, where nearby points are
fitted with a squared error loss and outlier points are fitted with a linear loss \citep{huber}.
We also investigated the asymmetrically re-weighted penalized least squares algorithm
\citep[arPLS;][]{baek_2015, zeng_2020}. While arPLS gives excellent results, its iterative matrix
algebra is much slower than other filters.
For this work, we settled on a rolling median filter. The kernel size is an adjustable parameter, which we provide a reasonable default of seven channels and time samples, which is used in this work. As shown by the residual plot in Fig.~\ref{fig:fit}, this is a robust to outliers fit, so it should port well to other
receiver systems. 
The median
filter has the advantage of being fast as well as being stable, where the polynomial and arPLS will sometimes
diverge, leading to a bad fit. Iterating the median filter to produce a cascade filter similar to \citet{galassi_2009} did not
produce any meaningful improvement.

Once the dynamic spectrum has been de-trended in frequency and time, we split the data into subbands of $n$ channels. 
As shown in the left most box of Fig.~\ref{fig:green_burst}, the noise level can change over band-pass. 
Calculating the measure of scale per subband is a constant approximation for changes in dynamic spectra noise level across 
the band-pass. 
We then calculate the MAD and median along the time axis of each of the subbands. We then use the rolling median filter
to find the median of median and the median of the MADs along the time axis. This procedure is a recursive approximation to finding the median and
MAD of two-dimensional blocks of spectra, while relaxing the requirement that the dynamic spectra dimensions are  an integer
number of sub-block dimensions. We perform two iterations. The first iteration de-trends the data using the median filter on the channel and
time medians, preserving any RFI spikes while removing the intrinsic band and time structure. The MAD filter is run and pixels that are 
flagged are replaced with Not a Number (NaN) values in the python implementation. Since the NaNs are not considered during the subsequent de-trending operations, this allows us to better model the good parts of the spectrum.  We de-trend the data again, now using only the medians along the two axes without the median
filter. This iteration allows the filter to be more sensitive to smaller scale structure. The MAD filter is run again and the
flagged values from both iterations are replaced by NaNs. A final de-trend is followed by replacing NaNs with the
inferred or user specified value, which is now the median along both axes of the dynamic spectra. This process is summarized in Fig.~\ref{fig:MAD_flowchart}.

\begin{figure}
\begin{center}

\begin{tikzpicture}[scale=1.2, transform shape, node distance=2cm]

\node (in1) [io] {Read file};
\node (pro1) [process, below of=in1] {De-trend - Rolling Median};
\node (pro2) [process, right of=pro1, xshift=1.5cm] {Calculate Median, MAD for Block};
\node (pro3) [process, right of=pro2, xshift=1.5cm] {Flag Outliers - Replace with NaN};
\node (pro4) [process, below of=pro3] {De-trend - Median};
\node (pro5) [process, left of=pro4, xshift=-1.5cm] {Calculate Median, MAD for Block};
\node (pro6) [process, left of=pro5, xshift=-1.5cm] {Flag Outliers - Replace with NaN};
\node (pro7) [process, below of=pro6] {De-trend - Median};
\node (pro8) [process, right of=pro7, xshift=1.5cm] {Replace Masked Data};
\node (pro9) [io, below of=pro8] {To File or FFT-MAD};

\draw [arrow] (in1) -- (pro1);
\draw [arrow] (pro1) -- (pro2);
\draw [arrow] (pro2) -- (pro3);
\draw [arrow] (pro3) -- (pro4);
\draw [arrow] (pro4) -- (pro5);
\draw [arrow] (pro5) -- (pro6);
\draw [arrow] (pro6) -- (pro7);
\draw [arrow] (pro7) -- (pro8);
\draw [arrow] (pro8) -- (pro9);

\node [fit=(pro1) (pro8) (pro3),draw,solid,blue] {};

\end{tikzpicture}
   \end{center}
    \caption{This flow chart shows the steps of the MAD filter. Data is read from file, de-trened, outliers are flagged, de-trened, more outliers are flagged. Then there is a final detrend followed by a replacement of flagged data.}
    \label{fig:MAD_flowchart}
\end{figure}

This replacement policy differs from the clipping policy
used by \citet[][]{chime_frb_2018, Ransom-2011, Mann-2021}. The reason for the choice will be discussed in Section~\ref{sec:composite}.
If the pulse has a low dispersion measure (DM), we can re-add the time trend, to preserve pulse
power.

Comparing the columns of Fig.~\ref{fig:green_burst}  shows the effects of the
MAD filter. The frequency axis has
been flattened, removing the effects of the signal chain. The time domain had likewise been
flattened; however, we re-added the total powers, to preserve the brightness of the low-DM pulses.
Almost all the bright salt-and-pepper noise visible in Fig.~\ref{fig:green_burst} has been removed. The narrowband aviation radar around 1550~MHz has been greatly reduced, but is still present. This calls for frequency-domain filtering, which we
discuss in the next section. Likewise, the bright narrow time RFI present around 1100~MHz has been greatly reduced, but is still present. 
The communication band around 1475~MHz has been completely removed. 
Importantly,  more pulses remain than when the measure of scale was calculated across the entire band-pass. Even with bright pulses, we see
better separation between the pulses and noise, shown by the larger pulse peak for unit noise (calculated via MAD). This reduction in 
noise level increased the detectability of pulses. 

For the raw 
dynamic spectra, the first pulse is the brightest. In the MAD block filter data, the final pulse is the brightest. This leads us to ask how filtering affects the pulse energy distribution. We investigate this question in Section~\ref{sec:filter_effects}.


For a given pulse morphology, larger DMs will cause the pulse to be spread out over
a larger section of the dynamic spectra. This causes the MAD sections to be larger
compared to the pulse. This can cause a DM dependence on the recovered pulse. 
However, it was shown in Section~\ref{sec:filter_effects} that for
($\mathcal{O}(100)$ S/N) candidates this effect is not measurable. 
As discussed in \citet[][]{chime_2021, Rajwade_332}, FRB DM and
pulse brightness are related. High DM FRBs will likely be less bright, and thus less likely to have their signal excised by the MAD filter. 

\subsection{Frequency Domain MAD}\label{sec:fft_mad}
As seen in Fig.~\ref{fig:green_burst}, we often encounter persistent periodic RFI signals which can survive the time domain MAD filter. 
We would like to remove these anthropomorphic cyclic signals while preserving the periodic pulsar signal. This cleaning
is best done in the frequency domain. We use the fast Fourier transform (FFT) along the time axis to make a representation of the dynamic spectra in Fourier
space and take advantage of the amplitude and phase information contained in the Fourier coefficients.
We are interested in excising powerful periodic signals in each of the channels, but do not need to know 
 when these signals start relative to the start of the dynamic spectra. Thus, we can ignore the phase
and flag the amplitudes. 
To flag the amplitudes, we follow a similar procedure described in Section~\ref{sec:mad_filter}, but taking into account the different structure of the Fourier domain vs the time domain.

We take the amplitudes of the Fourier transformed dynamic spectra, split these powers into frequency subbands. This split allows for changing noise and pulse
structure across frequencies.
As is the case in the time domain, there is significant structure in the Fourier domain. The origin of these trends include changes in instrument 
sensitivity and standing waves in the analogue chain. Red noise will leave a large spike in the Fourier domain at low frequencies. The DC level will show up at the lowest Fourier bin.
To remove the red noise spike, we fit a robust to outliers 5th order polynomial.  This is high enough order to fit most subbands, while not being so 
high that fitting becomes unstable. 
For this paper, we de-trend the frequency
bins by subtracting the median of the frequency bin calculated across
the subband. 

Once we have de-trended the amplitudes, we compute MAD across channels for each of the subbands. As discussed by \citet[][]{Gary-2007}, for Gaussian noise the real and imaginary parts
of the Fourier transform will be in turn be Gaussian distributed. The amplitudes will be $\chi^2$ 
distributed. The filter will excise more data in this $\chi^2$ distributed data than pure Gaussian noise of a given standard deviation. Outliers are flagged and replaced with zeros, the data is then inverse FFT'd. This is summarized in Fig.~\ref{fig:FFT-MAD_flowchart}.

\begin{figure}
\begin{center}

\begin{tikzpicture}[scale=1.2, transform shape, node distance=2cm]

\node (in1) [io] {Read file/MAD Filter};
\node (pro1) [process, below of=in1] {FFT Along Time Axis};
\node (pro2) [process, right of=pro1, xshift=1.5cm] {Split into Subbands};
\node (pro3) [process, right of=pro2, xshift=1.5cm] {Polynomial De-trend};
\node (pro4) [process, below of=pro3] {Calculate Median, MAD Across Chans};
\node (pro5) [process, left of=pro4, xshift=-1.5cm] {Flag Outliers - Replace with Zero};
\node (pro6) [process, left of=pro5, xshift=-1.5cm] {Inverse FFT};
\node (pro9) [io, below of=pro6] {To File or Highpass};

\draw [arrow] (in1) -- (pro1);
\draw [arrow] (pro1) -- (pro2);
\draw [arrow] (pro2) -- (pro3);
\draw [arrow] (pro3) -- (pro4);
\draw [arrow] (pro4) -- (pro5);
\draw [arrow] (pro5) -- (pro6);
\draw [arrow] (pro6) -- (pro9);

\node [fit=(pro1) (pro6) (pro3),draw,dotted,red] {};

\end{tikzpicture}
   \end{center}
    \caption{This flow chart shows the steps of the FFT-MAD filter. Data is read from file or accepted post MAD filtering. The data is FFTed along the time axis. Median and MAD is calculated along across channels. Outliers are flagged and replaced with zero. The data is inverse FFTed and written to file or given to the highpass filter.}
    \label{fig:FFT-MAD_flowchart}
\end{figure}

Broadband signals (such as FRBs and pulsars) will be present
across many channels. When only considering the amplitudes, broadband signals will be aligned across the band, since
only the phase component of the Fourier transform controls  their arrival time.  These astronomical signals will increase the median and the noise
level uniformly and will be protected from the filter.
Narrowband RFI will be limited to a few channels
and will be excised. This filtering methodology contrasts with that of \citet[][]{Mann-2021}, who also filter in the Fourier domain.
\citet[][]{Mann-2021} use statistics on per-channel blocks in Fourier space, flagging blocks that have outlying differences in mean or
standard deviation. 
\begin{figure*}

	\includegraphics[width=\textwidth]{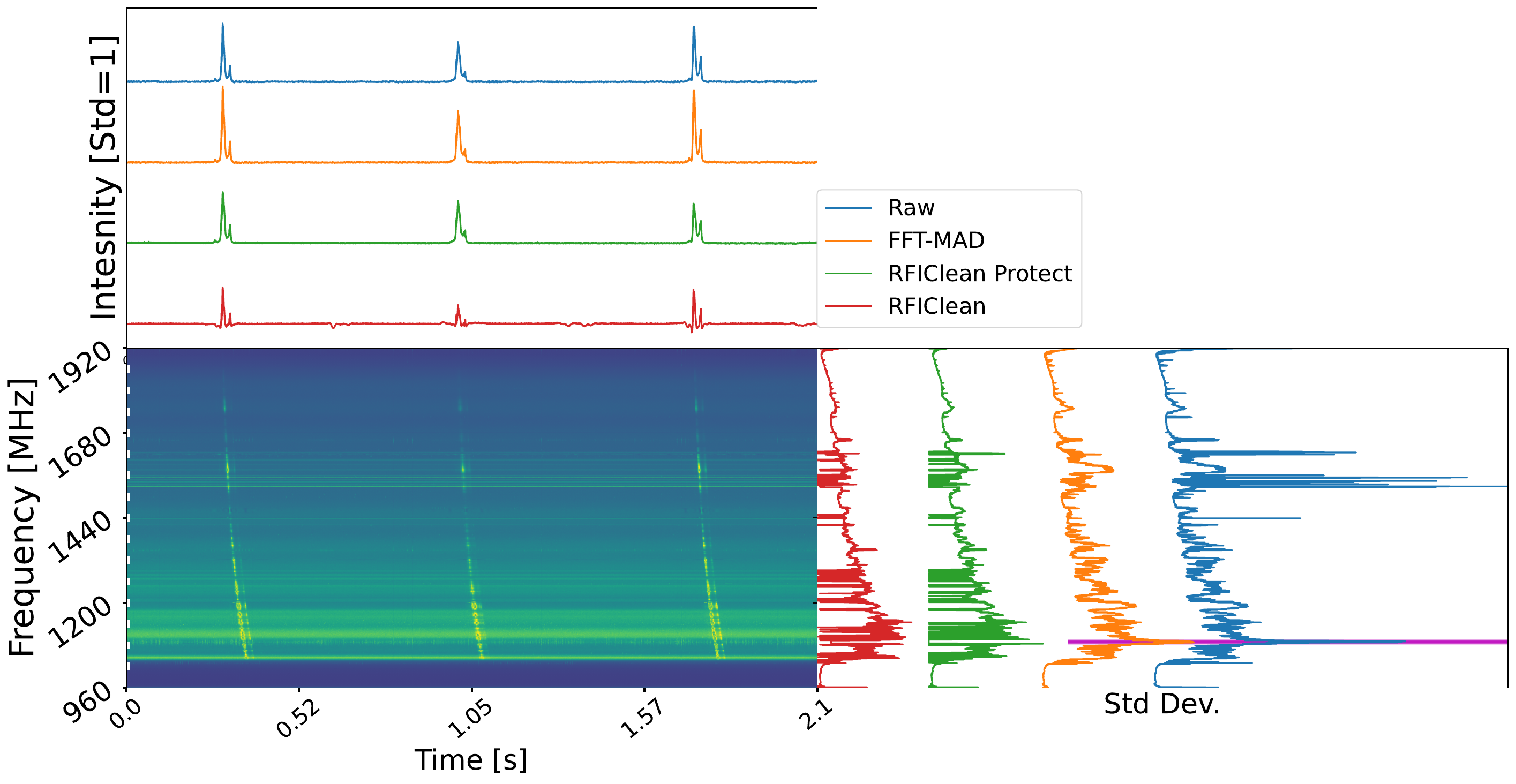}
    \caption{The dynamic spectrum of PSR B0329+54 shown in Fig.~\ref{fig:green_burst} cleaned using the FFT-MAD filter, which
    flagged 3.3\% of the Fourier domain. The left-side white blocks indicate subbands of 256 channels. The top plot shows the de-dispersed time series. The right plot shows the standard deviation of each channel. We compare three filters, \textsc{RFIClean} \citep{Mann-2021} run on defaults, \textsc{RFIClean} run with the pulsar frequency protected, and FFT-MAD. In the dynamic spectra, we see that the periodic RFI has been removed, but the pulse is intact. The large spikes in standard deviation at 1500~MHz, due to commutation and RADAR have been removed. The magenta block shows a band with large variance that is reduced but not completely de-trended by FFT-MAD. We also 
    show the effects of \textsc{RFIClean} with and without protecting the
    pulsar's frequency. 
    Unlike the time-domain MAD, we do not flatten the dynamic spectra.
    }
    \label{fig:fft_clean}
\end{figure*}

As shown in Fig.~\ref{fig:fft_clean}, 
their flagging method can remove signal from bright pulsars. To combat this, \textsc{RFIClean} \citep{Mann-2021} accepts a frequency that will be safe 
from filtering, preserving the pulsar signal. 
Our method is more forgiving to broadband signals.  Fig.~\ref{fig:fft_clean} shows that we achieve performance close the \textsc{RFIClean} without needing to know the frequency before cleaning. Our filtering methodology is advantageous in situations where we do not know if a pulsar will be present, as in blind searches or commensal observations. While the FFT-MAD filter is advantageous for bright pulsars, at the cost
of excising broadband periodic signals (such as power lines). We will filter these broadband signals with our high-pass filter, described in Section~\ref{sec:highpass}. The zero Standard Deviations of channels in Fig.~\ref{fig:fft_clean} indicates the entire channel is flagged by \textsc{RFIClean}. The MAD filter by default flags whole channels also. After the work in this paper, we started flagging channels at 16384 sample cadence. Not using this feature keeps in the spirit of this paper, flagging at the highest cadence.


Frequencies that are used for communication 
contain quasi-periodic signals that are based on the clock cycle of the transmitter. This leads to substantial power spikes in Fourier
space.
We do not flag the first Fourier bin, which contains the average power of the channel. This is done so that we do not need to find a replacement value. The average power of a channel does not affect transient search sensitivity.
The subsequent Fourier components contain zero average
power (with numerical limits of the FFT), so by removing them we smooth the channel but maintain the average power.
This smoothing can be seen between frequencies 1475--1520~MHz in Fig.~\ref{fig:fft_clean}. After the filtering, these frequencies have a much lower standard deviation.
Following experimentation, we chose to replace flagged amplitudes with zeros. This did not result in any unwanted ringing effects in the time domain and was much simpler than
other techniques, e.g., rescaling the powers or using least-squares spectral analysis  \citep[see, e.g., Chapter 13 of][]{press2007}.

We see the effects of this filter on PSR~B0329+54 in Fig.~\ref{fig:fft_clean}. The bright radar pulses have been removed, and the communication RFI has been greatly reduced. Bright salt-and-pepper
noise remains. The three pulses are largely the same, losing only a little power in the sub-pulses that precede and
follow the main pulse. We also see that the band-pass shape is
preserved. 
We did not use the time-domain MAD filter because it provided no useful information and makes replacement values difficult to determine. 
We did not have to do the flattening step with this filter because the total power in a channel
is determined by the first component for the FFT. We preserve this component and thus do not require the use of a replacement value. In this sense, the Fourier MAD filter can be seen as a smoothing filter, spreading out
RFI power over the channel. This is acceptable because we are interested in the change in power level, not the absolute
level.

\subsection{High-pass filtering}
\label{sec:highpass}
The first high-pass filter for pulsar searches was described by \citet[][]{Eatough-2009}, in which
the dynamic spectra is collapsed over frequency to form a zero-DM time series. This time series is
then subtracted off the dynamic spectra, thereby removing signals that are common across all channels at
a given time sample. This is very effective at removing broadband, smooth RFI sources, such as spark plugs
and unshielded electrical equipment. Variations on this technique are described in \cite{Lazarus_2015, Men_2019}.

\begin{figure}
\begin{center}
    
   \includegraphics[width=\columnwidth]{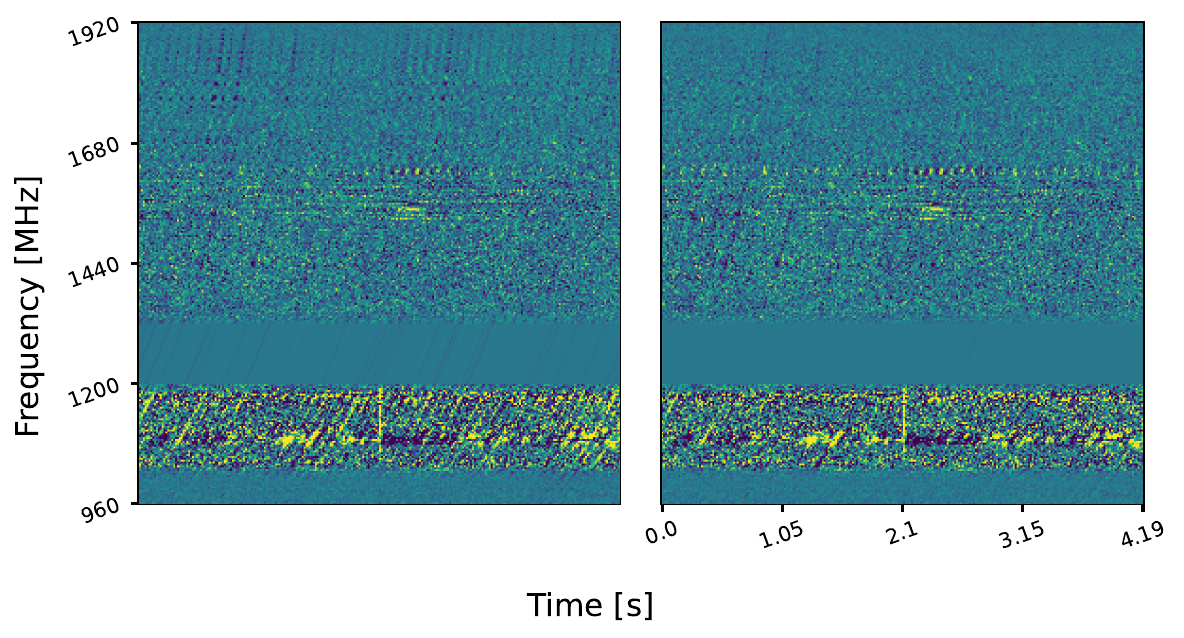}
   \end{center}
    \caption{Left: a section of dynamic spectra that has been cleaned by the MAD and the FFT-MAD filters, then the mean of the
    time series subtracted, as described by \citet[][]{Eatough-2009}. We integrated to 3.75~MHz $\times$ 16.4~ms per
    pixel.  The vertical structures are due to broadband
    zero-DM RFI and zero-DM time series subtraction. The RFI is brighter at lower frequencies. Subtracting the mean over
    frequencies leads to over subtraction in frequencies where the RFI is weak 
    and under subtraction  where the RFI is bright (1000--1200~MHz). \newline
    Right: the same data with the subtraction weighted by median filtered band-pass adapted from \citet[][]{Lazarus_2015}. The weighting prevents subtraction in low sensitivity areas 
    (such as the band-stop filter between frequencies 1200-1300 MHz) but we still see residual broadband RFI structure. The similarity between these two plots hints that a more aggressive 
    high pass filter is needed to remove this broadband structure.
    }
    \label{fig:zero_dm}
\end{figure}

This high-pass filter worked well for the Parkes Multibeam Pulsar Survey \citep[][]{parkes-multibeam}, which utilized one-bit sampling
over a 288~MHz band. 
Limited dynamic range of early receivers limited RFI structure that could be detected. 
More modern receivers can provide more than a GHz of bandwidth and often have more than eight bits of dynamic range. These combined give rise to
more complex zero-DM RFI than was seen in \citet[][]{Eatough-2009}. An example is shown in Fig.~\ref{fig:zero_dm}. Band-stop
filters and band roll off, coupled with changing RFI intensity, leads to the RFI being under-subtracted in some
frequencies and over subtracted in others. To combat changing sensitivity across the band-pass, \citet{Lazarus_2015} 
introduced a weighted subtraction, where the amount subtracted from the channel is weighted by the channel 
sensitivity.

This weighted subtraction is shown in the right panel of Fig.~\ref{fig:zero_dm}, where we see improved 
performance; however, over- and under-subtraction can still be seen, hinting at RFI evolution across the frequency band. To account for this, we extend the methodology of \citet[][]{Eatough-2009}
into a more general high-pass filter. For a given time sample, we FFT along the frequency axis, replace the lowest $n$ Fourier components
with zeros and then perform the inverse FFT, this filtering is equivalent to \citet[][]{Eatough-2009} for $n=1$ Fourier components as both then remove the DC component across frequencies. This process is summarized in Fig.~\ref{fig:highpass_flowchart}.

\begin{figure}
\begin{center}

\begin{tikzpicture}[scale=1.2, transform shape, node distance=2cm]

\node (in1) [io] {Read File/FFT-MAD};
\node (dec1) [decision, below of=in1, yshift=-1cm] {Freq. Highpass};
\node (pro11) [process, right of=dec1, xshift=2cm] {Subtract Means};
\node (pro9) [process, below of=dec1, yshift=-1cm] {FFT Across Freq.};
\node (pro10) [process, right of=pro9, xshift=1.6cm] {Remove lowest $n$ Fourier Components};
\node (pro12) [process, right of=pro10, xshift=1.6cm] {Inverse FFT};

\node (out1) [io, right of=in1, xshift=4cm] {Write File};

\draw [arrow] (in1) -- (dec1);
\draw [arrow] (dec1) -- node[pos=0.5,above]{$n=1$}(pro11);
\draw [arrow] (dec1) -- node[pos=0.5,left]{$n>1$}(pro9);
\draw [arrow] (pro11) -- (out1);
\draw [arrow] (pro9) -- (pro10);
\draw [arrow] (pro10) -- (pro12);
\draw [arrow] (pro12) -- (out1);

\node [fit=(dec1) (pro12),draw,loosely dashed,black] {};

\end{tikzpicture}
   \end{center}
    \caption{This flow chart shows the steps of highpass. The data are first read from a PSRFITS or filterbank file, or given post FFT-MAD filtering. The data are highpass filtered, either using the \cite{Eatough-2009}'s Zero DM ($n=1$ Fourier component removed by subtracting the mean accross frequencies), or the Section~\ref{sec:highpass} methodology  ($n>1$, Fourier components removed). The filter data are then written to a file.}
    \label{fig:highpass_flowchart}
\end{figure}

As discussed by \citet[][]{Eatough-2009}, this filtering will reduce the pulse energy as a function of DM. Low DM pulses
occupy a large portion of the dynamic spectra. These pulses can then lose a significant portion of their energy
to the filter. As discussed in \citet[][]{Eatough-2009}, this energy loss
can be partially recovered by using a convolution kernel that accounts for
how the filter changes the shape of the time series. We are not aware of any
search pipelines that have implemented these kernels.

This ideal high pass filter's jump discontinuity in frequency space will introduce Gibbs ringing \citep[i.e., oscillation across the band-pass;][]{wilbraham, gibbs},  in the time domain. We tried using tapered transition windows \citep[][]{blackman, hann}, where the transition is 
a smooth function, but
the improvement was minimal, and this made the filter less intuitive to use.
As discussed in the next section, this filter is
used almost exclusively with the
MAD filter, which de-trends frequency structure,  reducing the size of the low Fourier components before this
filter is run. As shown in the right most panel of Fig.~\ref{fig:composite}, some ringing is visible, however
it is near the noise level, and thus not of great concern.

\subsection{Composite Filter}
\label{sec:composite}

\begin{figure*}
   \includegraphics[width=\textwidth]{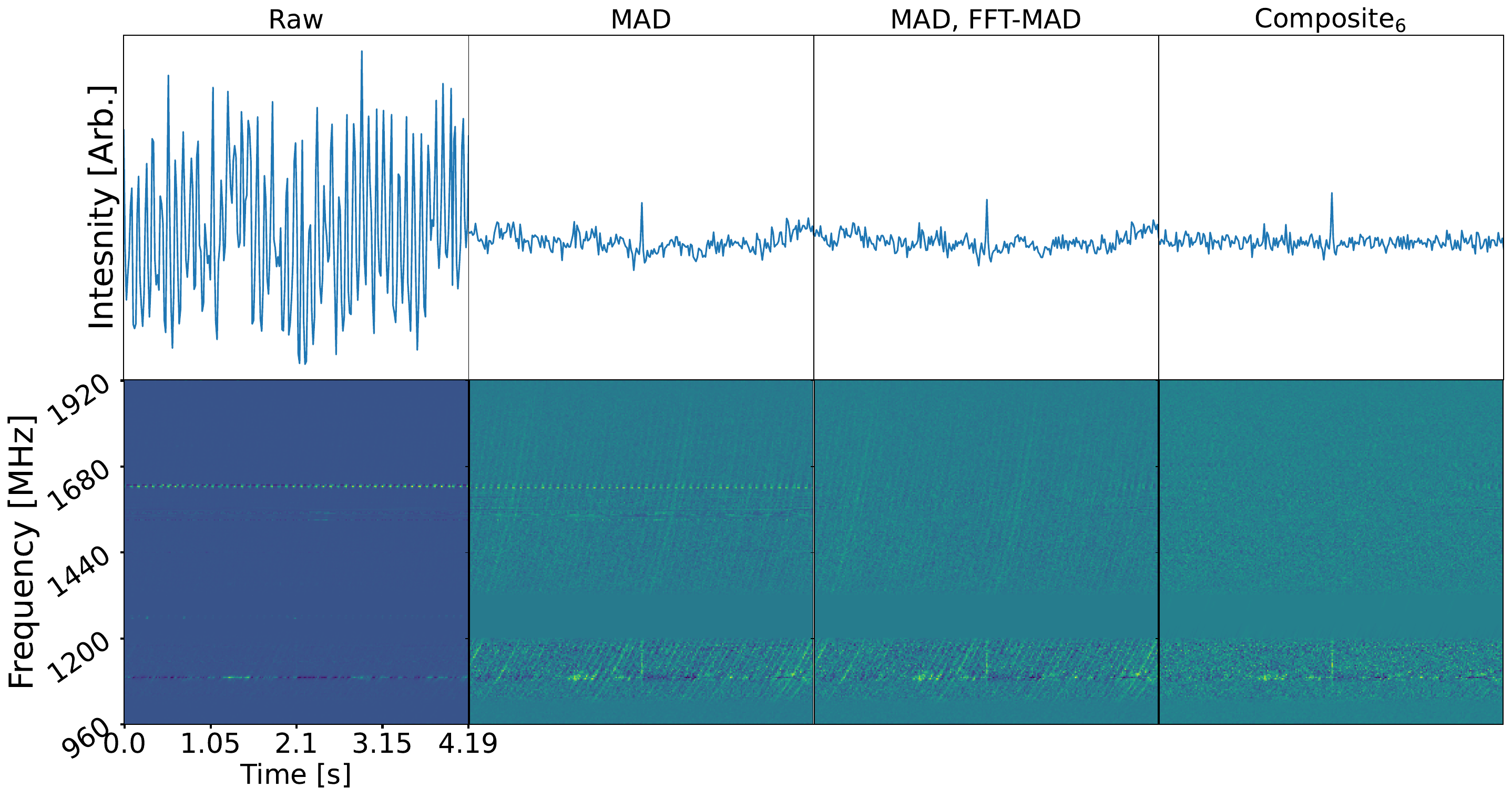}
    \caption{A section of a GREENBURST dynamic spectrum and its corresponding time series as it is being cleaned by the
    Composite filter$_6$. The bottom row shows the dynamic spectrum integrated to 59~MHz $\times$ 16.4~ms per pixel. The top row
    shows the intensity by integrating over all the frequencies. These are de-dispersed to a DM of 435~cm$^{-3}$~pc. The initial block has a standard deviation of 0.775. The MAD filter flags 7.16\% of the data,
    reducing the standard deviation to 0.100. FFT-MAD filter flags 1.84\%, reducing the standard deviation to 0.090.
    The high-pass filter removed 0.49\% of the Fourier Space, and }reduced the standard deviation to 0.082. The final result is a 7.8 S/N candidate
    that is not above S/N 6 in the other plots. As discussed in Section~\ref{sec:composite}, the subscript 6 denotes the high-pass filter is
    removing the lowest 6 Fourier elements. 
    \label{fig:composite}
\end{figure*}

Each of the above filters operates in a different domain, so it is often useful to combine all three of them.
We show such a case in Fig.~\ref{fig:composite} where we have multiple bright RFI sources. We begin by running the
time domain MAD filter. This removes salt-and-pepper RFI and de-trends the data in time and frequency.
As shown in the second panel of Fig.~\ref{fig:composite}, periodic RFI is still bright and present after the MAD filter. We then apply the FFT-MAD filter to remove this persistent narrowband RFI.
The third box of Fig.~\ref{fig:composite} shows the remaining periodic RFI has been removed. The long sweeping pulses are
zero-DM RFI. They show a quadratic sweep because these blocks are dispersed. As shown in Fig.~\ref{fig:zero_dm}, mean subtraction \citep[][]{Eatough-2009}
does not fully remove the effects of this signal.
Next we applied the high-pass filter, zero-weighting the first six Fourier components to remove the
broad-band RFI. For this pulse, removing six components maximizes the detected S/N. Shown in Fig.~\ref{fig:composite}, this process greatly reduced the structure in the dynamic spectra, revealing a candidate that 
only has a S/N above 6 (which is lower cutoff of the Heimdall search pipeline) when all three filters are used.
This signal path is summarized in Fig.~\ref{fig:composite_flowchart}.

\begin{figure}
\begin{center}

\begin{tikzpicture}[scale=1.2, transform shape, node distance=2cm]

\node (in1) [io] {Read File};
\node (pro1) [process, below of=in1] {Robust De-trend};
\node (pro2) [process, right of=pro1, xshift=1.5cm] {Flag Outliers};
\node (pro3) [process, right of=pro2, xshift=1.5cm] {De-trend};
\node (pro4) [process, below of=pro3] {Flag Outliers};
\node (pro5) [process, left of=pro4, xshift=-1.5cm] {De-trend};
\node (pro6) [process, below of=pro5, xshift=3.5cm] {FFT Across Times};
\node (pro7) [process, left of=pro6, xshift=-1.5cm] {Flag Ouliers};
\node (pro8) [process, left of=pro7, xshift=-1.5cm] {Inverse FFT};
\node (dec1) [decision, below of=pro8, yshift=-1.5cm] {Freq. Highpass};
\node (pro11) [process, right of=dec1, xshift=2cm] {ZeroDM};
\node (pro9) [process, below of=dec1, yshift=-1cm] {FFT Across Freq.};
\node (pro10) [process, right of=pro9, xshift=1.6cm] {Remove Low Freq.};
\node (pro12) [process, right of=pro10, xshift=1.6cm] {Inverse FFT};

\node (out1) [io, below of=pro6, yshift=0.75cm] {Write File};

\draw [arrow] (in1) -- (pro1);
\draw [arrow] (pro1) -- (pro2);
\draw [arrow] (pro2) -- (pro3);
\draw [arrow] (pro3) -- (pro4);
\draw [arrow] (pro4) -- (pro5);
\draw [arrow] (pro5) -- (pro6);
\draw [arrow] (pro6) -- (pro7);
\draw [arrow] (pro7) -- (pro8);
\draw [arrow] (pro8) -- (dec1);
\draw [arrow] (dec1) -- node[pos=0.5,above]{$n=1$}(pro11);
\draw [arrow] (dec1) -- node[pos=0.5,left]{$n>1$}(pro9);
\draw [arrow] (pro11) -- (out1);
\draw [arrow] (pro9) -- (pro10);
\draw [arrow] (pro10) -- (pro12);
\draw [arrow] (pro12) -- (out1);

\node [fit=(pro1) (pro4),draw,solid,blue] {};
\node [fit=(pro6) (pro8),draw,dotted,red] {};
\node [fit=(dec1) (pro12),draw,loosely dashed,black] {};

\end{tikzpicture}
   \end{center}
    \caption{This flow chart shows the steps of the composite filter. The data are first read from a PSRFITS or filterbank file, cleaned by the Time Domain MAD, described in Section~\ref{sec:mad_filter}, the blue solid box. The data are then cleaned by Time Domain MAD, Section~\ref{sec:fft_mad}, red dotted box. Finally, the data are highpass filtered, either using the \cite{Eatough-2009}'s Zero DM ($n=1$ Fourier component removed), or the Section~\ref{sec:highpass} methodology  ($n>1$, Fourier components removed), the black dashed box. The filter data are then written to a file.}
    \label{fig:composite_flowchart}
\end{figure}

\begin{figure*}
    \begin{center}
   \includegraphics[width=0.9\textwidth]{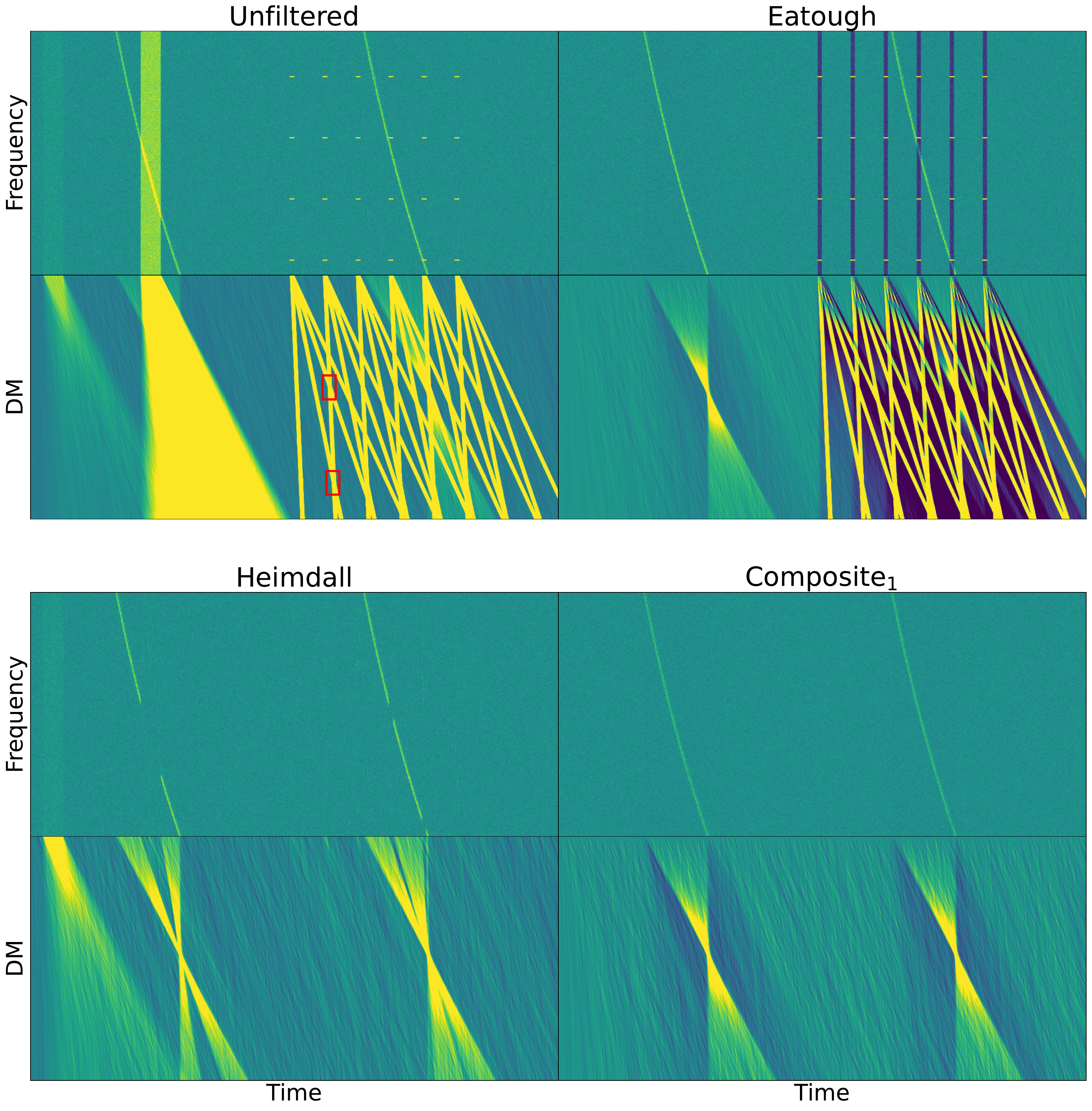}
    \caption{
    Comparison of the Composite$_1$ filter with two alternative broadband RFI removal algorithms. The top row of each column shows the dynamic spectra, the bottom shows DM--time space, transformed from the top using the Fast Dispersion Measure Transform \citep[FDMT;][]{fdmt}. The top right column shows the unfiltered dynamic spectra with two types of RFI: broadband zero-DM across all channels 
    and narrowband clusters correlated across the dynamic spectra. Both types increase the noise in the DM-time space, 
    making detecting pulses more difficult. The narrowband RFI can make bow-ties in the DM-space (two of which are highlighted by the red boxes) that can be mistaken by the search and classification algorithms as an astrophysical pulse. \citet{Eatough-2009} removes broadband RFI by subtracting the mean of all frequencies. This removes broadband RFI, but the narrowband RFI is also subtracted, spreading the effect across frequency. \textsc{Heimdall} \citep{Barsdell} removes dynamic 
    spectra whose mean value across frequency exceeds 5$\sigma$ times the noise, and replaces the values with randomly chosen nearby slices. This does a better job at removing the 
    narrowband correlated RFI. The bright broadband RFI is removed, however, so is much of 
    the bright pulse. The less bright broadband RFI remains, since it is below the cutoff.
    The Composite filter removes the narrowband features, allowing for high-pass filtering
    without the effects of narrowband features.
    }
    \label{fig:zero-dm}
    \end{center}
\end{figure*}

The ordering of the filters was determined by experimentation. The final order discussed above was found to work 
well because of the way the MAD filter de-trends the dynamic spectra and removes large spikes. This is a non-linear, low-pass filter. This filter removes the most outlying non-linear components of the dynamic spectra,  making the operations in Fourier space more stable. We noticed that zero-DM subtraction
alone would often increase the standard deviation of de-dispersed time series. This is possibly due to very bright
narrowband RFI, which, as we see in Fig.~\ref{fig:composite}, can be significantly brighter than broadband RFI, being smeared out in frequency by the zero-DM subtraction. This effect is shown in 
in Fig.~\ref{fig:zero-dm}, the second column shows  \citet{Eatough-2009}
does not reduce the noise in the DM--time space. 

Hence, we put the high-pass filter last, after all the narrowband RFI is removed.
As we will see in the following sections, the composite filter is sensitive to changes in the high-pass filter. For low DM sources, we need to choose less aggressive filtering. When referring to 
the composite$_n$ filter, the subscript $n$  denotes the number
of Fourier components excised by the high-pass filter. This was described
in Section~\ref{sec:highpass}

\begin{figure}
  \centering
 \raisebox{-0.5\height}{\includegraphics[width=\columnwidth]{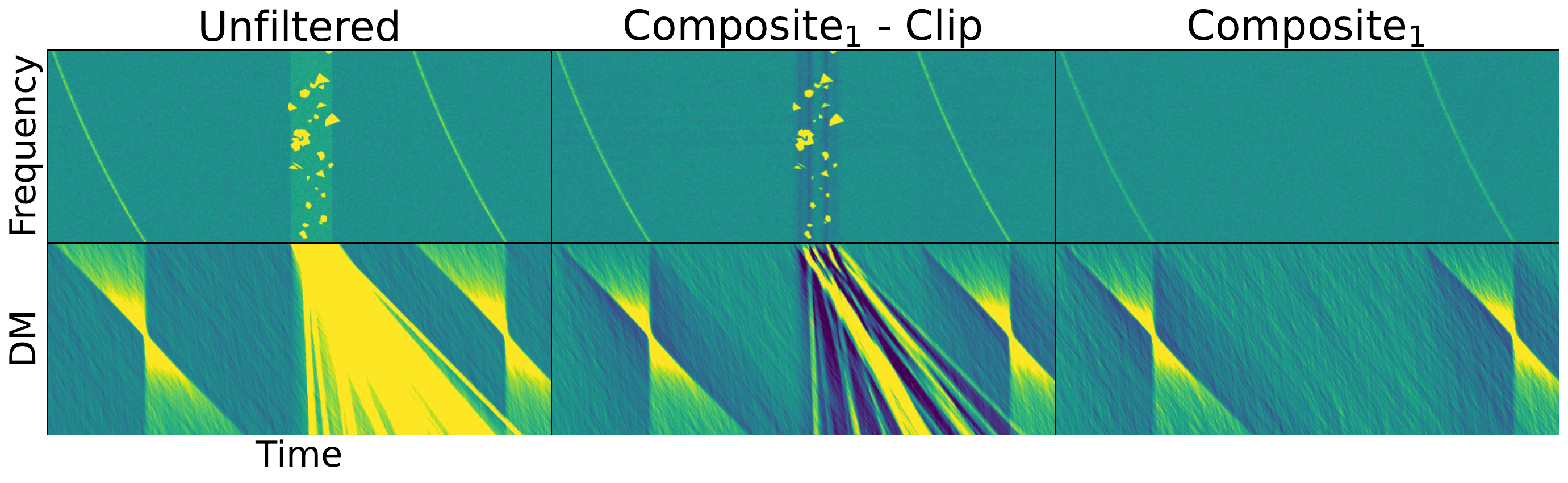}}
 \raisebox{-0.5\height}{\includegraphics[width=\columnwidth]{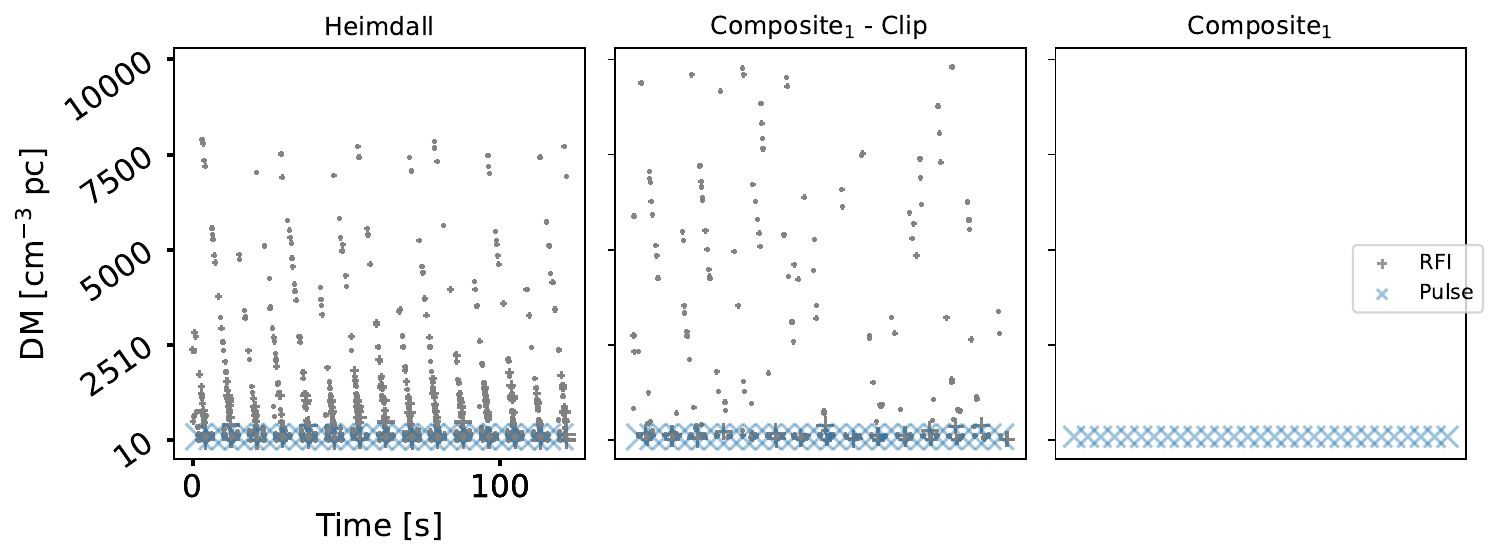}}
  \caption{A comparison of clipping versus replacement with central value (median) for the time domain MAD filter. Top panel: synthetic dynamic spectra with RFI that is composed of a broadband component coupled with bright randomly generated polynomials. We clean this dynamic spectra in 
  two ways. In the second panel, the time domain MAD clips the vales to 4$\sigma$ then the frequency domain MAD and high-pass filter are applied. 
  The third panel, the MAD replaces the flagged values with the central value (the median). For the former, the high-pass filtering causes broadband structure. This structure extends in along lines in the DM-time space, shown by the lower row of the top plot. Bottom panel: taking the above dynamic spectra and increasing the spacing between the interference to make the interference wider, and repeated the sequence for 120 seconds. The former makes the  following plots clearer,
  and the latter allows for structure at high DM. We processed this dynamic spectra with \textsc{Heimdall} \citep{Barsdell}, with the default RFI filter engaged. For both the unfiltered and clipped cases, the 
  structure in the DM-time plot shows up as lines on candidates over time.
  This effect is not seen when the flagged values are replaced with the central value (here median).
  }
  \label{fig:clip_vs_replace}
 \end{figure}

The smearing across frequency of narrowband RFI is why we choose to replace flagged data with the
central value (median) instead of a clipped value. If we used winsorization \citep{Winsor_1947,Dixon_1960} to replace outlying values the frequency spreading would
persist, but be limited to the clipped value, as shown in Fig.~\ref{fig:clip_vs_replace}.

The smearing effect is why we coupled the high-pass filtering with
the MAD filter. For optimal high-pass results, we must remove all the RFI corrupted pixels, or have a filter that can ignore flagged values,
as discussed in \citet{chime_rfi}. We must
flag on the same timescale as the data were recorded. Flagging on longer timescales helps reduce the presence
of narrowband RFI. However, longer timescale filtering might miss single bright points, so this problem can
persist. We further investigate the effectiveness of these filters on observed broadband RFI 
in Section~\ref{sec:dm-time}.

\section{Filter Effects}
\label{sec:filter_effects}

In this section we will use synthetic pulses to quantify filter effects. The first Subsection \ref{sec:pulse_creation} describes how we create these synthetic pulses. We then inject pulses into Gaussian Noise (without RFI) in Subsection \ref{sec:gauss_noise} In the next Section \ref{sec:sim} we inject pulses into a GREENBURST observation (with RFI). In Sections~\ref{sec:gauss_noise}, \ref{sec:sim}, and \ref{sec:astrophysical_pulses} we will use \textsc{Will} to preform pulse detection. This will allow us to report noise levels pre and post filtering. Section~\ref{sec:single_pulse_searches} will describe a Heimdall-Your-Fetch search, reflecting a standard single pulse search.

\subsection{Synthetic Pulse Creation}\label{sec:pulse_creation}

\begin{deluxetable}{lcr}
    \centering
    \caption{Distributions sampled for the pulse parameter space.}
    \label{table:paramater_space}
    \tablehead{
        \colhead{Parameter} & \colhead{Distribution} & \colhead{Range}}
        \startdata
        Brightness Median & Log-Normal & ($5\times10^2$, $5\times10^6$) Samp.\\
        Brightness Stand. Dev. & & 20\% of median \\
        Time Width & Uniform &(0.5, 30) ms \\
        Spectral Index & Uniform &(-3, 3) \\
        Frequency Width & Uniform &(350, 1000) MHz \\
        Center Frequency & Uniform &(1056, 1728) MHz \\
        \# Scintles & Uniform&(0, 3) \\
        Scatter Time & Uniform& (0.2, 5) ms\\
        Dispersion Measure & Uniform& (10, 3000) pc $\text{cm}^{-3}$\\
        \enddata
\end{deluxetable}

We create pulses that are Gaussian along the time and the frequency axis. We convolve 
the time profile with an exponential function, $e^{-t/\tau}/\tau$, where $\tau$ is the scattering time in 
Table~\ref{table:paramater_space}. For the frequency profile, we multiply the Gaussian by a spectral index 
$(F/F_{\text{center}})^{\alpha}$,  where $\alpha$ is the spectral index, $F$ is the channel frequency and  
$F_{\text{center}}$ is the reference 
frequency, which we choose the as the center channel. The radio light travels through ionized gas on its path from 
the source to the telescope. The non-uniformities in the plasma structure causes amplitude variations across frequency \citep[][]{intro_radio}.   We simulate scintillation by multiplying the frequency profile with the factor
\begin{equation}
    S \equiv \Big|\cos\Big(2\pi N_{s}(F/F_{\text{center}})^2 \Big)\Big|,
\end{equation}
where $N_s$ is half of the number of bright patches. We injected all the scintles at the same phase, since we 
do not anticipate this having a significant effect. We do not include intra-channel DM smearing \citep{galactic_position} in our synthetic pulses in this paper.  This omission stops the pulse shape from varying with DM, simplifying our analysis. For GREENBURST, the intra-channel smearing is minimal. At a DM of $3000$~cm$^{-3}$~pc it is 3 time samples. For 10,000~cm$^{-3}$~pc it is 9 samples at the lowest frequencies channels. Finally, we account for the receiver roll-off and the bandstop filter. These are frequencies where the analogue chain has low sensitivity, where we are unlikely to see an astrophysical signal.
We model the roll off
by calculating the medians along the band-pass, we then smooth these medians with a median filter. We calculate the
standard deviation of this smoothed band-pass. We flag channels that are within 
2/3 of a standard deviation of zero power. The transition between on and off regions 
is made using a Gaussian filter. An example of this process is shown in the \textsc{will}'s example
notebooks.\footnote{\url{https://github.com/josephwkania/will/blob/master/examples/inject_pulse.ipynb}}

The previous paragraphs describe the continuous functions used to make the pulse profile.  We need discrete values to inject into the
the dynamic spectra. To transform from continuous to discrete, we sample the continuous function for 
a given number of samples. Each one of these samples is where a dynamic spectra pixel will be increased by one.   Thus, the number of samples will be proportional to the pulse brightness
and is reported in
Table~\ref{table:paramater_space}. We reference this number and not S/N because we do not have a way to know the noise
level without RFI because we cannot perfectly remove all RFI.  
Pulse amplitude distributions of canonical pulsars are often thought to be log-normal, sums of log-normals, or a log-normal with 
exponential tails \citet[][]{Ritchings_1976, Brylyakova_2021, parkes_pulsar}. We used a log-normal distribution of pulse powers, but did not include an exponential tail.
Since we precisely know the injected pulse powers, we can see how the recovered pulse energy distribution is affected by both RFI and any filters. Understanding the pulse energy distribution is useful in constraining emission mechanisms. For example, giant pulses have
a power-law distribution, while canonical pules are log-normal. This indicates that canonical pulses and giant pulses have different
emission mechanisms \citep{Mickaliger_2012}.

\subsection{Synthetic Gaussian Noise}\label{sec:gauss_noise}
We used \textsc{will}\footnote{\url{https://github.com/josephwkania/will}} \citep[][]{will} to create synthetic pulses and inject them
into Gaussian noise that has the same MAD and median as a GREENBURST dynamic spectra. This data set does not include RFI, only measuring the filter effect on noise and the pulse.
For this section, we used the pulse detection in \textsc{will}.  
To detect pulses we first select a chunk of dynamic spectra containing the pulse. Second we dedisperse this dynamic spectra at the desired DM. Third, we sum all frequencies. Fourth, we perform a boxcar convolution at the desired pulse width. Fifth, we perform a linear de-tread. Finally,  we calculate the ratio of the maximum point to the noise level to arrive at S/N.
This noise measurement
is an additional reason why we used \textsc{will}, as  common single pulse search pulse packages do not typically report noise levels. 
This noise measurement is needed because we
are interested in measuring the interaction between noise level and recovered pulse characteristics. 
A pulse search that does not know pulse characteristics a priori will search over time, DM, and pulse width.
Candidates
are then clustered in DM--time--pulse width space, with one value reported for the cluster.
RFI can affect this
clustering \citep[see][for a discussion of clustering methodologies]{aggarwal2021robust}. Since we
know the injected pulse DM and width, we know the optimal width
and DM to search, this allows us to skip the clustering step and simplify our
analysis. (An example of this pulse misidentification is discussed in Section~\ref{sec:dm-time}.)
A blind single pulse will also produce many RFI false positives, which need to be classified 
as astrophysical or RFI. Commonly, machine learning is leveraged for classification. We will see in 
Section~\ref{sec:single_pulse_searches}, this classification has a selection function that would need to be 
corrected.

We cleaned the dynamic
spectra with our four filters. The effects of filters are shown in Fig.~\ref{fig:filter_response}. We see that over
DM and pulse intensity, the filters behave as expected. The time domain MAD filter removes parts of 
very bright pulses. These pulses are much brighter than the sky and instrument noise background, and thus flagged as RFI. This is predominant when a small part of the pulse is in the MAD filter's dynamic spectra window, thus the pulse is not reflected in the median and scale statistics. However, some part of the pulse persists and is recoverable. Note that Power response plot of Fig.~\ref{fig:filter_response} goes to very high signal to noises, far beyond the S/N of a few hundred we see with bright pulsars. We chose this range to fully visualize the response curve. While we lose significant power at bright pulses, we should still be able to detect them post filtering, since they even power filtering they contain significant power. The MAD filter drives 
the response of the Composite
filter in the bright pulse regime. The FFT-MAD filter had no dependency on pulse power, the pulse amplitudes line up and the
pulse is preserved. The high-pass filter also  behaves as expected. The signal removed by this filter is a function of 
the fraction of the total bandwidth the pulse occupies. This, of course, is not a function of pulse power.

The response over DM similarly behaves as expected. Low DM pulses occupy a higher percentage
of the band, and more of the pulse gets removed by the high-pass filter. At higher DM, the pulse occupies a smaller fraction of the 
frequency channels,  and less of the pulse is removed. This DM-response filter shows us that filtering introduced
some variability, on the order of 10\%, to the recovered S/N. This variability was expected, but  we are surprised by the size
of the effect. This variability comes from two sources. First, the background and the pulse are randomly generated.
These realizations will have different random bright points both on and off the pulse that will get flagged. The flagging
will cause random variations around the mean injected S/N. The second effect, for the MAD and Composite filters, is
as the pulse changes in DM, the pulse is spread out over different numbers of MAD sub-arrays. If the pulse only
occupies a small portion of the sub-array, more of the pulse could be flagged.
If the pulse occupies a large portion of the sub-array, then the pulse statistics will drive the
sub-array statistics and none of the pulse will be flagged. This filter-pulse interaction is determined by the
characteristics of the pulse, the pulse phase relative to the filter sub-arrays, and DM.
We partially marginalize over these last two qualities in Fig.~\ref{fig:filter_response} by using
multiple pulses randomly phase shifted relative to the filter and multiple DM averaged into a bin. We use the standard deviation of S/N in the DM bins to produce $1~\sigma$ errors.

We can use these two plots to determine which filtering scheme would be useful. If we are interested in low DM pulses,
we should use a less aggressive high-pass filter, or forgo this filtering
step. If we expect bright pulses, we would forgo MAD filtering, which has a poor response to a pulse where S/N is above a few hundred.
The former decision can be made by even during blind searches by consulting estimated DM modules,
such as \citet[][]{ne2001} and \citet[][]{Yao_2017} for a given pointing direction and estimating a minimum reasonable
DM. Observations that need to discern between low DM astrophysical pulses and broadband RFI would
be better served with an RFI reference and subtraction scheme, as described by \cite{Briggs_2000}.

\subsection{Synthetic Pulses in Observed Noise}\label{sec:sim}

To inject pulses into observed data, we used three hours of GREENBURST data from a single pointing that had been searched and did not have any bright detected astrophysical pulses. 
GREENBURST records filterbank files (i.e., a discretely sampled array of frequency and time data from the telescope) that are 597~s long. 
We took twenty-two of these files and used them as the base to inject pulses. 
For each file, we sample the parameter space described in 
Table~\ref{table:paramater_space}, all the pulse properties except for the individual pulse powers 
are the same for a given  run on a single file. See \citet[][]{greenburst} for the physical insight behind the parameter ranges. Each file had 288 pulses interjected. We repeated this processes 115 times
for a total of 33120 pulses. These base files were used multiple times, but our analysis is focused on
the effects on burst morphology and not the effects of sky position. 

\begin{figure}
  \centering
  \raisebox{-0.5\height}{\includegraphics[width=0.75\columnwidth]{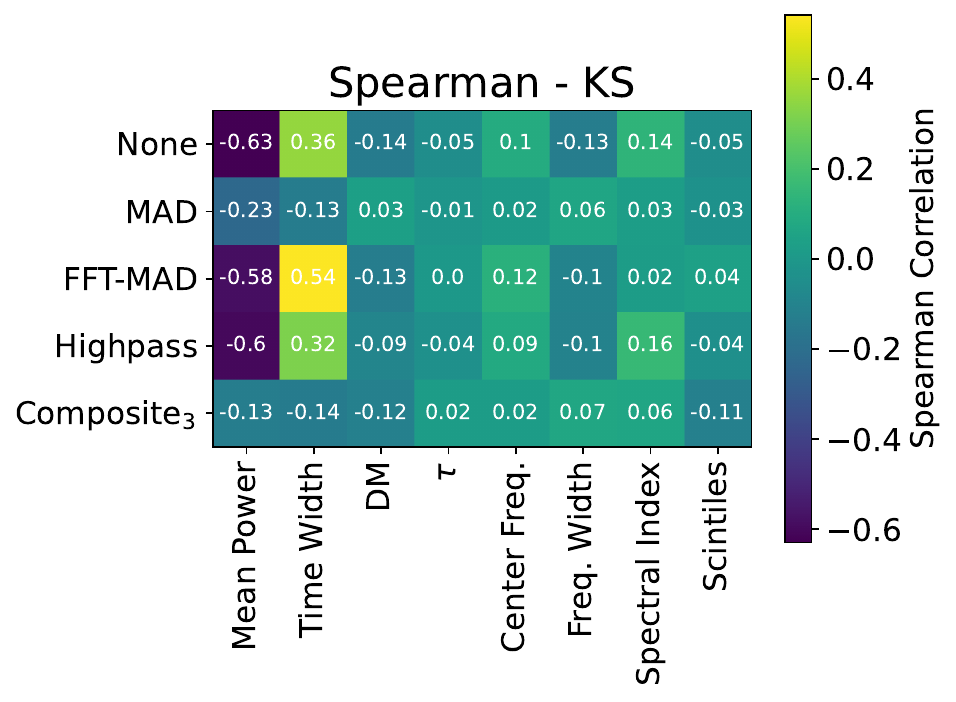}}
  \raisebox{-0.5\height}{\includegraphics[width=0.75\columnwidth]{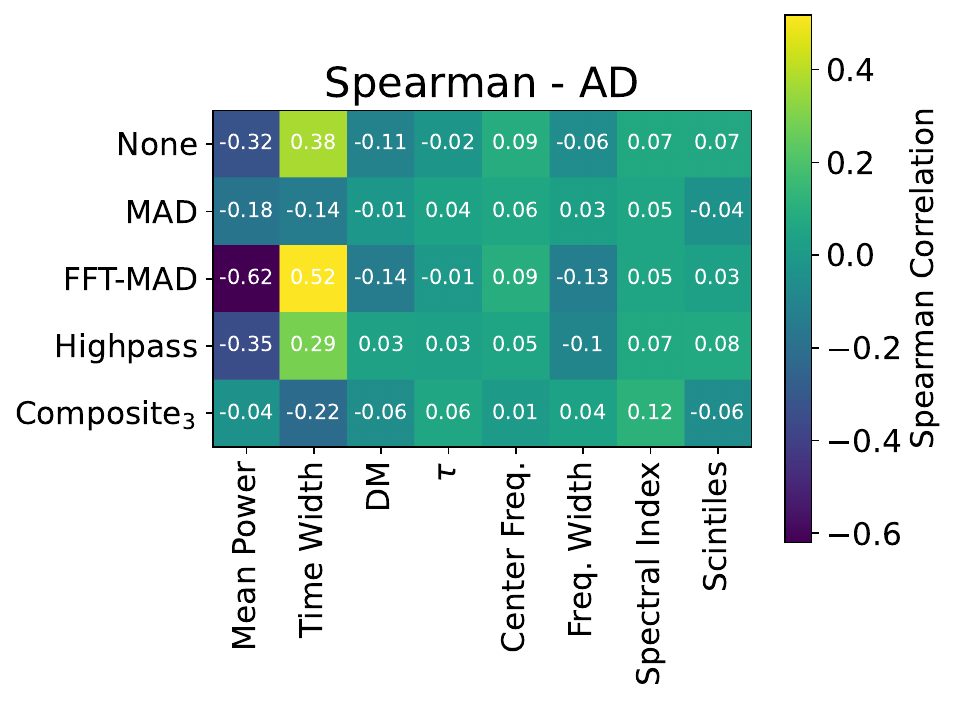}}
  \caption{The Spearman rank correlation coefficient \citep{spearman} between 
  the injected pulse characteristics and Kolmogorov–Smirnov \citep{KOLMOGOROV, Smirnov} (KS)—Top or Anderson-Darling \citep{anderson-darling} (AD) test—Bottom between the injected and recovered power distribution.
  }
  \label{fig:ks_ad}
 \end{figure}

First, we sample the space described by Table~\ref{table:paramater_space}, which we use to create a pulse profile. 
From this profile, we calculate the ideal boxcar width. We sample this profile and inject this into the file. For more details about the above process, see \textsc{will}'s documentation\footnote{\url{https://josephwkania.github.io/will/}}.
 We sequentially clean this file with each one of the filters, and perform the same pulse detection 
as described above. We record these values and provide a summary in Table~\ref{tab:synthetic_pulses}. An additional breakdown is given in Fig.~\ref{fig:ks_ad}.

\begin{deluxetable*}{lllllllll}
     \centering
     \caption{
     A comparison of the four filters on synthetic pulsars. $\overline{\text{Amplitude}}$
     is the average amplitude of the convoluted pulse. $\overline{\text{Noise}}$ is the
     average noise level of the de-trended time series around each pulse. $\overline{\text{S/N}}$ is the mean of the S/N of each pulse. The Folded S/N is the weighted mean
     of the folded S/Ns across epochs. KS is the modified Kolmogorov-Smirnov test between and injected power and recovered S/N. AD is 
     the modified Anderson-Darling test for the aforementioned distance. 
     Occupancy is the percent of pulse windows with a
     pulse above S/N 6. \% Flagged is the percentage of the data excised by each filter.
     Values in parentheses are the standard error of the least significant digits, computed using bootstrap techniques \citep{bootstrap0, bootstap1}. 
     }
     \label{tab:synthetic_pulses}
\setlength\tabcolsep{0pt}
  \tablehead{
         \colhead{Filter} \hspace{1cm}
         & \colhead{$\overline{\text{Amplitude}}$} \hspace{1cm}
         & \colhead{$\overline{\text{Noise}}$} \hspace{1.5cm}
         & \colhead{$\overline{\text{Folded S/N}}$} \hspace{1cm}
         & \colhead{S/N} \hspace{1cm}
         & \colhead{KS} \hspace{1.75cm}
         & \colhead{AD} \hspace{1.5cm}
         & \colhead{Occup.} \hspace{1cm}
         & \colhead{\% Flag}\\   }   
         \startdata
         None   &      10.33$\pm$0.98    & 0.263$\pm$0.010      & 51.5$\pm$7.3 & 353  & 0.109$\pm$0.013     & 1.29$\pm$0.71  & 78.1 & None \\
         MAD    &       7.36$\pm$0.55    & 0.0918$\pm$0.0013     & 82.9$\pm$6.5 & 829  & 0.0813$\pm$0.0069   & 1.94$\pm$0.80  & 91.7 & 2.36 \\
         FFT-MAD    &   8.89$\pm$0.85    & 0.0880$\pm$0.0016    & 96$\pm$10    & 1352 & 0.0625$\pm$0.058    & 0.342$\pm$0.071 & 91.1 & 2.29 \\
         High-pass  &  9.73$\pm$0.97     & 0.253$\pm$0.011      & 49.6$\pm$7.2 & 351  & 0.108$\pm$0.012     & 2.4$\pm$1.4  & 75.6 & 0.293 \\
         Composite$_3$ & 6.71$\pm$0.53   & 0.0864$\pm$0.0013    & 78.4$\pm$6.3 & 1100 & 0.0824$\pm$0.069    & 1.93$\pm$0.81  & 89.5 & 4.10 \\
     \enddata
\end{deluxetable*}

The first column is the average pulse amplitude, $\overline{\text{Amplitude}}$, i.e., the mean 
amplitude of the  boxcar convolved pulse. We see that each filter has some loss, and the Composite filter
inherits these losses. The next column, $\overline{\text{Noise}}$ is the average of the 
convolved noise levels at each pulse window. We see that both outlier filters provide a substantial 
reduction in the noise level. The average S/N over all the pulses is $\overline{\text{S/N}}$. Folded S/N adds all the pulses together and
then runs the boxcar detection on the resulting pulse.

To test the distribution of injected to recovered pulses, we use two statistical tests,
the two sample Kolmogorov-Smirnov (KS) \citep{KOLMOGOROV, Smirnov} and the Anderson-Darling (AD) \citep{anderson-darling} tests. The goal of both tests
is to quantify the difference between the injected and recovered pulse distributions. In the ideal case, 
the RFI filtering should bring these distributions closer together by removing spurious signals that distort
our view of the recovered pulse without degrading the pulses. Since we are injecting pulse samples, we 
need to find a way to map these values to S/Ns. We take inspiration from the Lilliefors test \citep[][]{Lilliefors1967OnTK}, a normality test for when the mean and variance is unknown. 
The Lilliefors test makes the population variance one and the population mean zero before performing the normality tests, thus testing if the samples come from a normal distribution, but necessarily one with unit variance and zero mean.
Likewise, we can make the powers and the recovered S/N unit variance and zero mean, and then apply both
two sample tests. This will change the degrees of freedom, so we cannot report significance values without numerical
simulation of these tests. 

Using the above approach, we find mixed results. For some RFI filters, particularly the FFT-MAD,  we see the values of both test statistics
go down, indicating that the two distributions are closer to each other. We see that this is not the case 
for the MAD filter, the Anderson-Darling value increases, while the Kolmogorov-Smirnov value decreases.
The Anderson-Darling statistic, $A^2$ for two continuous distributions $F$ and $F_n$ is defined \citep[][]{anderson-darling} as 
\begin{equation}
    A^2 = n \int^{\infty}_{-\infty} \frac{(F_n(x)-F(x)))^2}{F(x)(1-F(x))}dF(x).
\end{equation}
Comparing the above equation to the two sample Kolmogorov-Smirnov distance 
\begin{equation}
    D = \sup_x|F_1(x) - F_2(x)|, 
\end{equation}
with $F_1$ and $F_2$ being a cumulative distribution functions and $\sup$ being the supremum function; shows that the Anderson-Darling test is weighted toward the tails of the distribution. In our test, these
tails will be high and low S/N pulses. This gives us an explanation why the KS and AD statistics diverge
when applying the MAD filter. The MAD filter stabilized the noise level and  separates low S/N candidates
from the noise, this improves the KS statistic, which is equally weighted among all the pulses. This
filter will clip bright pulses, reducing the returned fidelity of the S/N high tail and increasing the AD 
static. Which filter is most useful partially depends on the characteristics of the pulses of interest. We see that the outlier filters do a good job of increasing the
occupancy, that is, the percent of pulse windows that have a S/N above six. However, in
in this test, the Composite filter is too aggressive, which is indicated by the increased KS and AD distance. 
This indicates the Composite filter is useful to finding pulses, but care must be given when deriving pulse statistics. We should investigate other ways to remove RFI, such as machine learning and Cyclostationary methods, to remove RFI  while preserving more of the pulse for this task. We can further investigate the the pulse characteristics change by fitting burst bursts with burstfit \citep{burstfit} and reinject with \textsc{will}, filter and re-extract with \textsc{burstfit}. This will be an involved process that we will leave for future work. The \% flagged tells us the amount
of data that is flagged in each domain, the Composite \% flagged is the sum of the 
percent flagged of reach filter. The high cadence filters can make significant 
improvements to the recovered pulse characteristics while flagging only a small amount of 
of the data. 
 
Fig.~\ref{fig:ks_ad} shows a more nuanced view of the values reported in Table~\ref{tab:synthetic_pulses}. We see 
that increasing pulse power is often negatively correlated to the 
recoverability of injected distribution. The 
opposite is generally true for pulse width. Together, these two 
quantities make an energy density. Brighter pulses increase the energy 
density, wider pulses spread the energy over more time samples. Comparing 
the top and bottom plots of Fig.~\ref{fig:ks_ad} indicates that this 
correlation is stronger for the body of the pules (KS) than the tail of
of the energy distribution (AD). This difference is due to the difference
in sensitivity to the tails of the KS and AD tests. Interestingly, we do not also see a strong correlation with the frequency width of the pulse. The frequency width controls the energy density of the pulse, with wider widths spreading the pulse energy over more channels. The interesting dichotomy between these two is that the pulse is always summed over all frequencies, for time the pulse integrated only up to the width of the pulse.
This means RFI that is present across the bandpass is always summed, while across time is only summed up to the width of the pulse. This dichotomy hints that either this is a difference in how RFI acts across time or that our test is sensitive to the how the means of both distributions are made to be zero before being compared. The former is almost certainly a contributing factor. 
Gaussian noise is expected to integrate down with the square root on the number of samples summed, RFI is non-Gaussian \citep[][]{Gary-2007, Gary_2010, Nita_2010}.
So, we do not expect RFI to integrate in the same way as Gaussian noise. If we wanted to apply corrections to observed pulse populations, we should also futher investigate additional statistical tests to refine these results. Interestingly the characteristics other than power and width have little impact on the correlation.

\subsection{Pulsars —  Single Pulses — WILL}\label{sec:astrophysical_pulses}
The final test makes use of real pulses in GREENBURST data. Pulsars have more complex pulse structures than the synthetic ones previously modeled. To investigate this, we have chosen four pulsars that cover a range of DMs and brightnesses, as shown in Table~\ref{tab:psr_info}.  

\begin{deluxetable}{l|ccccccr}
 	\centering
 	\caption{Basic parameters for the four pulsars considered. For each pulsar, we list the period,
 	DM, and the mean flux density at 1400~MHz, S$_{1400}$. These values are 
 	as reported in \citet[][]{psr_cat}. Windows are the
 	number of pulse windows considered in the analysis.
 	}
 	\label{tab:psr_info}
 		\tablehead{
 		\colhead{Pulsar} & \colhead{Period [ms]} & \colhead{DM [pc $\text{cm}^{-3}$]} & \colhead{S$_{1400}$ [mJy]} & \colhead{Windows}
 		}
        \startdata
  		J1935$+$1616 & 359 & 159 & 58 & 8276\\
  		J0742$-$2822 & 167 & 73.7 & 26 & 7482\\
  		J0358$+$5413 & 156 & 57.1 & 23 & 10438\\
  		J1705$-$1906 & 300 & 22.9 & 5.66 & 10754\\
 	\enddata
\end{deluxetable}

As in the previous subsection, Table~\ref{tab:pulsar_response} reports the $\overline{\text{Amplitude}}$, $\overline{\text{Noise}}$,  $\overline{\text{S/N}}$, and folded S/N. 

\begin{deluxetable}{lllllclcr}
     \centering
     \caption{
     A comparison of the four filters on four pulsars. $\overline{\text{Amplitude}}$
     is the average amplitude of the box car convoluted pulse. $\overline{\text{Noise}}$ is the
     average noise level of the de-trended time series around each pulse. $\overline{\text{S/N}}$ is the mean S/N of each pulse. Folded S/N is the weighted mean
     of the folded at the pulsar period S/Ns across epochs. Spearman is the Spearman coefficient between the pulse
     S/N and the standard deviation. Occupancy is the percent of pulse windows with a
     pulse above S/N 6. \% Flagged is the percent of the data flagged by each filter. The 
     boxcar widths, in samples, used in this table in descending order are 16, 8, 8, 16.      Values in parentheses are the standard error of the least significant digit(s), computed via bootstrap methods \citep{bootstrap0, bootstap1}.
     }
     \label{tab:pulsar_response}
         \tablehead{
         \colhead{PSR} & \colhead{Filter} & \colhead{$\overline{\text{Amplitude}}$} & \colhead{$\overline{\text{Noise}}$} & \colhead{$\overline{\text{S/N}}$} &
         \colhead{Folded S/N} & \colhead{Spearman} & \colhead{Occupancy} & \colhead{\% Flagged}\\
         }
         \startdata
          & None    & 10.964$\pm$0.044      & 0.10040$\pm$0.00027   & 117.65$\pm$0.64 & 1640 & -0.664$\pm$0.011 & 98.1 & None\\
          & MAD     & 6.930$\pm$0.017       & 0.04508$\pm$0.00047   & 154.66$\pm$0.42 & 2000 & -0.395$\pm$0.011 & 98.3 & 4.48\\
          & FFT-MAD & 10.473$\pm$0.042      & 0.06200$\pm$0.00014   & 177.69$\pm$0.86 & 2000 & -0.593$\pm$0.011 & 98.2 & 4.82\\
          & High-pass & 10.957$\pm$0.045    & 0.10162$\pm$0.00027   & 116.13$\pm$0.63 & 2040 & -0.659$\pm$0.011 & 98.1 & 0.293\\
          \rot{\rlap{~J1935+1616}}
          & Composite$_3$ &  6.850$\pm$0.017 & 0.04509$\pm$0.00054 & 153.29$\pm$0.43  & 1780 &  -0.391$\pm$0.011 & 98.3 & 7.06\\
          \hline
          & None    & 3.307$\pm$0.021 & 0.11040$\pm$0.00024 & 31.20$\pm$0.22 & 133 &  -0.340$\pm$0.012 & 99.0 & None \\
          & MAD     & 2.533$\pm$0.013 & 0.05881$\pm$0.00089 & 43.60$\pm$0.23 & 199 & -0.186$\pm$0.012 & 99.9 & 5.15 \\
          & FFT-MAD & 3.200$\pm$0.020 & 0.08004$\pm$0.00017 & 41.37$\pm$0.28 & 167 & -0.321$\pm$0.011 & 99.7 & 2.92\\
          & High-pass & 3.309$\pm$0.021 & 0.11101$\pm$0.00024 & 31.01$\pm$0.22 & 153 & -0.335$\pm$0.012 & 98.9 & 0.293\\
          \rot{\rlap{~J0742-2822}}
          & Composite$_3$ & 2.514$\pm$0.013 & 0.05893$\pm$0.00091 & 43.22$\pm$0.23 & 202 & -0.194$\pm$0.012 & 99.9 & 5.72\\
          \hline
            & None  & 1.277$\pm$0.018   & 0.08149$\pm$0.00016 & 17.22$\pm$0.28 & 20.8 & -0.2853$\pm$0.0098 & 66.7 & None \\
            & MAD   & 0.7140$\pm$0.0050 & 0.04272$\pm$0.00069 & 16.45$\pm$0.11 & 11.2 & 0.1741$\pm$0.0098 & 84.9 & 4.59 \\
            & FFT-MAD & 1.219$\pm$0.018 & 0.06029$\pm$0.00089 & 21.05$\pm$0.32 & 23.2 & -0.1915$\pm$0.0099 & 75.7 & 4.85 \\
            & High-pass & 1.276$\pm$0.018 & 0.08188$\pm$0.00016 & 17.09$\pm$0.27 & 19.5 & -0.2821$\pm$0.0097 & 66.4 & 0.293 \\
          \rot{\rlap{~J0358+5413}}
          & Composite$_3$ & 0.7329$\pm$0.0053 & 0.04252$\pm$0.00066 & 17.00$\pm$0.12 & 14.7 & 0.1630$\pm$0.0098 & 85.0 & 6.55\\
          \hline
          & None & 2.996$\pm$0.014   & 0.21140$\pm$0.00038 & 14.543$\pm$0.071 &  95.1 & -0.2691$\pm$0.0096 & 93.5 & None  \\
         & MAD & 2.727$\pm$0.012& 0.07386$\pm$0.00016 & 38.05$\pm$0.17 & 218 & -0.2424$\pm$0.0097 & 99.9 & 4.82\\
         & FFT-MAD & 2.901$\pm$0.013 & 0.13564$\pm$0.00027 & 22.19$\pm$0.11 & 131 & -0.3432$\pm$0.0096 & 98.6 & 3.18 \\
         & High-pass & 2.688$\pm$0.013 & 0.21074$\pm$0.00037 & 13.043$\pm$0.065 & 34.2 & -0.2290$\pm$0.0097 & 90.2 & 0.293\\
         \rot{\rlap{~J1705-1906}}
         & Composite$_3$ & 2.487$\pm$0.012 & 0.07898$\pm$0.00012 & 31.76$\pm$0.14 & 61.5 &  -0.0770$\pm$0.0098 & 99.9  & 6.16\\
         \enddata
\end{deluxetable}

\begin{figure}
  \centering
 \raisebox{-0.5\height}{\includegraphics[width=0.45\columnwidth]{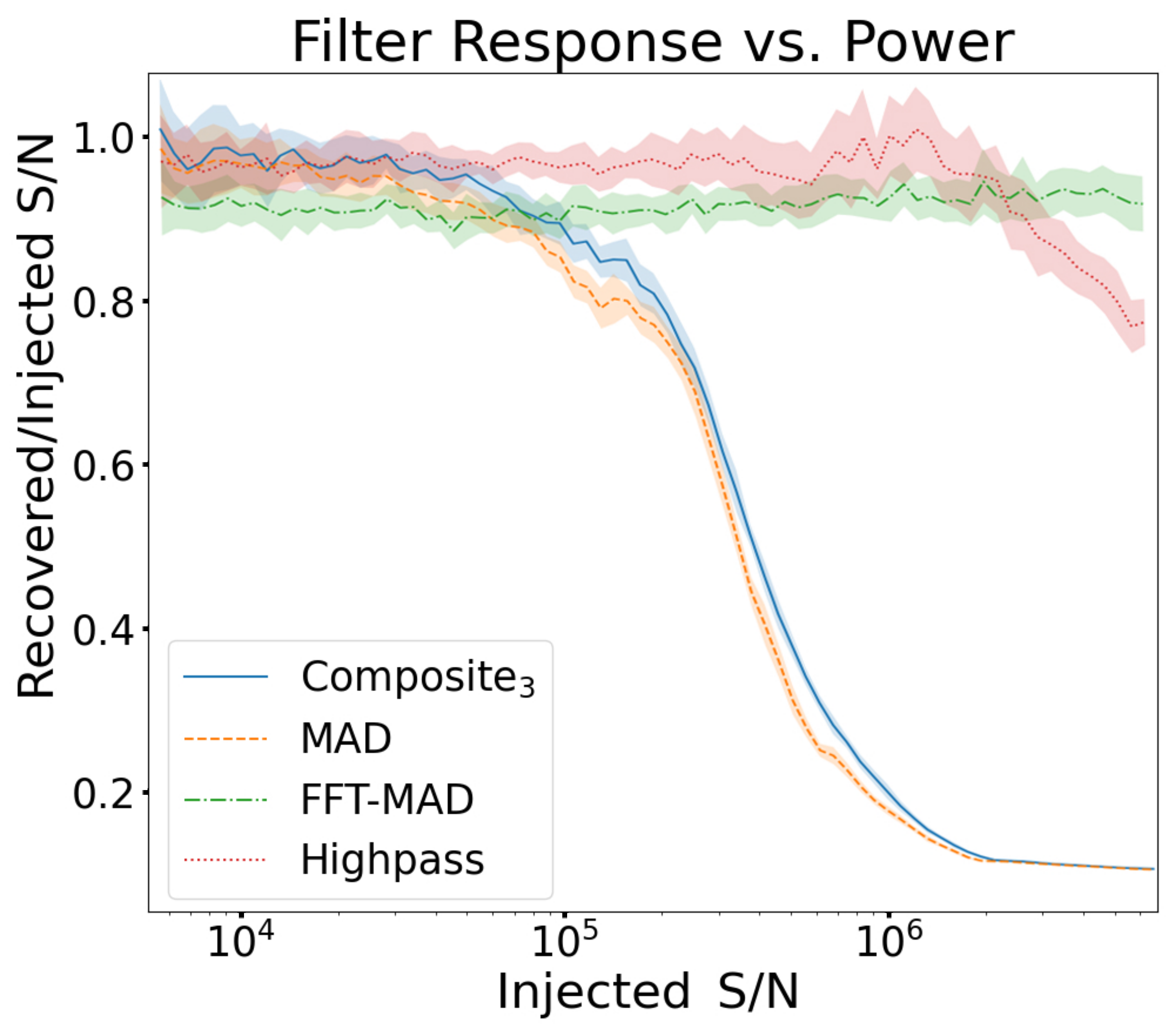}}
 \raisebox{-0.5\height}{\includegraphics[width=0.45\columnwidth]{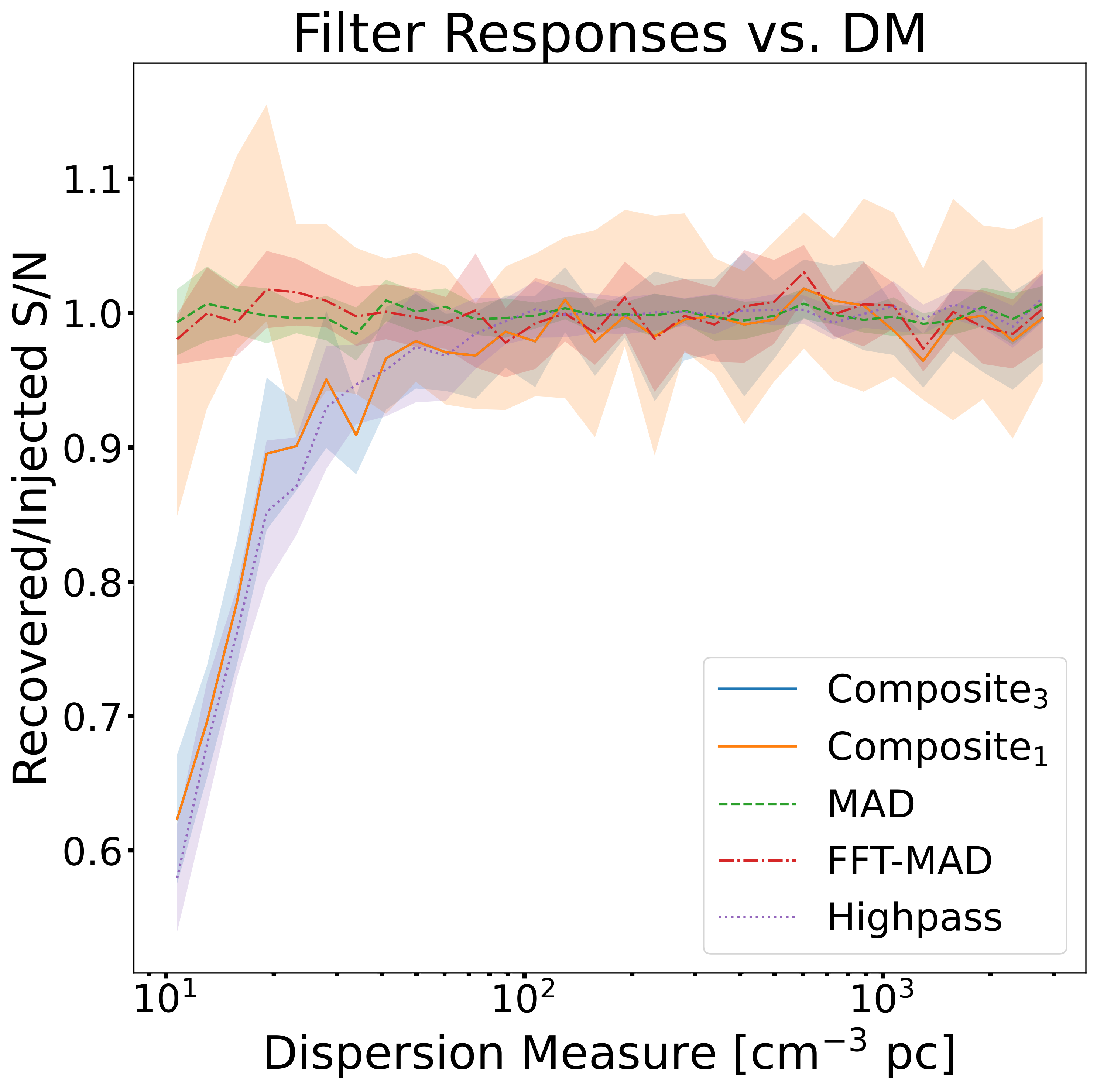}}
  \caption{Wideband, canonical, pulse injected into Gaussian data that has been scaled to resemble a GREENBURST band-pass
  in standard deviation and median. 
  The left plot shows a DM$=100$~cm$^{-3}$~pc
  injected with various intensities. The right plot shows the response for a pulsar
  with S/N$=50$. Both pulses are Gaussians in time and frequency with time sigma of 1 ms, $\tau=2$, center frequency 
  of 1440~MHz, frequency sigma of 300~MHz, a spectral index of 0.5, and 2 scintles. One sigma
  error bars are shown. Composite$_3$ is the composite filter with the first three Fourier components removed,
  Composite$_1$ removes the first Fourier component. 
  }
  \label{fig:filter_response}
 \end{figure}

\begin{figure}
    \centering
    \gridline{
    \fig{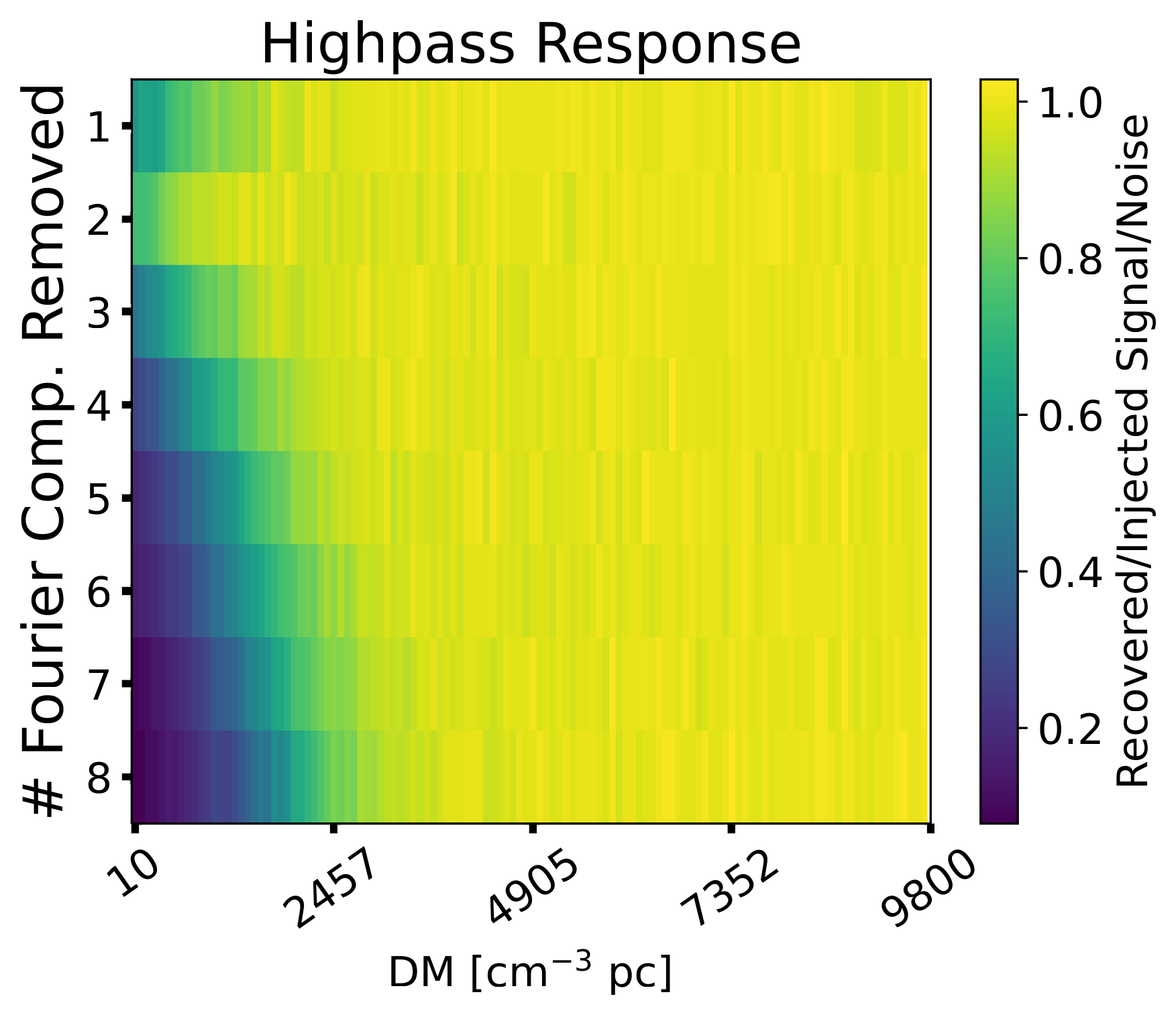}{0.40\columnwidth}{}
    \fig{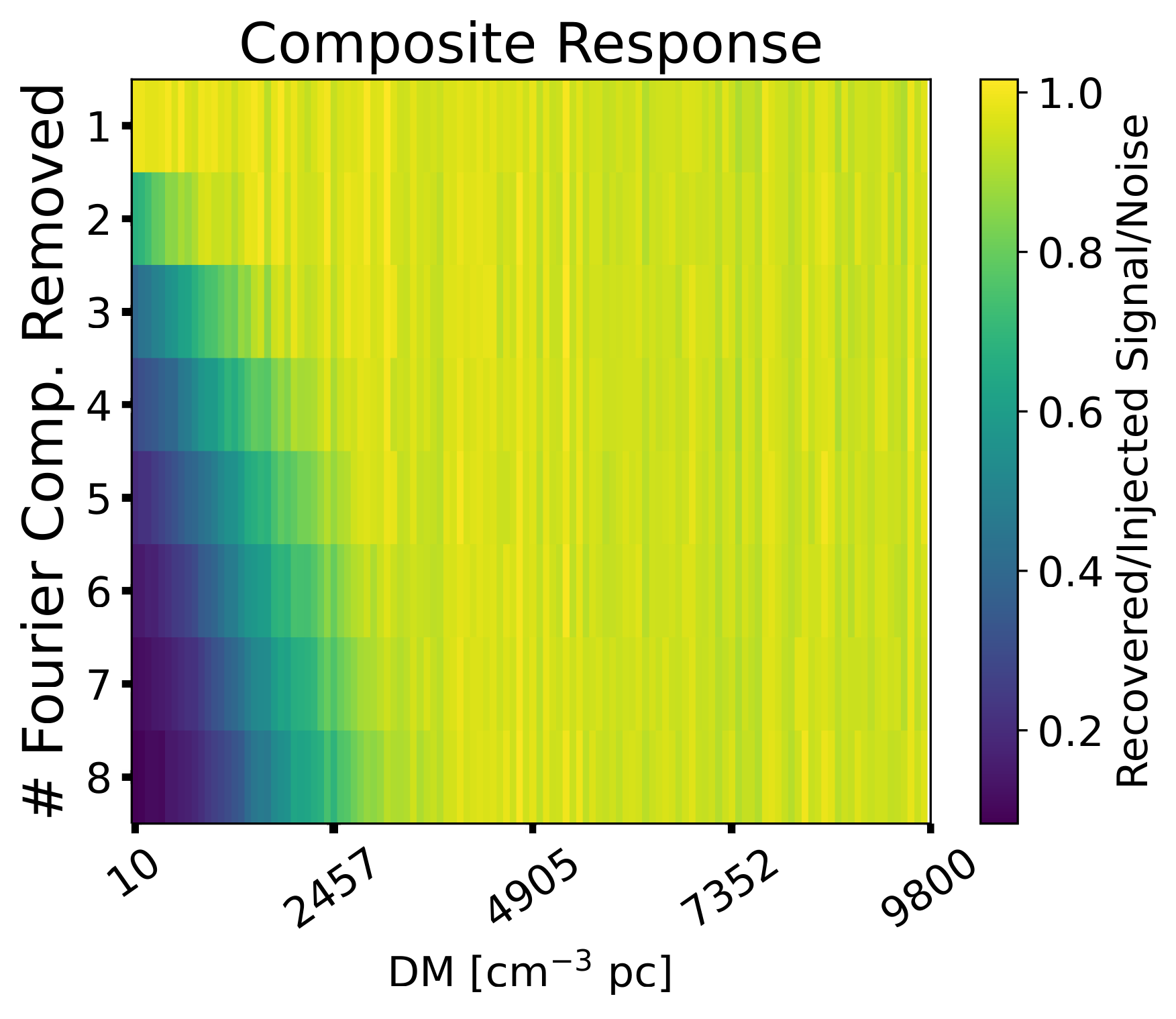}{0.40\columnwidth}{}
    }
    \gridline{
    \fig{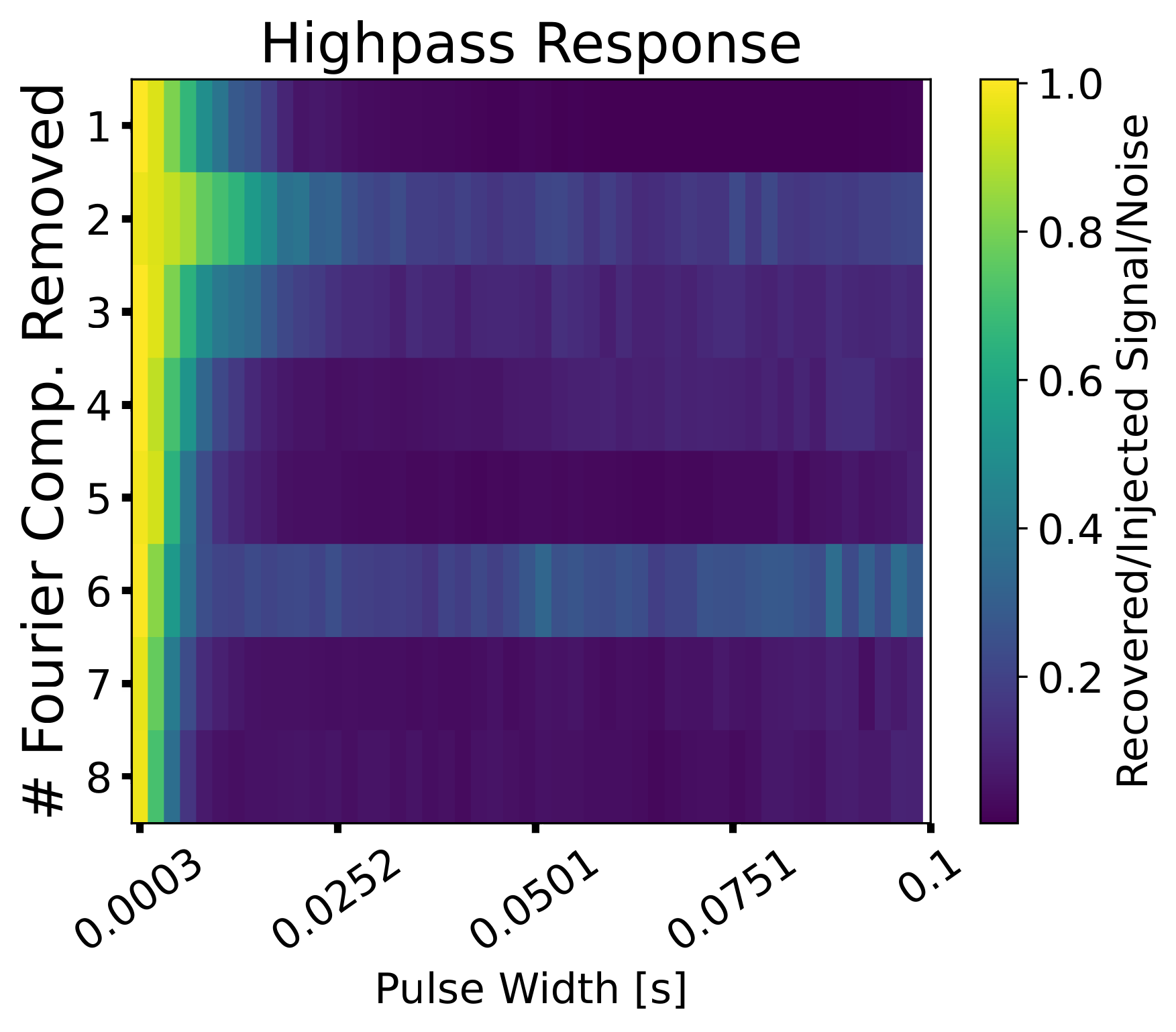}{0.40\columnwidth}{}
    \fig{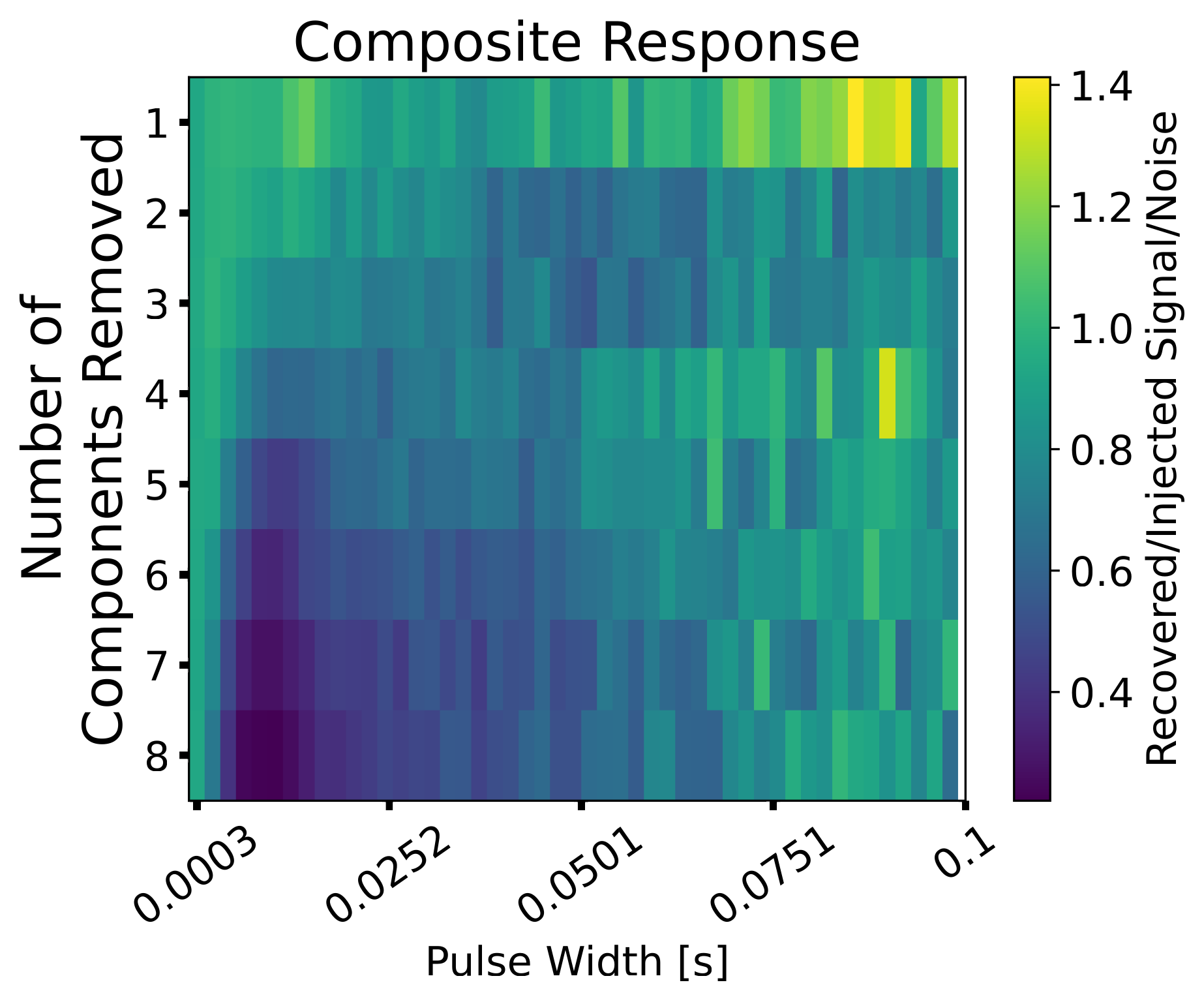}{0.40\columnwidth}{}
    }
    \gridline{
    \fig{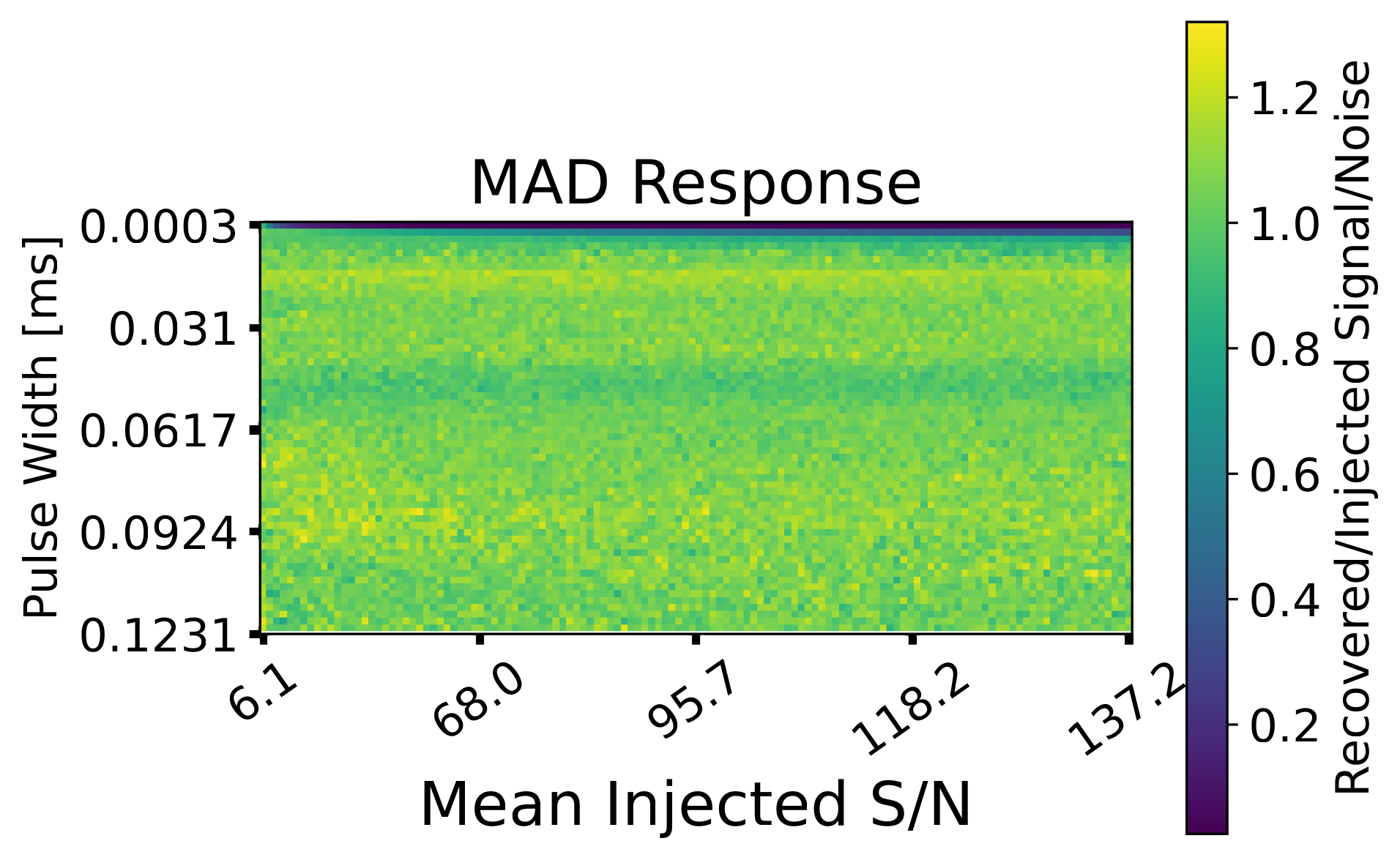}{0.40\columnwidth}{}
    \fig{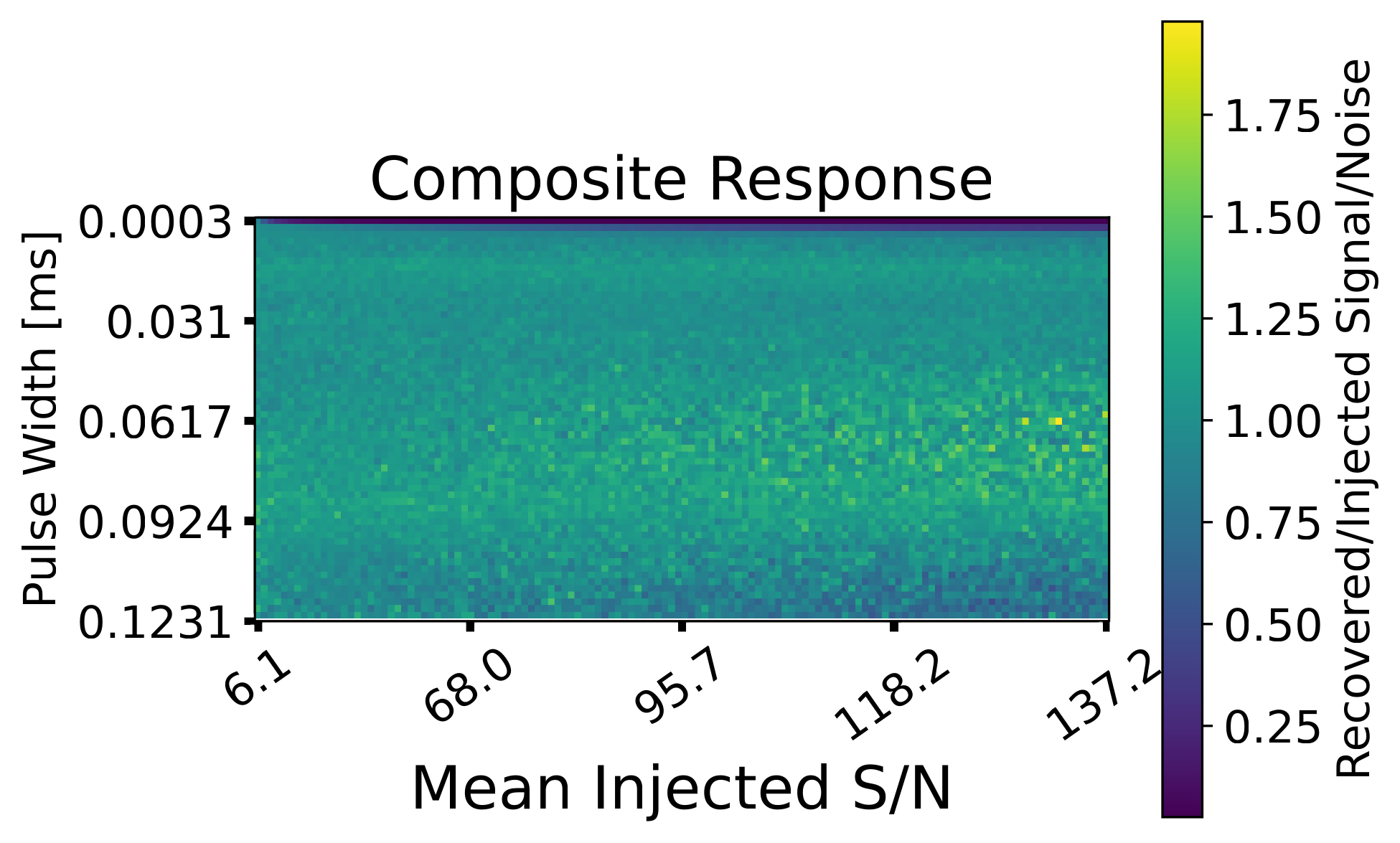}{0.40\columnwidth}{}
    }
    \caption{The same pulse as in Fig.~\ref{fig:filter_response}, varied over highpass filter strength, DM, pulse width, and power. Top — The pulse varied over DM and highpass filter strength. Middle — The pulse at DM = 500 pc $\text{cm}^{-3}$ varied over pulse width for the highpass filter (Left), and Composite Filter (Right) when removing Fourier elements one to eight. Bottom — The pulse with changing width and power. 
    As hinted by Fig.~\ref{fig:ks_ad}, pulse width has a significant impact on filter response. In the top row, wider pulses occupy more of the bandpass, and are more heavily removed by the highpass filter. Wider pulses are less affected by the MAD filter because they occupy more the of the square of dynamic spectra used to calculate the median and MAD. The top right plot shows a complex interaction between the MAD and highpass filters.}
    \label{fig:filter_transfer}
\end{figure}

We again see that
the filters
can significantly improve both the individual pulse S/N and the stacked S/N. However, there is no one-size-fits-all solution. For PSR~J1935+1616, for example, the FFT-MAD filter performs the 
best, but the time domain MAD filter removes more of the pulse than is optimal. This result is shown in Figs~\ref{fig:filter_response}\&\ref{fig:filter_transfer}. 

Fig.~\ref{fig:filter_response} maps the the response of an ensemble of filters over pulse injected S/N and Dispersion Measure.  On the left we see that very bright pulses lose energy to the MAD (and thus also composite filter). On the right plot we see low DM pulses get filtered by the Highpass (and also the composite filter). Fig.~\ref{fig:filter_transfer} shows 
the filter response based on how aggressively the dynamic spectra is highpass filtered (top four plots). The top two plots show that by increasing the number of Fourier components removed, more of the pulse is removed.
The middle two plots show that wider pulses lose more pulse energy to the highpass filter, and this increased with highpass filter strength. The middle right plot shows more complex behavior than expected. The bottom two plots show filter response across pulse width and S/N. This shows that pulses bright, narrow band pulses loose significant power to the MAD filter.

\begin{figure}
  \centering
  \raisebox{-0.5\height}{\includegraphics[width=0.75\columnwidth]{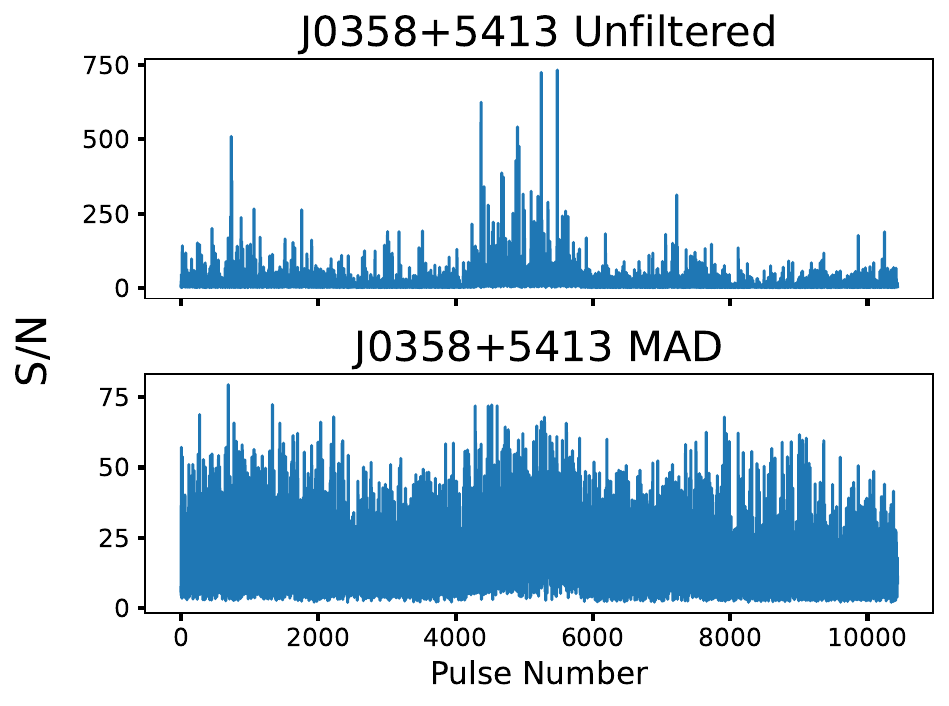}}
  \caption{Pulses from J0358+5413, the top panel shows the pulse S/Ns without filtering. 
  The bottom panel shows the S/Ns after MAD filtering. The very bright pulses (S/N$>100$) are substantially reduced after filtering.
  }
  \label{fig:J0358}
 \end{figure}

A less expected result is PSR~J0358+5413, J0358
is about half as bright as J1935+1616, but exhibits the same effect. This is because
much of J0358+5413's energy is in very bright (above 200 S/N) pulses. This is shown in Fig.~\ref{fig:J0358}.

The less bright pulses are not removed, and their S/N increases as expected.
This gives a mixed result for the MAD filter, the number of windows with a pulse above six sigma
increases while the average pulse S/N decreases. We see that
  some very bright pulses make a significant portion of the total pulse energy. As expected by Fig.~\ref{fig:filter_response}, the composite filter greatly reduces the energy of these bright pulses.
  However, the S/N of the pulses closer to the noise level increases. These leads to the results in
  Table~\ref{tab:pulsar_response}, where we see an increase in the number of pulse 
  windows with a pulse above S/N of six. However, the average S/N decreases. This is also reflected
  in the stacked S/N as well. The bright pulses contain a significant portion of the total power, 
  and their removal adversely affects the combined S/N. 
 
PSR~J1705--1906 has a low DM of 23~cm$^{-3}$~pc, this causes the pulse to take up a significant fraction of the bandwidth. The high-pass filter subsequently removes 10\% of the pulse power. This result again 
shows the need to consider the source while choosing the filtering. Instead of three Fourier components, we
should have excised one or zero Fourier components. We will explore the former later in this section.

Unlike the case with synthetic pulses, we do not know the intrinsic power distribution, so we cannot perform the KS or AD tests.
Instead, we will use the Spearman's rank correlation coefficient \citep{spearman}  to measure the relationship between a pulse's S/N and 
the noise level. 

The Spearman correlation coefficient $r_s$ is defined by 
\begin{equation}
    rs = \frac{\text{cov[R[}X\text{],R[}Y\text{]]}}{\sigma_{\text{R[X]}}\sigma_{\text{R[Y]}}}
\end{equation}
Where $\text{cov[R[}X\text{],R[}Y\text{]]}$ is the covariance of the rank variables, the rank of $X$ or $Y$. $\sigma_{\text{R[X]}}$ $\sigma_{\text{R[Y]}}$ refer to the standard deviations of the rank variables R[X] and R[Y].
We choose the Spearman rank coefficient because it describes how well the relationship between two
variables can be described by a monotonic function. We expect for a given signal brightness that S/N and noise level will
be monotonically decreasing. In the ideal case, S/N and noise level will be uncorrelated, the noise will be 
Normal and any changes in pulse S/N will
reflect intrinsic changes to the source brightness. If these variables are correlated, then S/N is partially measuring
the evolving radio frequency environment, and the pulse measurements as reported by S/N will be less meaningful.

We see that this raw value is negative, as expected. With filtering, we can bring this metric
closer to zero. This trend to zero is an indication that the S/Ns better reflect the intrinsic
pulse energy and are less affected by the varying radio frequency environment. 
Even though these are bright pulses, the filtering noticeably increases the percentage of windows
with pulses. This means that search pipelines should be able to find more pulses. These pulses can then be 
used to better understand the source.
The percentage flagged is higher than for the simulated pulses. This reflects that these observations
took place at times throughout the day, leading to a worse RFI environment. The Composite
filter removes less than ten percent of the data. This small amount of flagging indicates an advantage of 
filtering on the shortest timescales. Flagging methods that remove blocks of data will often flag higher 
fractions of the data.

The above paragraphs have shown that we must be cautious when using the high-pass filtering.
For the remainder of this section, we will only remove the first Fourier component. This is equivalent to the high-pass filtering described in
\cite{Eatough-2009}. We therefore relabel high-pass as Eatough. Our 
Composite$_1$ filter will still be more effective than the classical 
implementation due to the reasons discussed in Section~\ref{sec:composite} and Section~\ref{sec:dm-time}. In this section, we have shown that High-pass filter alone is not particularly effective. As discussed in more detail in Section~\ref{sec:composite}, very bright narrow band RFI dominated the dynamic spectra. While the high-pass filter can remove the broadband RFI, it also spreads out the narrow band RFI unless we remove the narrow band RFI before highpass filtering.

\subsection{Single pulse searches}\label{sec:single_pulse_searches}
We also searched the pulsar observations with \textsc{Heimdall} \citep{Barsdell}, from DMs in the rage 10---10,000~cm$^{-3}$~pc. For all the \textsc{Heimdall} runs discussed in this paper, we will use the default \textsc{Heimdall} RFI filters on. As described in \citet{Barsdell}, there are two filters, a narrowband RFI filter and a broadband filter. The latter is shown in Fig.~\ref{fig:zero-dm}. The \textsc{Heimdall} filters are fast; so turning them off will not save much time. Therefore there is little motivation to turn these filters off, as in \citep{iqrm}, we will leave them on running Heimdall with its default settings. Thus, our Heimdall-Your-Fetch search pipeline only changes by adding a prefiltering step. The Heimdall RFI filters will also still be useful after preprocessing with the filters being discussed in this paper. The FFT-MAD does not intentionally remove non-periodic broadband interference, and the high-pass does not remove narrowband. When we apply the highpass filter before Heimdall, the Heimdall Zero-DM filter will not have any effect. Likewise, the MAD filter removes most of the narrow-bad RFI, leaving little for the Heimdall filter to remove. In Section~\ref{sec:dm-time}, we will see that even with the Heidmall zero-DM filter enabled, we see evidence of broad band RFI in the pulse candidates. We do not 
use additional RFI filtering during candidate creation, except when noted. 

\begin{figure*}
    \gridline{
        \fig{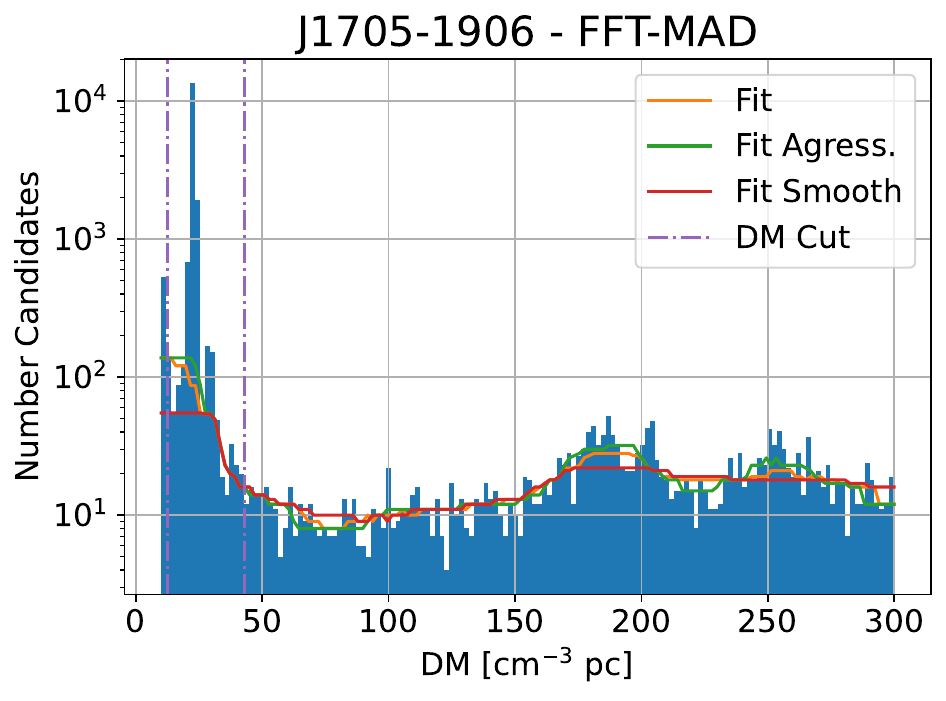}{0.46\textwidth}{}
        \fig{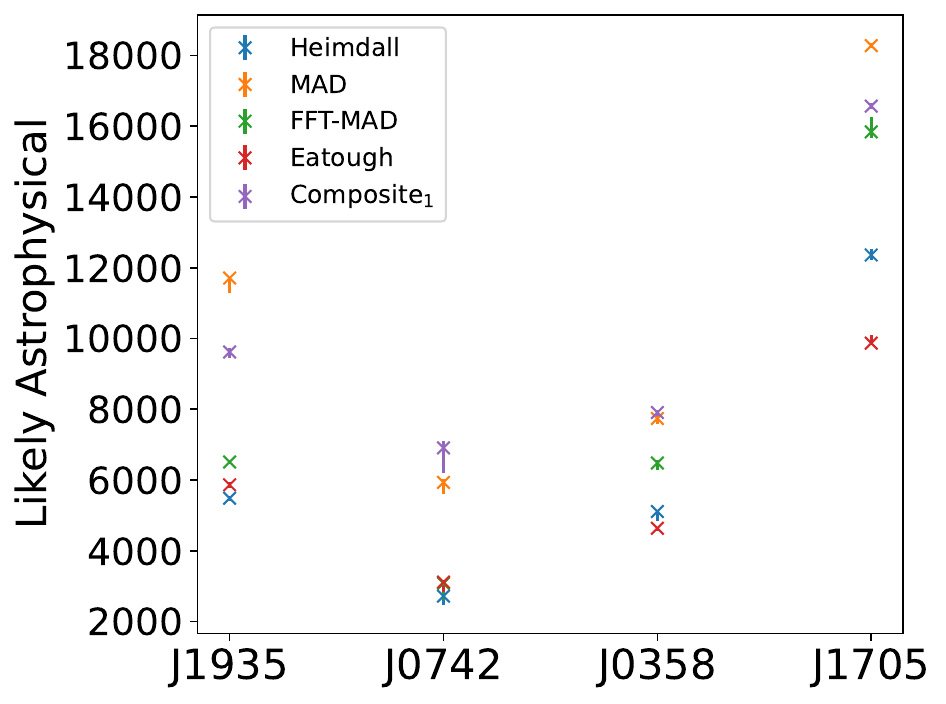}{0.46\textwidth}{}
        }
    \gridline{
        \fig{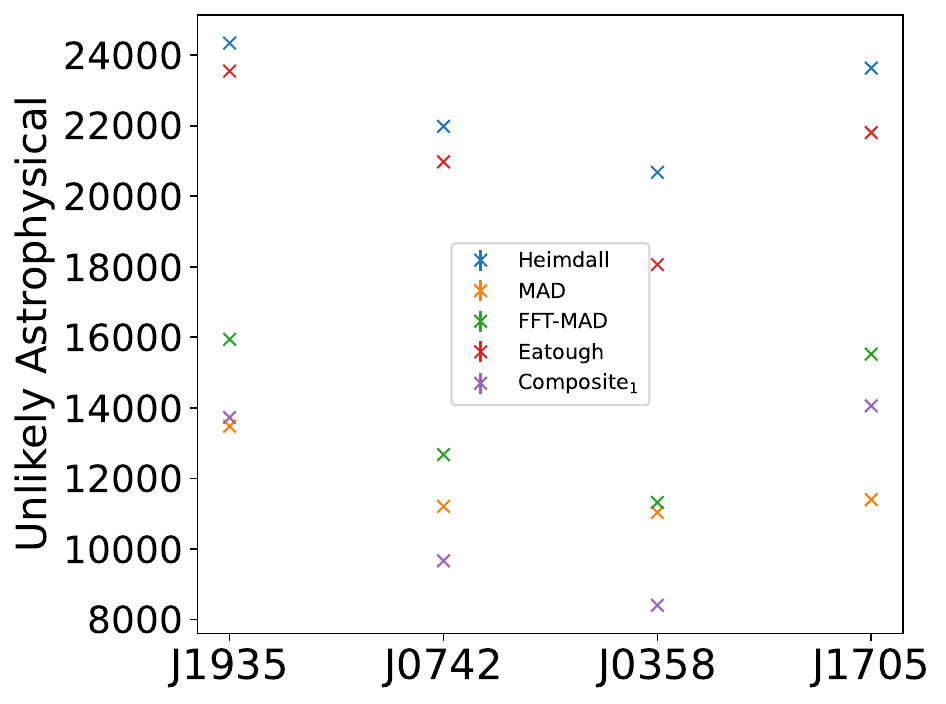}{0.46\textwidth}{}
        \fig{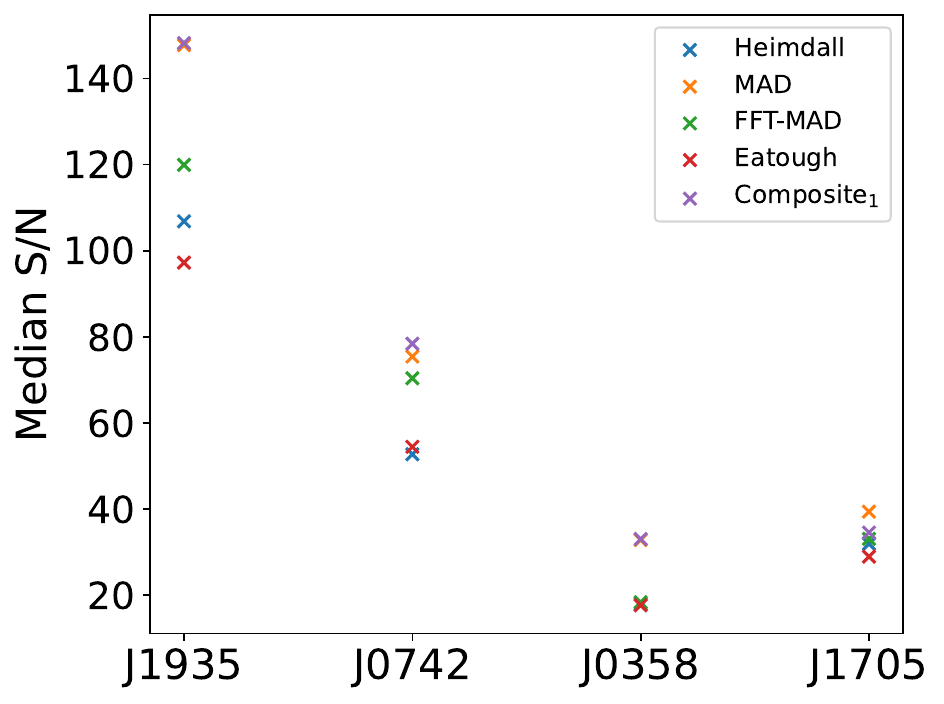}{0.46\textwidth}{}
     }
         \caption{
         \textsc{Heimdall} candidates from excess DM. Top left: candidate separation 
         in DM space for the FFT filter applied to J1705$-$1906.  Top right: excess candidates
         for each pulsar. Bottom left: number of candidates that are unlikely astrophysical. Bottom right: median S/N of candidates in the DM slice around the pulsar. The error bars are derived from the three smoothed fits, the center value 
         reported as the dot, with the other two making the high and low estimate.
    }
    \label{fig:heimdall_excess}
\end{figure*}

The outputs are processed in two ways. The first method is shown in 
Figure.~\ref{fig:heimdall_excess}. 

Note that there are more pulses there than windows in Table~\ref{tab:psr_info} because 
we initially limited our search to the best parts of the observations. This avoids time when the noise diode was on.  Instead, when we search the complete dataset, a  
plot of the candidates as a function of DM shows a large spike in candidates
at the pulsar's DM. We estimate the number of astrophysical pulses as follows. First, we fit a smooth curve to
the histogram bin using \textsc{jess}'s median filter, described in Section~\ref{sec:mad_filter}, with filter
lengths of 15, 25, 35 bins. Second, we consider DMs $\pm 20$ cm$^{-3}$~pc from the pulsar's DM (except for 
J1705$-$1906, where the lower limit is 12.5~cm$^{-3}$~pc  to avoid the spike shown in the top left of 
Fig.~\ref{fig:heimdall_excess}). Finally, we calculate the number of pulses above the three lines and report this
value as \# Likely Astrophysical in the top right plot of Fig.~\ref{fig:heimdall_excess}. We initially considered including a S/N cut. This looked promising for separating pulsar and RFI for PSRs J1935$+$1616 and J0742$-$2822, but 
not J0358$+$5413 and J1705$-$1906. We decided not to pursue this cut further. The point 
is the middle value of the three fits, and the error bars are the remaining two values. Like Table~\ref{tab:pulsar_response}, we see that there is no one size fits all solution. We also see that
MAD and Composite filter do a good job of increasing the number of likely astrophysical pulses and 
decreasing the number of RFI candidates, shown by the bottom left of Fig.~\ref{fig:heimdall_excess}. 
Since the number of pulses exceeds of background RFI candidates within all the pulse 
windows, the median pulse S/N is informative, and we see even with the bright J1935$+$1616 there is an appreciable increase in median S/N.

\begin{figure*}
    \gridline{
        \fig{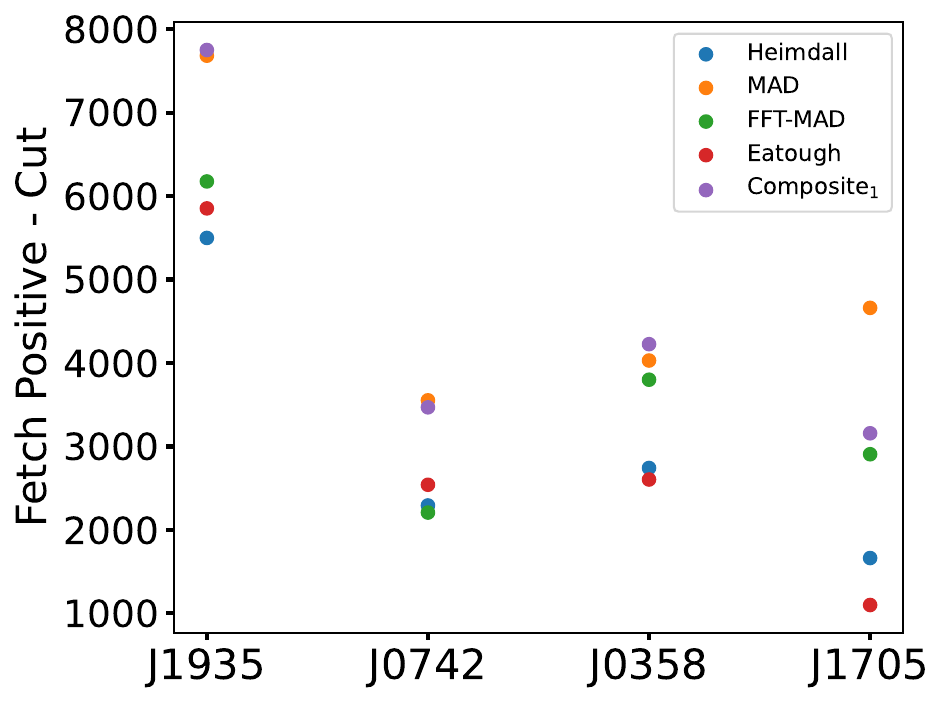}{0.46\textwidth}{}
        \fig{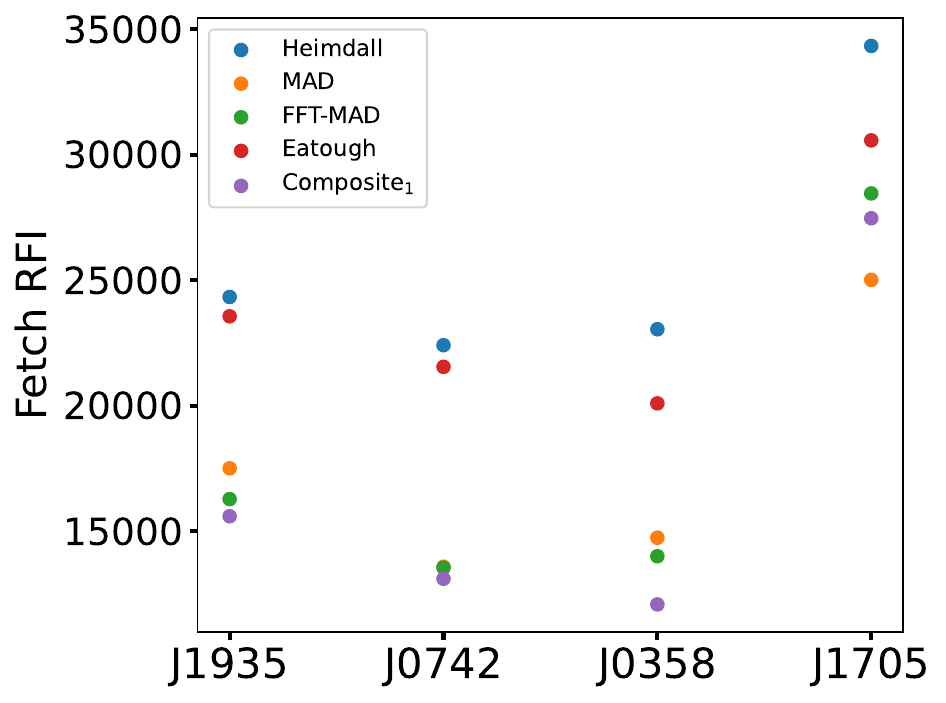}{0.46\textwidth}{}
    }
    \gridline{
        \fig{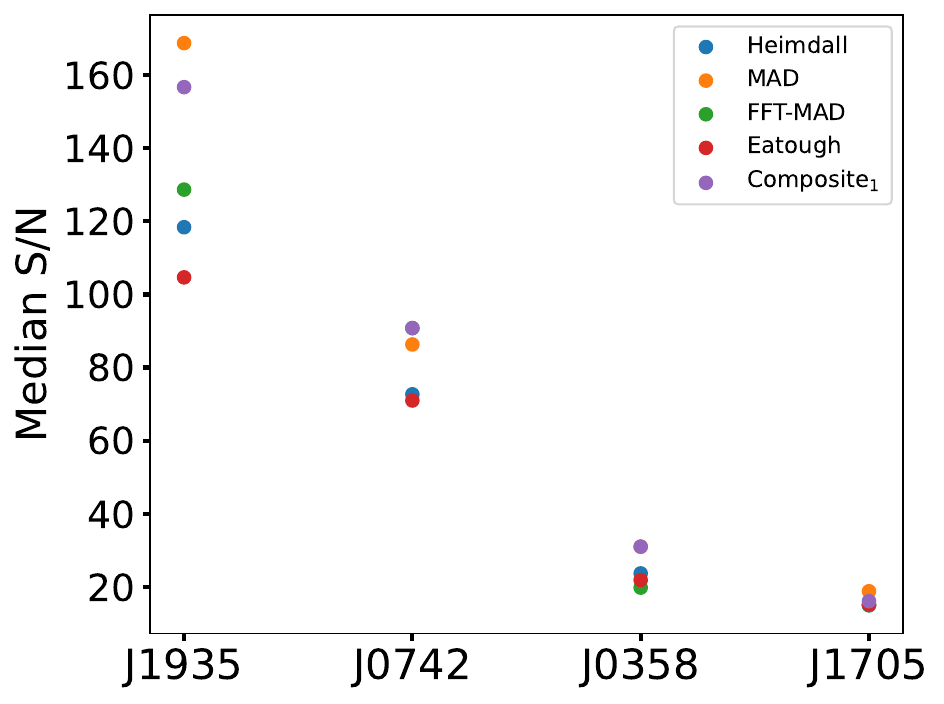}{0.46\textwidth}{}
        \fig{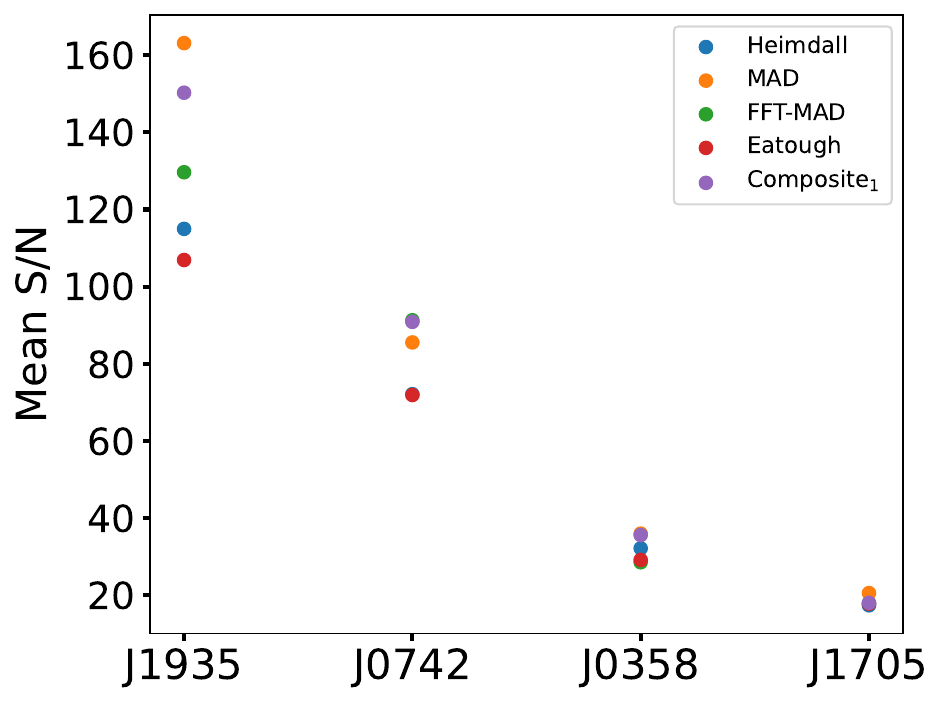}{0.46\textwidth}{}
     }
         \caption{
         The 
         \textsc{Heimdall}
         candidates analyzed  using \textsc{Fetch} \citep{fetch}. Top left: number of
         positive candidates, within the DM range, $\pm 20\text{ pc } \text{cm}^{-3} $ of the expected pulsar DM (except for J1705 witha  lower cut off of 12.5 $\text{pc }\text{cm}^{-3} $ Top right: number of candidates classified as RFI
         by \textsc{Fetch}. Bottom left: median S/N of the positive candidates. Bottom right: mean S/N of the positive candidates. 
    }
    \label{fig:heimdall_fetch}
\end{figure*}

We used \textsc{Fetch} \citep{fetch} \textsc{model a} to classify the \textsc{Heimdall} candidates, \textsc{model a} as an Accuracy of 99.88\%, Recall of 99.92\% and an Fscore 99.87\%. \citep[][]{fetch}. The numbers of positive candidates are reported in the top left of Fig.~\ref{fig:heimdall_fetch}. To check if \textsc{Fetch} is not falsely classifying RFI as pulses, we checked if there are any \textsc{Fetch} outside $\pm 20\text{ pc } \text{cm}^{-3} $ of the expected pulsar DM (except for J1705, which we used a lower cut off of 12.5 $\text{pc } \text{cm}^{-3} $ ). We to not have any \textsc{Fetch} positive candidates outside this DM window, indicative that our classifier is working well.
Comparing the top left of Fig.~\ref{fig:heimdall_fetch} and top right panel of Fig.~\ref{fig:heimdall_excess}
hints a severe selection effect of the classifier, especially for J1705$-$1906. 
A tabular form of Figs.~\ref{fig:heimdall_excess} \& \ref{fig:heimdall_fetch} is given in 
Table~\ref{tab:heimdall}. From Table~\ref{tab:heimdall}, we see \textsc{Fetch} has a selection effect, favoring brighter pulses. This brightness selection effect has been noted by others, including \cite{Nimmo_2023}. 
This unknown selection function is another reason (in addition to the need to report noise values)
that we used \textsc{Will}
in Sections~\ref{sec:sim}  \& \ref{sec:astrophysical_pulses}, and not the \textsc{Heimdall} search pipeline.

Instead of machine classification, we relied on DM, time, and width filtering.
As shown by 
Fig.~\ref{fig:heimdall_excess}, DM filtering alone can give us regions where
$> 80$\% of the candidates are from the astrophysical source. 
The Composite filter does well at 
limiting the number of pulses that  \textsc{Fetch} will classify as negative. At low DM, turning off the high-pass filtering would be beneficial, as expected. Even with high-pass filtering, we see a significant increase over \textsc{Heimdall}.  This is important because 
making the candidates for classification is often the most time consuming step in a 
single pulse search pipeline, with higher DM candidates taking longer
to create. Again, the bottom row of Fig.~\ref{fig:heimdall_fetch} shows increases in mean and median
S/N.

\begin{deluxetable*}{llcccccccc}
\rotate
    \tablehead{
         PSR & Filter & \# Likely Pulses & Median S/N$_{\text{slice}}$  & Unlikely Astrophysical & \textsc{Fetch} Pulses & \textsc{Fetch} RFI & Median S/N &  $\overline{\text{S/N}}$   \\
    }
    \startdata
          & Heimdall    &   5486 +92/-73    &   106.8   &   24344 +73/-92   &   5499    &   24331   &   118.4   &   115.0   \\
          & MAD         &   11708 +418/-25  &   147.7   &   13479 +25/-418  &   7683    &   17504   &   168.7   &   163.1   \\
          & FFT-MAD     &   6511 +51/-16    &   119.9   &   15942 +16/-51   &   6176    &   16277   &   128.7   &   129.7   \\
          & Eatough    &   5863 +92/-67    &   97.2    &   23550 +67/-92   &   5853    &   23560   &   104.7   &   106.9   \\
          \rot{\rlap{~J1935+1616}}
          & Composite$_1$   &   9620 +171/-101  &   148.2   &   13724 +101/-171 &   7751    &   15593   &   156.7   &   150.3   \\
          \hline
          & Heimdall    &    2718 +260/-43  &   52.7    &   21984 +43/-260  &   2293    &   22409   &   72.7    &   72.1    \\
          & MAD         &   5933 +335/-101  &   75.4    &   11202 +101/-335 &   3553    &   13582   &   86.3    &   85.5    \\
          & FFT-MAD     &   3079 +471/-65   &   70.4    &   12672 +65/-471  &   2208    &   13543   &   90.8    &   91.3    \\
          & Eatough    &   3111 +277/-64   &   54.5    &   20977 +64/-277  &   2541    &   21547   &   71.0    &   71.9    \\
          \rot{\rlap{~J0742-2822}}
          & Composite$_1$ & 6911 +724/-194  &   78.4    &   9661 +194/-724  &   3469    &   13103   &   90.8    &   90.9    \\
          \hline
          & Heimdall    &   5109 +259/-44   &   18.0    &   20678 +44/-259  &   2742    &   23045   &   23.7    &   32.1    \\
          & MAD         &   7737 +140/-57   &   32.8    &   11030 +57/-140  &   4030    &   14737   &   31.0    &   36.0    \\
          & FFT-MAD     &   6482 +196/-87   &   18.4    &   11317 +87/-196  &   3800    &   13999   &   19.8    &   28.5    \\
          & Eatough    &   4635 +80/-15    &   17.7    &   18061 +15/-80   &   2604    &   20092   &   21.9    &   29.2    \\
          \rot{\rlap{~J0358+5413}}
          & Composite$_1$ & 7907 +213/-65   &   33.0    &   8400 +65/-213   &   4226    &   12081   &   31.0    &   35.6    \\
          \hline
         & Heimdall &   12361 +135/-178 &   31.9    &   23637 +178/-135 &   1663    &   34335   &   15.0    &   17.4    \\
         & MAD      &   18279 +74/-72   &   39.4    &   11397 +72/-74   &   4661    &   25015   &   18.8    &   20.5    \\
         & FFT-MAD  &   15839 +170/-428 &   33.1    &   15525 +428/-170 &   2907    &   28457   &   15.1    &   17.8    \\
         & Eatough &   9868 +52/-234   &   29.0    &   21804 +234/-52  &   1100    &   30572   &   15.2    &   17.8    \\
         \rot{\rlap{~J1705-1906}}
         & Composite$_1$ &  16570 +82/-80   &   34.6    &   14057 +80/-82   &   3158    &   27469   &   16.1    &   18.0    \\
         \hline
         & Heimdall &   25674 +746/-338 &   35.4    &   90643 +338/-746 &   12197   &   104120  &   73.9    &   75.0    \\
         & MAD      &   43657 +967/-255 &   49.5    &   47108 +255/-967 &   19927   &   70838   &   78.7    &   90.2    \\
         &  FFT-MAD &   31911 +888/-596 &   39.8    &   55456 +596/-888 &   15091   &   72276   &   70.7    &   77.0    \\
         & Eatough &   23477 +501/-380 &   35.1    &   84392 +380/-501 &   12098   &   95771   &   73.9    &   74.7    \\
        \rot{\rlap{~~~~~~~Total}}
        & Composite$_1$ & 41008 +1190/-440  &   49.0    & 45842 +440/-1190  &   18604   &   68246   &   89.5    &   90.7    \\
        \hline
     \enddata
     \caption{Tabular form of Figs.~\ref{fig:heimdall_excess} \& \ref{fig:heimdall_fetch}. \# Likely Pulses denoted the likely number of pulses based on excess DM candidates, as described in Fig.~\ref{fig:heimdall_excess}. Median S/N$_{\text{slice}}$ is the median S/N in the 40 DM units slice around pulsar DM.  Unlikely astrophysical sources are likely RFI based on the excess DM method. \textsc{Fetch} Pulses are \textsc{Fetch} classified as astrophysical. \textsc{Fetch} RFI are candidates classified as RFI. Median S/N and $\overline{\text{S/N}}$ is the median and mean S/N of \textsc{Fetch} astrophysical candidates, respectively. Comparing the median S/N$_{\text{slice}}$ and median S/N indicates that the \textsc{Fetch} selection function is S/N dependent.}
     \label{tab:heimdall}
\end{deluxetable*}

\subsection{DM-time candidate plots}\label{sec:dm-time}
\begin{figure}
\gridline{
         \fig{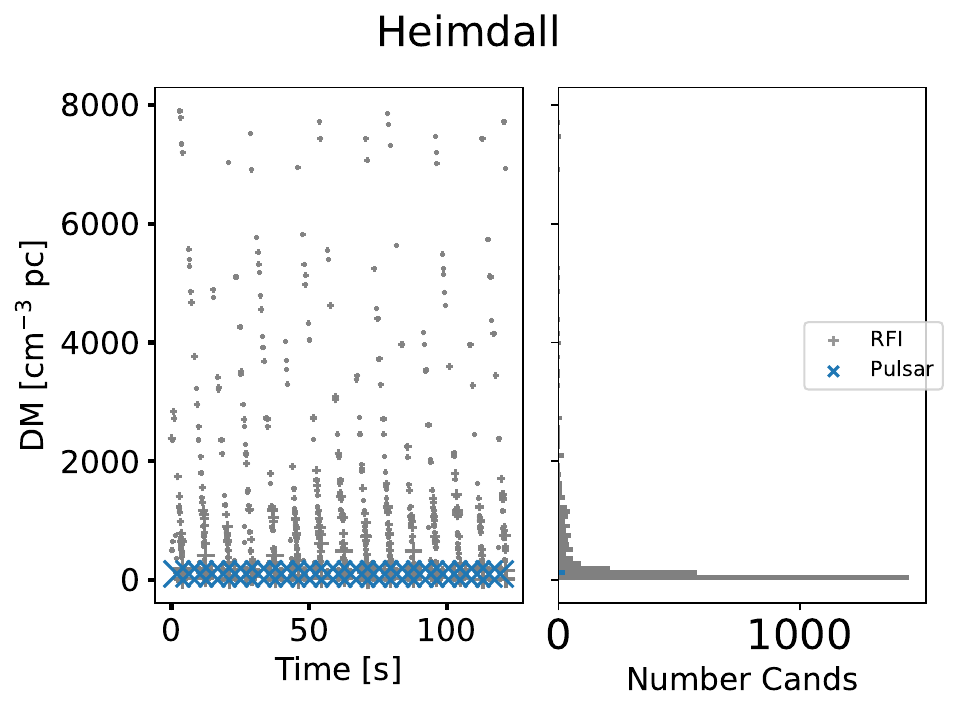}{0.45\textwidth}{}
         \fig{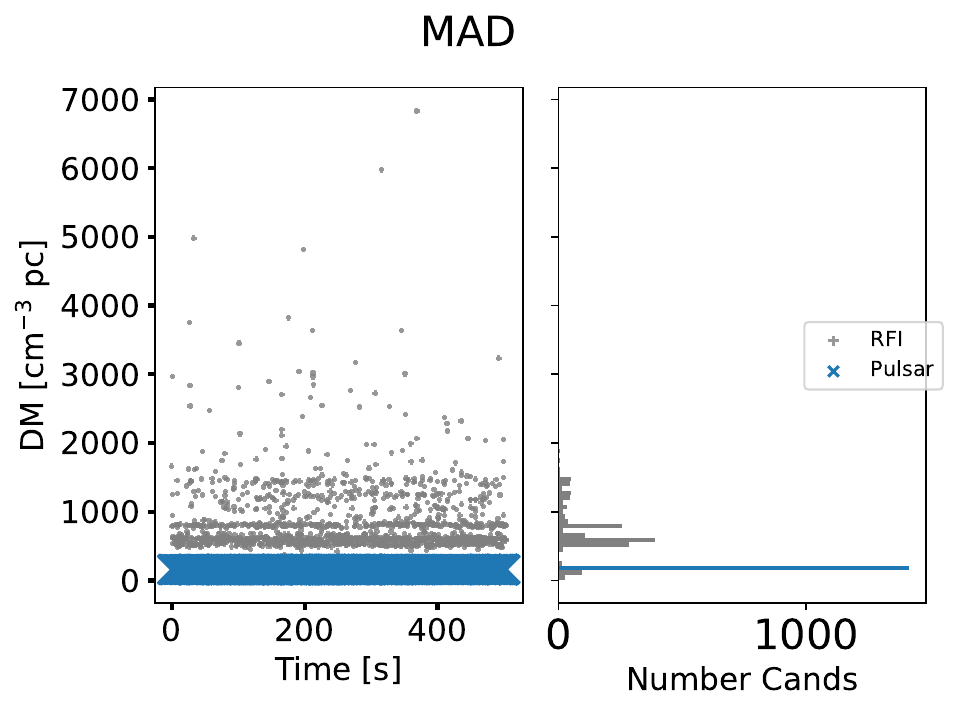}{0.45\textwidth}{}
         }
\gridline{
         \fig{Data_2021-02-09_11-10-48_fft_dm.pdf}{0.45\textwidth}{}
         \fig{Data_2021-02-09_11-10-48_Eatough_dm.pdf}{0.45\textwidth}{}
         }
\gridline{
         \fig{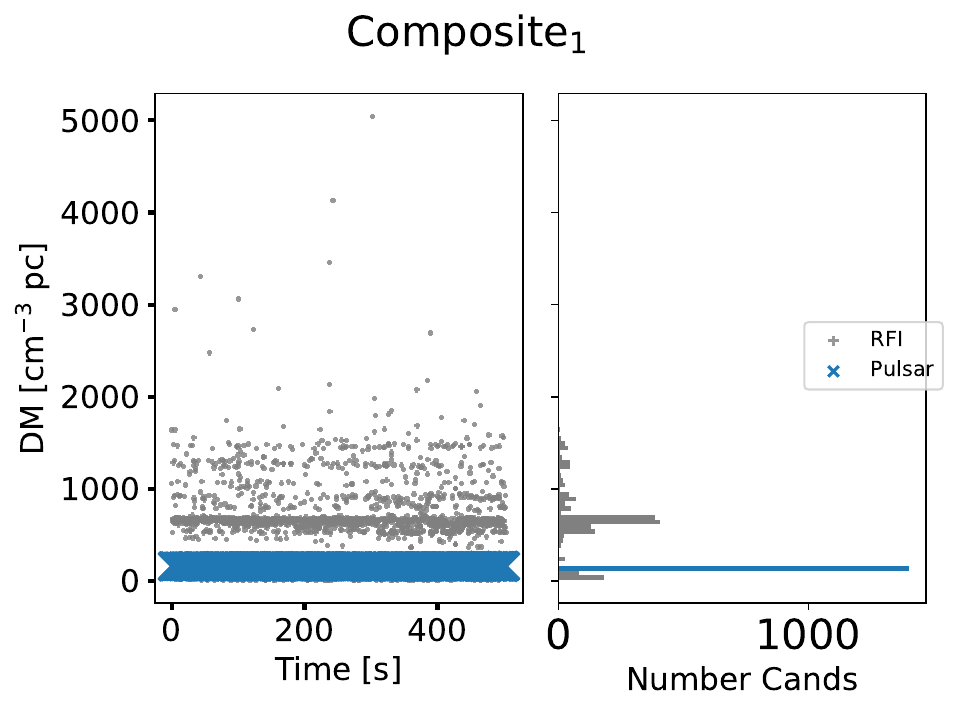}{0.45\textwidth}{}
         \fig{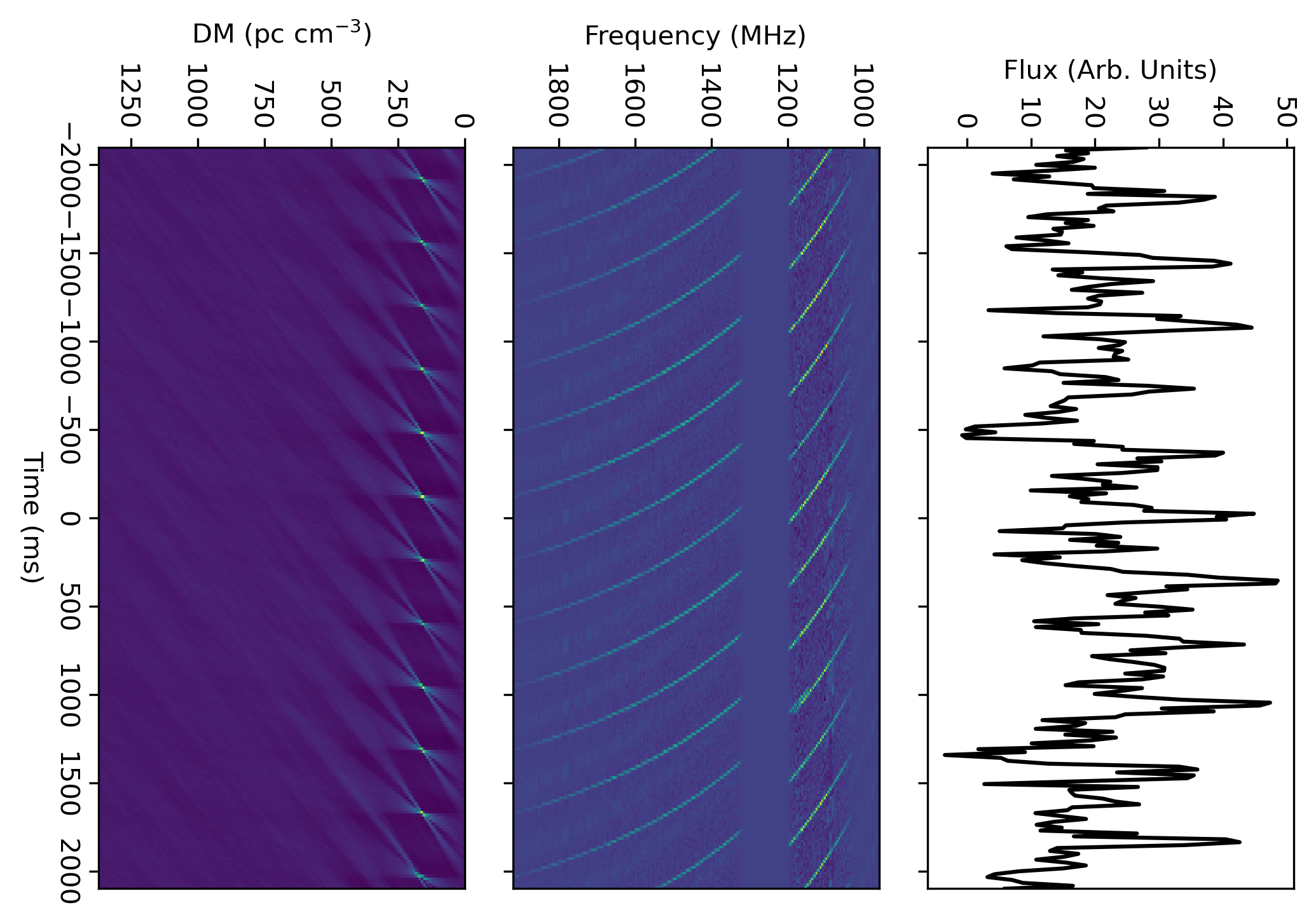}{0.45\textwidth}{}
         }
         \caption{
         DM–time candidate plots for \textsc{Heimdall} and the four filters for an 538-second observation of J1935+1606 that started on 2021-02-09 11:10:48 UTC. Left plots: candidates on the DM-time plane, where marker size is scaled to S/N. Right plots: Histograms of the left plot.
         The \textsc{your} candidate plot on the bottom right is a typical candidate from 
         the spike in Composite$_1$ at  DM $\approx$650~cm$^{3}$~pc.
    }
    \label{fig:dm-distrabution}
\end{figure}
\begin{figure*}
        \gridline{
         \includegraphics[width=0.46\textwidth]{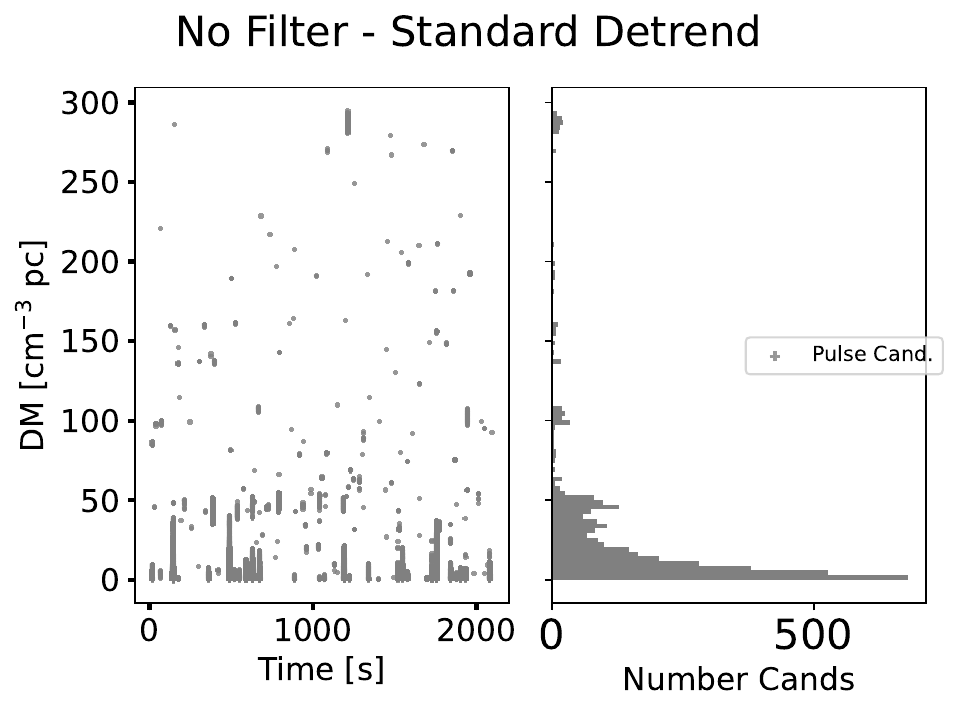}
         \includegraphics[width=0.46\textwidth]{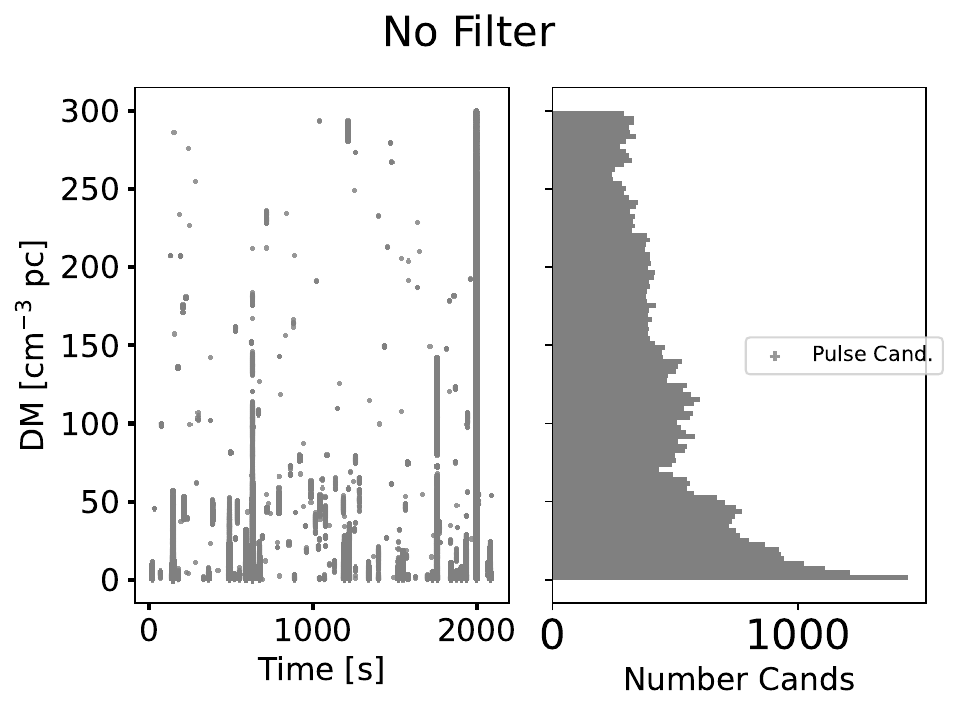}
     }
    \gridline{
         \includegraphics[width=0.46\textwidth]{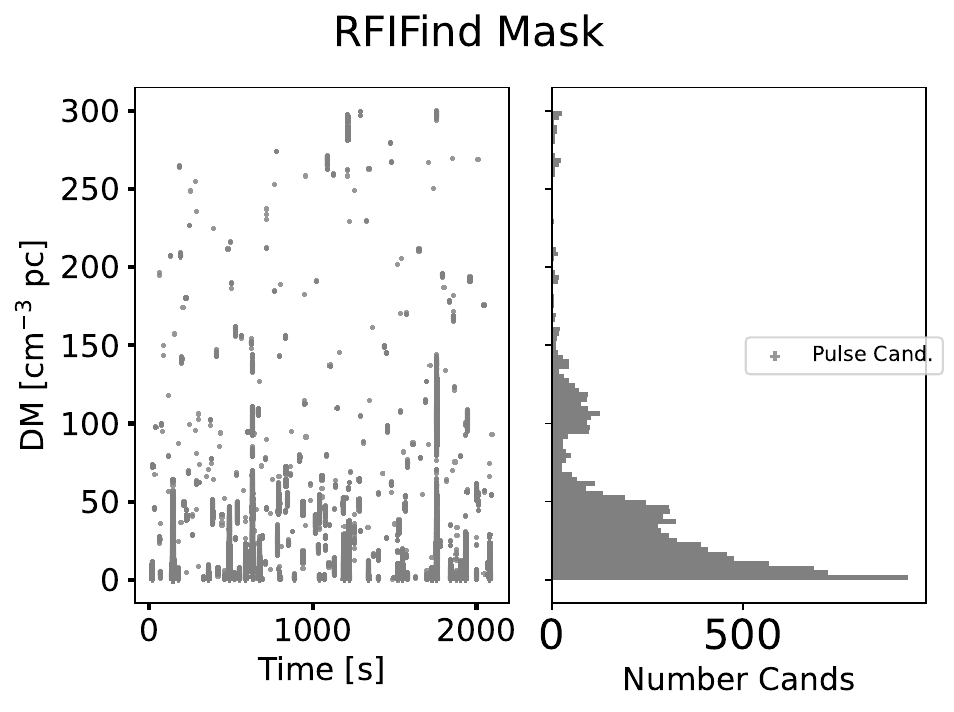}
         \includegraphics[width=0.46\textwidth]{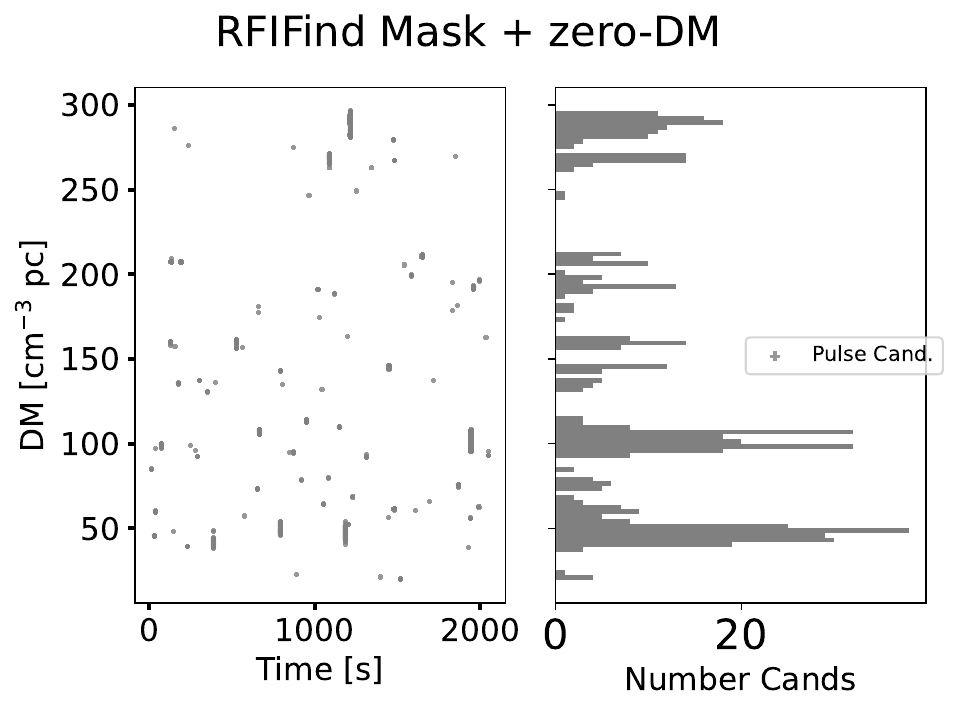}
     }
         \caption{
         DM–time candidate plots from the Parkes Multibeam Pulsar Search \citep{parkes-multibeam}. The 1-bit data 
         from this survey was used to by \citet{Eatough-2009} to show the effectiveness of the zero-DM filter.  We searched these 
         data with \textsc{Presto}'s \citep{Ransom-2011} \textsc{single\_pulse\_search.py} with a threshold of $6\sigma$. The top left 
         panel is when no RFI filtering is applied. This plot 
         looks significantly different from Fig.~7 of \citet{Eatough-2009}, which is also an observation of PSR J1738$-$3211 
         from the PMPS survey. This is because \textsc{single\_pulse\_search.py} detreneding removes a portion of the broadband RFI. 
         Applying the \textsc{--fast} flag (as recommended for 1-bit data), which uses a less accurate de-trend. We also changed the 
         de-trend length from the default of 1000 samples  to 8000 samples. These changes give us the top-right plot, which has similar 
         RFI structure to what is shown in  \citet{Eatough-2009}. We then processed the data with \textsc{Presto}'s RFI find mask, 
         computed with 1 second blocks. This is shown in the bottom-left plot. The number of detected pulses is reduced, however lines 
         that extend along the DM axis are still present. These lines show that broadband RFI is present. Finally, we apply the zero-DM 
         filter in addition to the aforementioned mask, producing the plot on the bottom right. The DM lines of candidates are gone, 
         there are also very few candidates at low DM. Together, these effects indicate that all the broadband interference has been 
         removed. A significant fraction of pulses are from the pulsar, which has a DM of 49.6~cm$^{-3}$~pc.
    }
    \label{fig:dm-distrabution-pmps}
\end{figure*}

Fig.~\ref{fig:dm-distrabution} shows the distribution DM–time space for a single observation.
The long lines along DM for a given frequency are due to broadband interference, likely of 
terrestrial origin. The top left panel of Fig.~\ref{fig:dm-distrabution} shows that \textsc{Heimdall}'s broadband RFI filter lets through significant zero-DM RFI. The middle right
panel of Fig.~\ref{fig:dm-distrabution} shoes that preprocessing with \citet{Eatough-2009}'s
zero-DM filter does not mitigate the situation. This contrasts with Fig.~\ref{fig:dm-distrabution-pmps}, in which  \citet{Eatough-2009}'s filter has completely removed
the broadband interference.

\begin{figure*}
     \centering
    \gridline{
         \fig{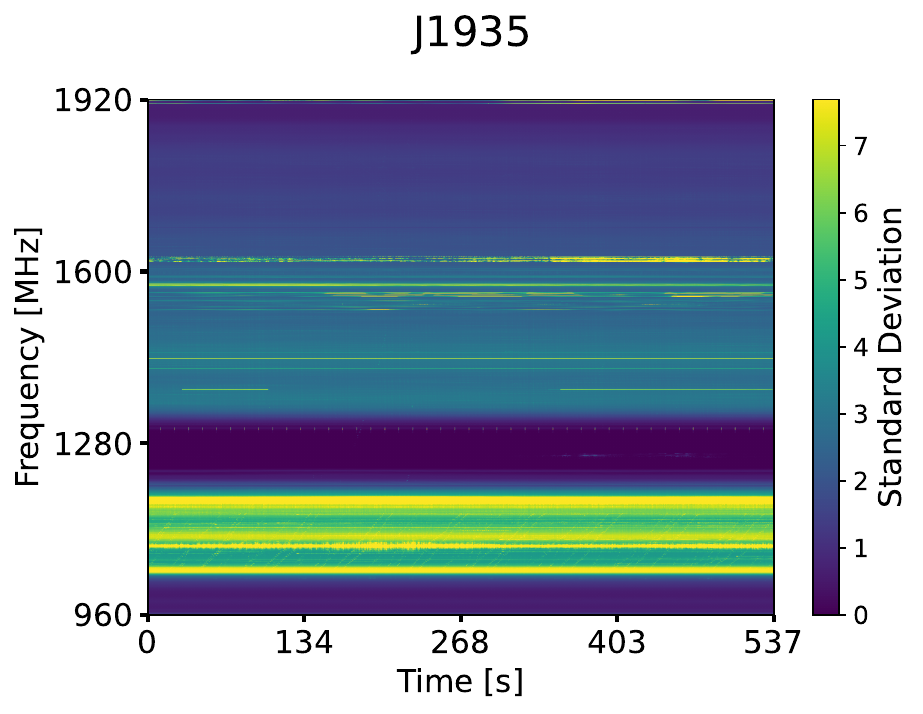}{0.45\textwidth}{}
         \fig{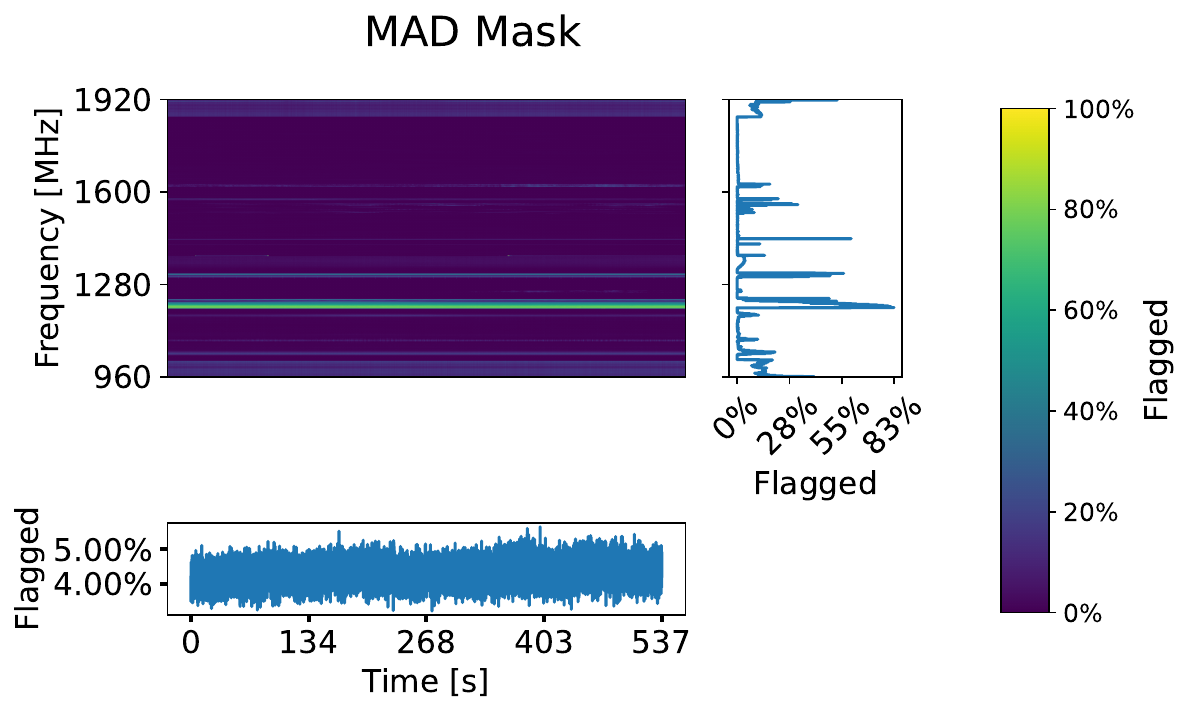}{0.45\textwidth}{}
     }
    \gridline{
         \fig{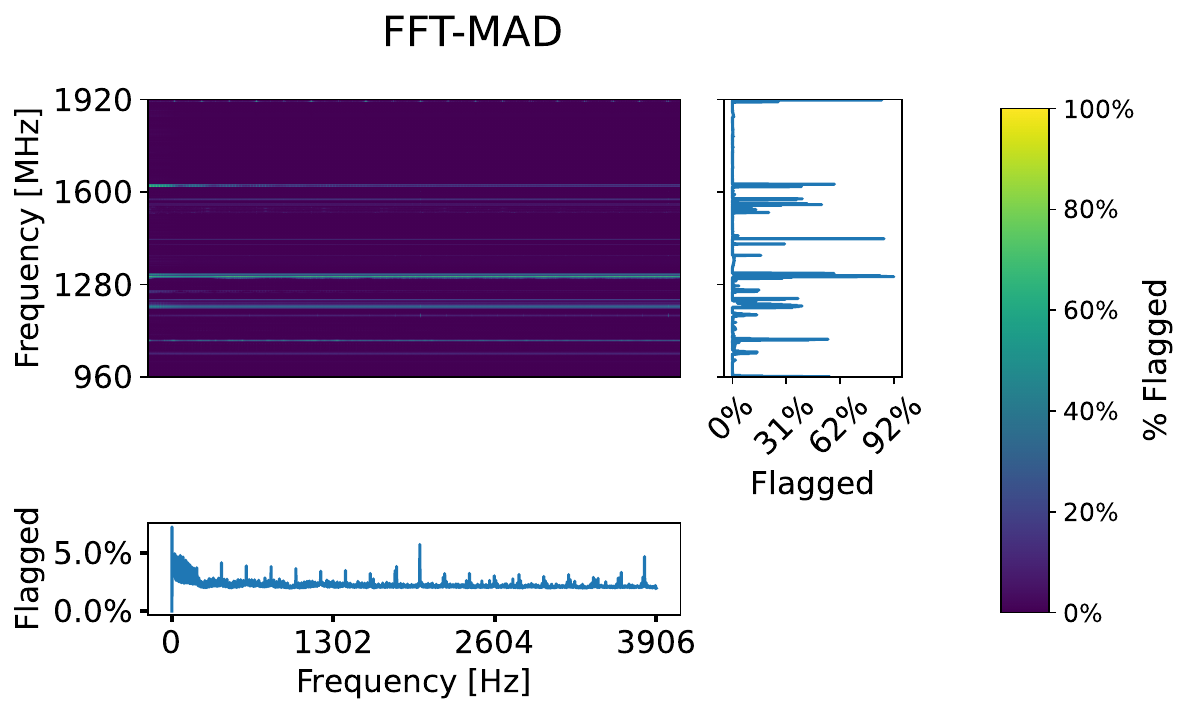}{0.45\textwidth}{}
         \fig{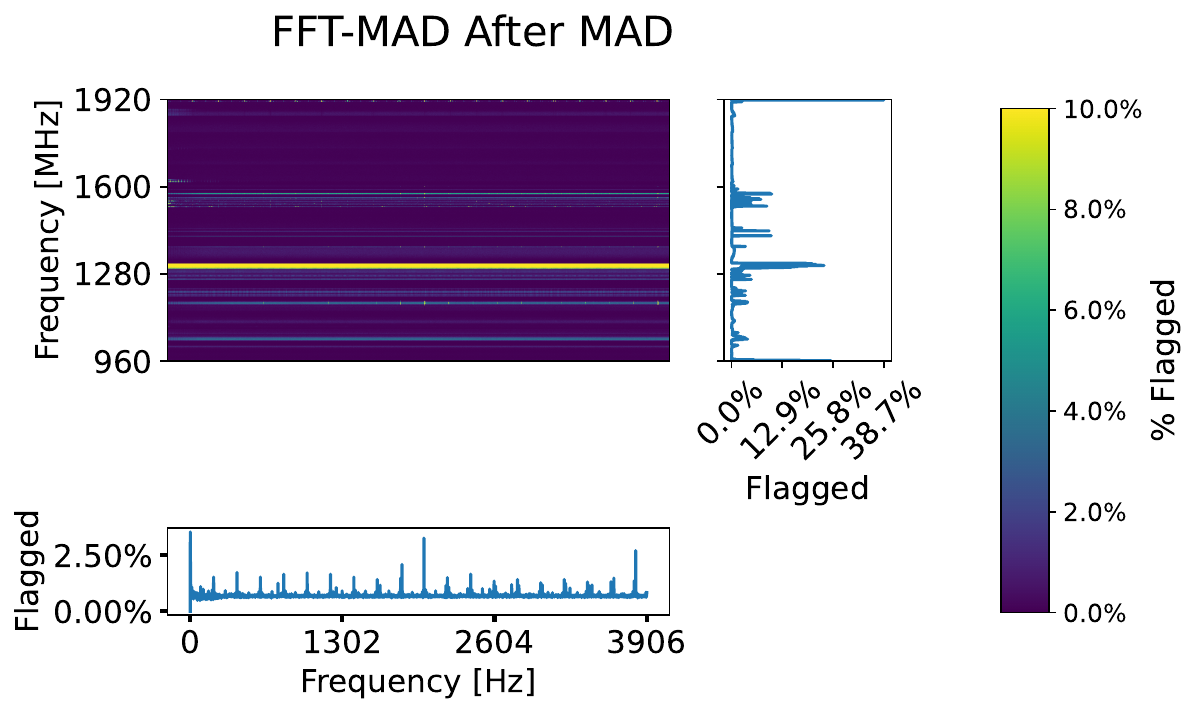}{0.45\textwidth}{}
     }
\gridline{
         \fig{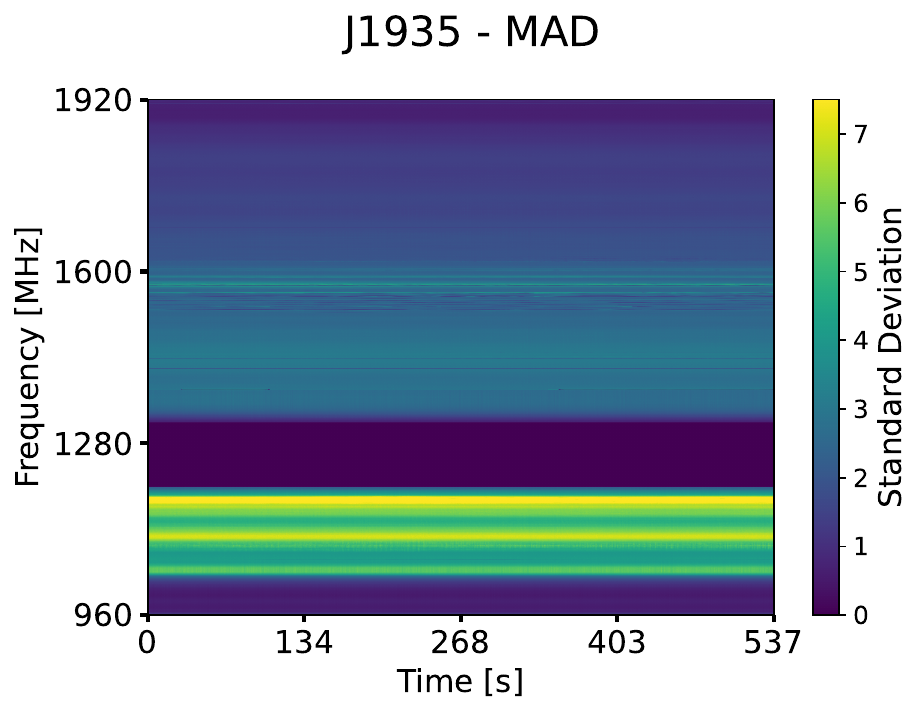}{0.32\textwidth}{}
         \fig{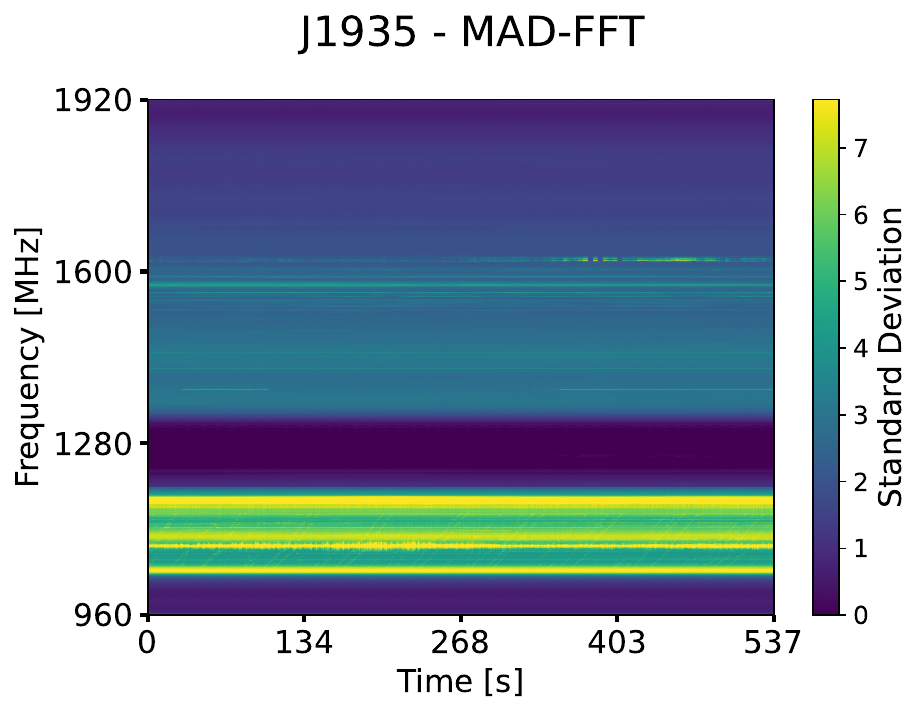}{0.32\textwidth}{}
         \fig{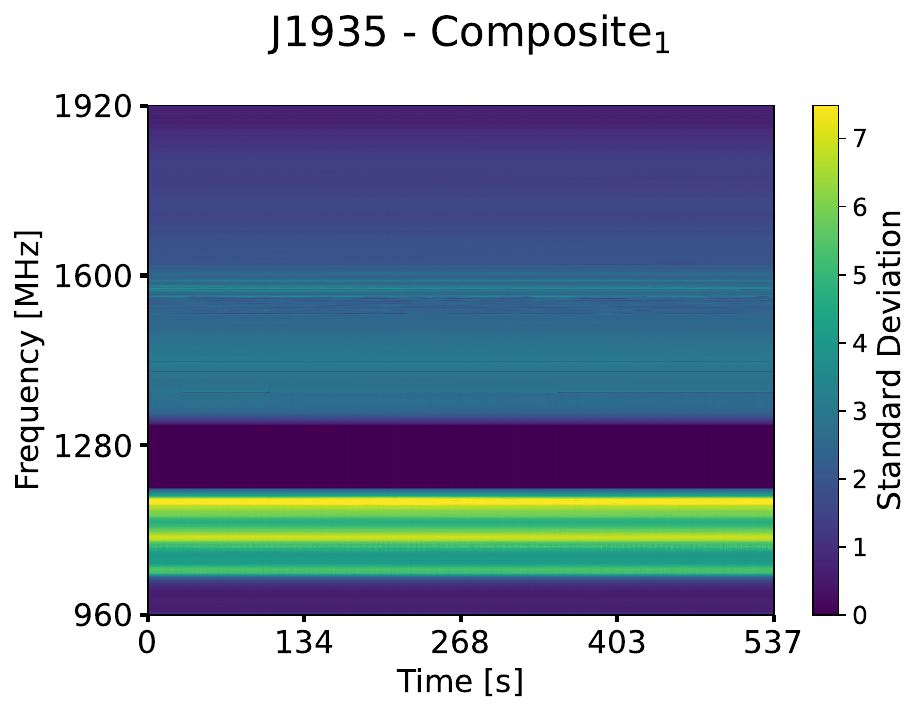}{0.32\textwidth}{}
     }
         
         \caption{
    The masks of the filters applied to the filterbank shown in Fig.~\ref{fig:dm-distrabution}.
    Top left: standard deviation of 64 times samples for a single channel. We see that the standard 
    deviation more clearly shows the location of RFI than intensity. Top right: the top-left subplot shows the mean mask of the MAD filter,
    taken over 64 time samples. The top-right subplot shows the fraction of each channel that is flagged. The bottom plot shows the fraction of each time block (or frequency) that is flagged. This makes the mask the same resolution as the plot on the left. 
    2nd left: mask of the FFT-MAD filter applied directly to the filterbank. 2nd right: FFT-MAD filter applied after the data has been cleaned with the MAD filter. The range of 
    mask has been clipped at 10\%, otherwise none of the mask detail is visible.
    Bottom rows: standard deviation of the filterbank cleaned with the three filters, the areas of high standard deviation have been excised. We see that the FFT-MAD filter misses substantial noise around 1600~MHz in the 
    last third of the observation.
    }
    \label{fig:filter_masks}
\end{figure*}

As we discussed in Section~\ref{sec:composite}, an increasingly complex 
RF environment coupled with high dynamic range receivers has limited the effectiveness
of older zero-DM filtering techniques. Our MAD filter, Fig.~\ref{fig:dm-distrabution}, 
does a good job at removing small, complex, RFI. The default of this filter is to remove
the median, and we see that this leads to a formidable broadband filter. This effectiveness
is promising.  As we approach Gaussian noise, the median, and mean should become the same.
The FFT-MAD filter further removes structure across the band-pass. 
After the final step of the composite$_1$ filter, zero DM-Subtraction, the bottom of 
Fig.~\ref{fig:dm-distrabution} shows we have
broadband RFI removal performance comparable to the bottom right of Fig.~\ref{fig:dm-distrabution-pmps} which is from 1 bit data. 
We investigated
the source of DM$ \approx 650$~cm${-3}$~pc candidates across all times, visible in the MAD
and Composite$_1$ filters, and show a typical candidate in Fig.~\ref{fig:dm-distrabution-pmps} bottom right. The band stop filter at frequencies 1200-1300 MHz removes a portion of the pulse. The search then finds a fake pulse
at when sections of two different pulses line up. This row of candidates is not 
from RFI, but an artifact of the pulse search. 
The masks for these data are shown in Fig.~\ref{fig:filter_masks}.

\begin{figure}
\centering
   \includegraphics[width=0.75\columnwidth]{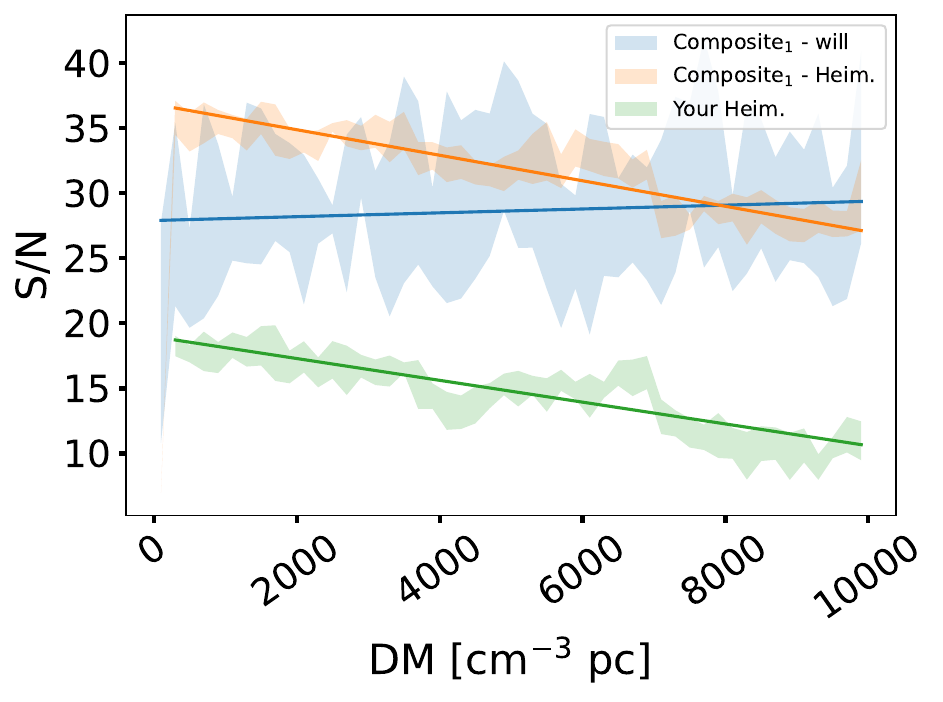}
    \caption{Filter response as a function of DM measured with  \textsc{Heimdall} \citep{Barsdell}. Composite$_1$ - will is the S/N after composite filtering.  \textsc{will} \citep{will} and Composite$_1$ - Heim. is the S/N as measured by \textsc{Heimdall}. Your Heim. is the measured S/N with \textsc{Heimdall} with a channel mask calculated by \textsc{your}. 
    }
    \label{fig:dm_response}
\end{figure}

\subsection{Pulsar time of arrival calculations}
\begin{figure}
\centering
   \includegraphics[width=0.75\columnwidth]{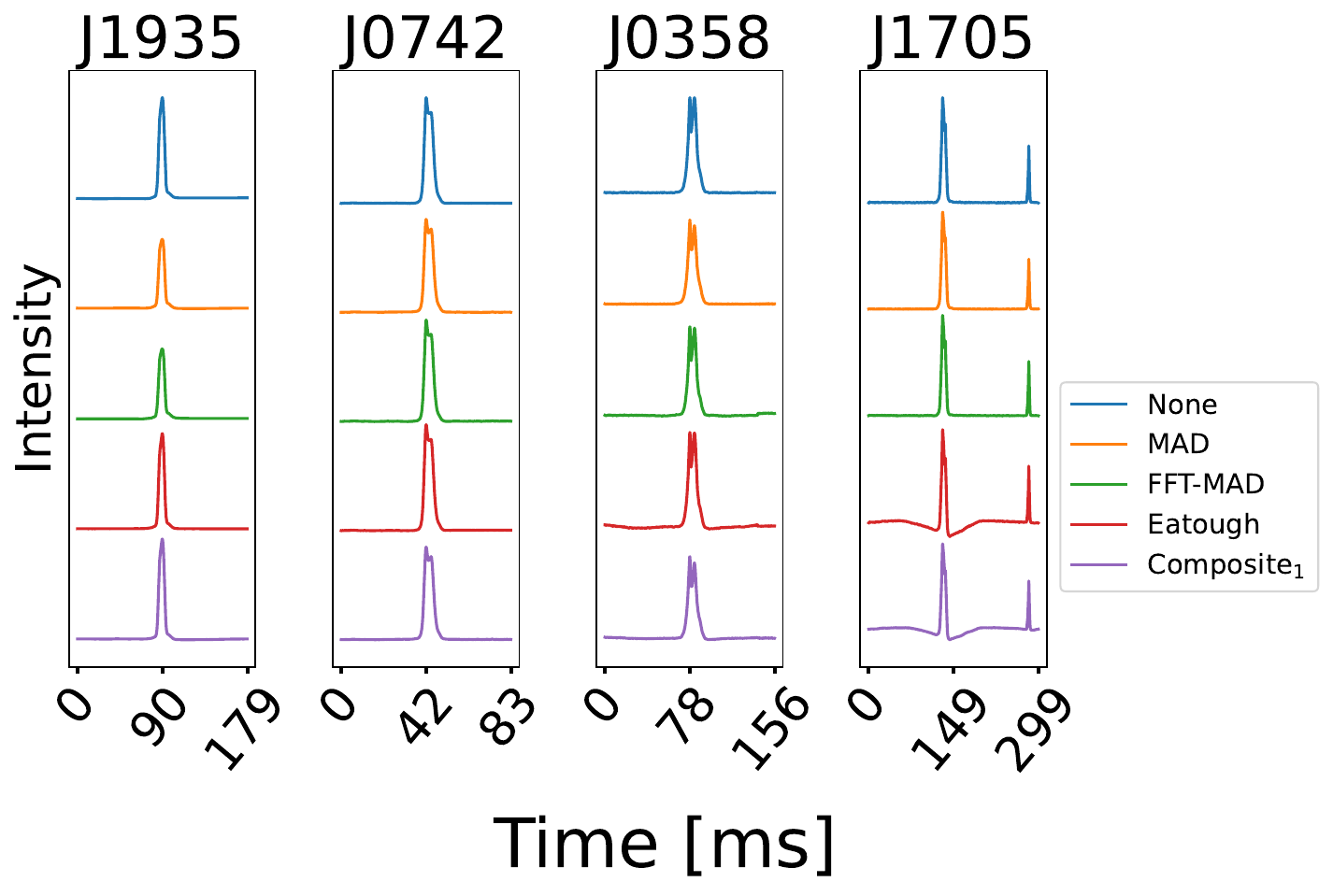}
  \caption{The pulse profiles from a single file for no filter, the component filters,
  and the Composite filters. Each was calculated with 1000 bins across the full period. These were extracted the folded profile from \textsc{Presto}'s .bestprof using
  \textsc{priwo} \citep{priwo}. For pulsars J1935$+$1616 and J0742$-$2822, we cut off half of the pulse phase to better show
  the pulse. Note that these profiles are as relayed as calculated by \textsc{prepfold} \citep{Ransom-2011}, and are not scaled.
  }
  \label{fig:pulse_profiles}
 \end{figure}

A data quality test we can perform is comparing times of arrival (TOAs) between the filters and 
\textsc{rfifind}  \citep{Ransom-2011}. If the filters are working as expected, the TOAs should be consistent. We will use
a Gaussian template. To find the Gaussian width as a fraction of pulse period, we folded our longest observation
of each pulsar using \textsc{prepfold} \citep{Ransom-2011}. Using \textsc{priwo} \citet{priwo} we extract the folded pulse profile in .bestprof. This profile is shown in Fig.~\ref{fig:pulse_profiles}. We then use \textsc{SciPy} \citep{2020SciPy-NMeth} to fit a Gaussian function. Using this width, we use \textsc{get\_TOAs.py} 
\citep{Ransom-2011} to generate TOAs for every five minutes of observation. We
also ran \textsc{rfifind} \citep{Ransom-2011} with zero-DM filtering turned on and using 4800 sample blocks. We  
report the mean differences between these TOAs and the ones obtained after application of our filters in Table~\ref{table:toas}.
In Table~\ref{table:toas}, we see that the TOAs generally agree well  with each other. The largest
differences in mean TOA relative to uncertainty are in J0358+5413 with the MAD and Composite filters.
The high-pass filter increases the uncertainty of J1705$-$1906 TOAs. Fig.~\ref{fig:pulse_profiles} shows the dip 
in the pulse profile as described in \citet{Eatough-2009}. The agreement between TOAs and visual similarity 
in pulse profiles show the filters behave for folded pulses.

\begin{deluxetable}{ l|cccc} 
 	\centering
  \label{table:toas}
 	\caption{Mean \textsc{rfifind} TOA minus the filter TOA and uncertainty in $\mu$s. 
 	}
  \tablehead{\colhead{} & \colhead{J1935$+$1616}         & \colhead{J0742$-$2822}           & \colhead{J0358$+$5413}        & \colhead{J1705$-$1906}
  }
 		\startdata
        \# TOAs         & 12               & 3                 & 7                & 11             \\
        \hline
  		\textsc{Presto} &      $\pm 88.1$  &        $\pm14.5$  &       $\pm 54.0$ &      $\pm 307$ \\
  		MAD             & $76   \pm 111$   &  $-29.5 \pm 17.0$ & $157.9 \pm 72.6$ & $-10   \pm 160$ \\
  	  FFT-MAD         & $41.1 \pm 90.1$  & $3.6    \pm 14.4$ & $-16.4 \pm 48.3$ & $-28  \pm 162$ \\
  	    Eatough	        & $36.5 \pm 89.7$  & $2.5    \pm 14.6$ & $-37.4 \pm 69.1$ & $83   \pm 299$ \\
        Composite$_1$       & $75   \pm 106$   & $-30.0  \pm 17.6$ & $171.6 \pm 76.4$ & $109  \pm 286$ \\
 		\enddata
\end{deluxetable}

A final test we perform is injecting the same pulse at a range of different DMs. We
take a 134-second chuck of dynamic spectra following the pulse shown in Fig.~\ref{fig:composite}.
A search on this cut revealed no bright pulses, but the large amount of broadband RFI remained. 
We use a $6\times10^4$ sample pulse with a time sigma of 50 milliseconds, $\tau=20$,
center frequency of 1440~MHz, frequency sigma of 650~MHz, flat spectral index, and three scintles. 
We injected 50 pulses between DMs in the range 100---9900~cm$^3$~pc, randomly adding between 0 and 1.05~s to where the pulse was injected. We repeated this process at 4 times separated by 16.4 seconds. The mean of these runs is denoted by the
line, and the standard deviation is shown by the error bars in Fig.~\ref{fig:dm_response}. For comparison, we also used the RFI mitigation in \textsc{your} \citep{your} for both \textsc{heimdall} and candidate creation. \textsc{your} flags 
spectral kurtosis and excess channel brightness. We see that the Composite filter has a significantly higher
S/N. Additionally, the Composite$_1$ missed 7 out of 200 pules while \textsc{your\_heimdall.py}, missed 11. Most of the missed pules were at  DM$=$100~cm$^{-3}$~pc. The downward slop in both S/Ns measured by \textsc{Heimdall} \citep{Barsdell} is due to adaptive time scrunching. Because intra-channel DM smearing becomes 
significant at higher DMs, \citet{Barsdell} reduces the time resolution by factors of two.
Turing off time scrunching, this effect largely goes away. The \textsc{Heimdall} slope remains slightly negative, while \textsc{will} shows a  positive slope. We defer tests across pulse width and pulse brightness for a future study.

\section{Performance}\label{sec:performance}
\begin{figure}
\centering
   \includegraphics[width=0.75\columnwidth]{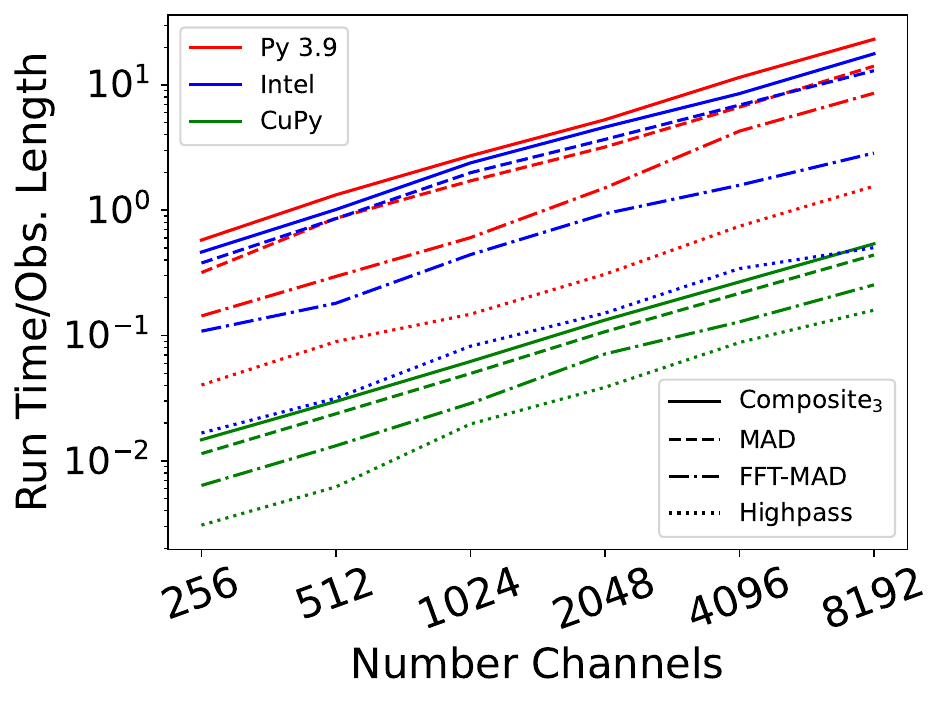}
    \caption{Filter run times for 256 $\mu$sec sampled 8-bit data for a 10-minute observation. 
    We test three \textsc{Python} backends, Anaconda Python, Intel Distribution for Python and \textsc{CuPy}. These tests were run on dual Intel Xeon Silver 4116 and NVIDIA RTX 3080 Ti. I/O is from a NVMe SSD.
    }
    \label{fig:run_times}
\end{figure}
We used \textsc{fake} \citep{sigproc} to create 10 minute observations with between 256 and 8192 channels. Fig.~\ref{fig:run_times} shows the filter run time normalized by the length of the observation. 
This shows that with GPU acceleration, we can maintain super real-time
filtering to at least 8192 channels. We also see that the \textsc{CuPy} acceleration decreases the run time by a factor of 30 over base \textsc{Python}. Fig.~\ref{fig:run_times} also shows that the Intel Distribution for
Python can provide considerable speedup when taking FFTs, which are shows improvements in the FFT-MAD and 
high-pass filters.

The majority of the compute time for the composite filter is spent on time domain MAD (80\%), while the FFT-MAD (17\%)
and the high-pass (1.5\%) takes significantly less. The remainder of the time is for data handling and I/O.
The MAD filter is more computationally expensive because it carries out multiple de-trendings, 
which requires calculating the medians along the time and frequency axis. By leveraging GPUs, the medians can be calculated massively in parallel. 

There are several steps we could take for better performance, which would be necessary for multiple beams or for
real-time CPU processing. We could replace the first iteration of the MAD filter with a Gaussianity test, such
as a Kurtosis filter. This would save both the MAD calculation and the initial de-trending, making this filter almost 
twice as fast. MAD is expensive because it requires two partitions of the data, one to find the median and another
to find the median of deviations. We could use the interquartile range which requires the 
partition in three places (20th, 75th for the range and 50th percentile for the replacement value) which tends to be faster than the double partition needed for MAD. A third consideration is we calculate the MAD along subbands on the time axis,
by taking the median of these medians, we can more cheaply calculate the medians for the time deterend. 
Another alternative to approximate the median for eight bit and below data are by using histograms.
This method is used by \citet{pulsartools}. 
A combination of these above in a higher performance language would greatly reduce the computational requirements. 








\section{Conclusions}
\label{sec:conclude}
We have investigated three RFI filters in this work that have advantages to their 
earlier implementations. These filters work in different domains, leading us to the conclusion that they would work in concert to remove a wide variety of RFI.
Understanding the transfer function on true pulses is vital to determine the true characteristics of the pulses.
We investigated the effects of the filters in three ways. First we inject synthetic pulses into
Gaussian noise. In the absence of RFI, an ideal filter would flag no data in this situation.
We see that the MAD filter 
removes very bright pulses. We also see that the high-pass filter removes power from low DM pulses. The words of  \citet{Fridman} remain true, ``There is no universal method of RFI mitigation.''

Next, we injected pulses onto an actual observation. This allowed us to see the effects of the 
filters on a wide variety of pulses. We see that the MAD and FFT-MAD filters greatly reduce the 
noise level, which is further reduced by the high-pass. We use two statistical tests to show that filtering can allow the recovered S/Ns to better reflect the intrinsic pulse energies. 

We filtered four pulses that range over brightness and DMs. This experiment indicates that there is no ideal universal filter. The optimal filtering depends on the RFI present, the 
characteristics of the pulse, and what we want to know about the pulse. The composite
filter has a robust to RFI overall response and can be used in a wide range of situations to greatly
reduce the noise level of an observation. As a data integrity check,  we derived compared TOAs for each pulsar to those derived from cleaning with \textsc{rfifind}. The TOAs agreed within the uncertainties. 
We ran a blind single pulse search over the pulsar observations. This search shows the FFT-MAD, and particularly the 
MAD and Composite filters, can increase the number of single pulses found and reduce the number of RFI candidates. These filters can also significantly increase
the typical pulse S/N. 

The filters presented here are not a unique combination, adding or removing a test may be useful in 
different radio frequency environments and for different sources. These filters can also be 
improved in both RFI excision and performance. For example, a better MAD filtering can be done 
at the de-dispersion stage. As the de-dispersed dynamic spectra are collapsed, the MAD filter can be iterative 
applied. Filtering along the frequency axis, sum nearby frequencies, and then filter again. This filter is more sensitive to large structures, but is costly. 



Our goal was to extend the set of  millisecond cadence available to work over 
a wide range of telescopes and sources,  while still  removing the minimal amount of signal, and run faster than real-time on GPUs. We believe that these filters meet these criteria. We designed a composite filter
that is effective for many types of RFI and is particularly suited for removing  broadband RFI. 
These filters are available at \url{https://github.com/josephwkania/jess} and 
can clean both filterbank and psrfits files. We hope that others find them useful in their pulsar and FRB searches. 

\section*{Acknowledgements}
JWK and KB are supported by a National Science Foundation (NSF) award 2006548.
DRL acknowledges support for GREENBURST from the NSF through awards AAG--1616042 and AAG--2406570. KB and DRL acknowledge the Directorate for Mathematical and Physical Sciences, Division of Astronomical Sciences, NSF Award Number 2307581.
The Green Bank Observatory is a facility of the National Science Foundation operated under cooperative agreement by Associated Universities, Inc.
We would like to thank Devansh Agarwal for his help
with GREENBURST data collection as well as Loren Anderson, Tapasi Ghosh, Emmanuel Fonseca, and the anonymous reviewer for their comments that have improved 
the clarity and discussion presented in this paper. 

\software{\textsc{CuPy} \citep[][]{cupy}, \textsc{jess} \citep[][]{jess}, \textsc{Matplotlib} \citep[][]{matplotlib}, \textsc{NumPy} \citep[][]{numpy}, \textsc{priwo} \citet{priwo}, Presto \citep[][]{Ransom-2011}, \citep[][]{python3},
\textsc{Rich} \citep[][]{rich}, \textsc{scikit-learn} \citep[][]{scikit-learn, sklearn_api}, \textsc{SciPy} \citep[][]{2020SciPy-NMeth}, \textsc{Sigproc} \citep[][]{sigproc}, \textsc{Sphinx} \citep[][]{sphinx}, and \textsc{Your} \citep[][]{your}}.
\section*{Data Availability}

 The output for the real and synthetic pulse searches are available 
 at Zenodo \citep{Zenodo} (\citealt{kania_2022_sql};\dataset[DOI: 10.5281/zenodo.6487651]{https://doi.org/10.5281/zenodo.6487651}, the observation of the four pulsars 
 are available at PSR J1935+1616 (\citealt{kania_2025_J1935};\dataset[DOI: 10.5281/zenodo.16737518]{https://doi.org/10.5281/zenodo.16737518}),  J0742-2822 (\citealt{kania_2025_J0742};\dataset[DOI: 10.5281/zenodo.16538733]{https://doi.org/10.5281/zenodo.16538733}), PSR J0358+5413(\citealt{kania_2025_J0358};\dataset[DOI: 10.5281/zenodo.16293240]{https://doi.org/10.5281/zenodo.16293240}), J1705-1906 (\citealt{kania_2025_J1705};\dataset[DOI: 10.5281/zenodo.16539186]{https://doi.org/10.5281/zenodo.16539186}). The corresponding author can make GREENBURST filterbanks used for the injection test
 available on reasonable request.

\bibliography{Millisecond_Cadence}{}
\bibliographystyle{aasjournal}



\end{document}